\documentstyle[preprint,aps]{revtex}
\tighten
\draft
\widetext
\input epsf
\preprint{YITP-SB-01-43, NYU-TH/01/08/11}
\bigskip
\bigskip

\begin{document}
\title{Orientiworld}
\medskip
\author{Zurab Kakushadze\footnote{E-mail: 
zurab@insti.physics.sunysb.edu}}
\bigskip
\address{C.N. Yang Institute for Theoretical Physics\\ 
State University of New York, Stony Brook, NY 11794\\
and\\
Department of Physics, New York University, New York, NY 10003}
\date{September 6, 2001}
\bigskip
\medskip
\maketitle

\begin{abstract} 
{}We propose a consistent
string theory framework for embedding brane world scenarios with
infinite-volume extra dimensions. In this framework the Standard Model fields
are localized on D3-branes sitting on top of an orientifold 3-plane. The
transverse 6-dimensional space is a non-compact 
orbifold or a more general conifold. The 4-dimensional gravity on
D3-branes is reproduced due to the 4-dimensional 
Einstein-Hilbert term induced at the quantum level. The
orientifold 3-plane plays a crucial role, 
in particular, without it
the D3-brane world-volume theories would be conformal due to the tadpole
cancellation. We
point out that in some cases the 4-dimensional Planck scale 
is controlled by the size of certain relevant (as opposed to marginal) 
orbifold blow-ups. We can then have a scenario
with the desirable 4-dimensional Planck scale, the string scale of 
order TeV, and the cross-over to 10-dimensional gravity 
around the present Hubble size. 
We discuss some general features as well as 
concrete models in this ``Orientiworld'' framework,
including those with D7-branes. We point out that the D7-brane gauge 
symmetry at the quantum level becomes part of the 4-dimensional gauge symmetry.
We present an ${\cal N}=1$ supersymmetric model with 3 chiral
generations of quarks and leptons, where the original gauge group 
(which contains an $SU(6)$ subgroup) 
can be Higgsed to obtain a Pati-Salam model with 3 chiral
generations, the Pati-Salam Higgs fields required to break the gauge group
further to that of the Standard Model, as well as the desired electroweak
Higgs doublets. The superpotential in this model is such that
we have precisely one heavy (top-like) generation.

\end{abstract}

\section{Introduction}

{}One motivation for considering extra dimensions comes from 
superstring theory (or M-theory), which is believed to be a consistent theory
of quantum gravity. Thus, the critical number of space-time dimensions in
superstring theory is 10 (while M-theory is an 11-dimensional theory). 
However, in order to model the real world with 
critical string theory (or M-theory), one must
address the question of why the extra dimensions have not been observed.
One way to make extra dimensions consistent with observation is to assume
that they are compact with small enough volume. If the Standard Model gauge
and matter fields propagate in such extra dimensions (as is the case in, say,
weakly coupled heterotic string theory), then their linear sizes
should not be larger than about inverse TeV \cite{anto}. On the other hand, 
in the Brane World scenario the Standard Model gauge and matter fields
are assumed to be localized on branes (or an intersection thereof), 
while gravity lives in a larger dimensional bulk of space-time. Such a scenario
with compact extra dimensions can, for instance, 
be embedded in superstring theory via
Type I$^\prime$ 
compactifications. Then the extra directions transverse to the branes
can be as large as about a tenth of a millimeter \cite{TeV}. 

{}Originally considering compact (or, more generally, finite
volume) extra dimensions was motivated by the requirement that at the distance
scales for which gravity has been measured one should reproduce 4-dimensional
gravity. However, in \cite{DGP,DG} a remarkable proposal was set forward.
In particular, according to \cite{DG} 4-dimensional gravity can be
reproduced on a 3-brane in infinite-volume bulk (with 6 or larger space-time
dimensions) up to ultra-{\em large} distance scales. Thus, in these scenarios
gravity is almost completely localized on a brane (which is 
almost $\delta$-function-like) with ultra-light modes penetrating into the 
bulk. As was explained in \cite{DG}, this dramatic modification of gravity
in higher codimension models with infinite volume extra dimensions
is due to the Einstein-Hilbert term on the brane, which
is induced via loops of non-conformal brane matter \cite{DGP,DG}.

{}In this paper we describe an explicit string theory framework where the
Standard Model gauge and matter fields are localized on (a collection of)
D3-branes embedded in infinite-volume extra space. In particular, we consider
{\em unoriented}
Type IIB backgrounds in the presence of some number of 
D3-branes embedded in an orbifolded 
space-time. The D3-brane world-volume
theory in this framework is non-conformal (at least for some backgrounds with
at most ${\cal N}=1$ supersymmetry). At the quantum level we
have the Einstein-Hilbert term induced on the branes, which leads to almost
complete localization of gravity on the D3-branes. In particular, up to an
ultra-large cross-over distance scale the gravitational interactions of the
Standard Model fields localized on D3-branes 
are described by 4-dimensional laws of gravity. 

{}One of the key ingredients in this framework is that
we have an unoriented background. Thus, if we consider orientable Type IIB 
backgrounds (that is, those without orientifold planes) in the presence of
some number of D3-branes embedded in an orbifolded space-time, 
then, as was explained in \cite{BKV}, the finiteness
of the theory (that is, tadpole cancellation conditions) 
implies that the D3-brane world-volume theory is necessarily 
conformal\footnote{More precisely, theories with at least ${\cal N}=1$ 
supersymmetry are conformal \cite{BKV}. As to non-supersymmetric
theories, to decouple bulk tachyons one considers the large $N$ limit, where
they too are conformal.}.
However, as was originally pointed out in \cite{orient}, we can obtain
{\em non-conformal} gauge theories if we consider orientifolds of such 
Type IIB backgrounds. In particular, this is the case for backgrounds with
at most ${\cal N}=1$ supersymmetry. Given the importance of orientifold planes
in our framework, we refer to it as ``Orientiworld''.

{}The orientiworld framework appears to have a rich structure for model
building. In particular, since the extra dimensions have infinite volume, the
number of D3-branes is arbitrary. Moreover, the number of allowed orbifold
groups is infinite. Thus, {\em a priori} the orbifold group can be an
arbitrary\footnote{More precisely, there is a mild restriction on allowed
orbifold groups if we require modular invariance of the closed string sector
in the corresponding
oriented Type IIB background.} 
subgroup of $Spin(6)$, or, if we require ${\cal N}=1$ supersymmetry
to avoid bulk tachyons, of $SU(3)$.
To obtain a finite string background, we
still must impose {\em twisted} tadpole cancellation conditions. However, 
twisted tadpoles must also be canceled in compact Type IIB orientifolds. Then
the number of consistent solutions of the latter 
type is rather limited \cite{class}
as we can only have a finite number of D3-branes, and, moreover, the number of
allowed orbifold groups is also finite as they must act crystallographically
on the compact space. On the other hand, as we already mentioned, in the
orientiworld framework the number of consistent solutions is {\em infinite},
which is encouraging for phenomenologically oriented model building.

{}This richness of the orientiworld framework can be exploited to construct
various models for phenomenological applications. We discuss some examples
of orientiworld models in detail. In particular, we give an example with
three chiral generations. In this model, which is
${\cal N}=1$ supersymmetric, the gauge group contains a phenomenologically
appealing $SU(6)$ subgroup\footnote{A compact version of this model, which is
the first orientifold model with three generations of
quarks and leptons, was originally 
constructed in \cite{3gen} (and a non-compact version was subsequently 
constructed in \cite{orient1}).}.
Albeit the gauge group in this model contains
an $SU(6)$ subgroup, 
as was pointed out in \cite{3gen}, it is not a grand unified model
as we have no Higgs field in an appropriate higher dimensional
representation. However, as was pointed out in \cite{KT}, 
one can still break the gauge group down to that of the Standard Model.
We discuss this model in detail, and point out the importance of the
gauge subgroups coming from D7-branes, which are also present in this model.
In particular, the D7-brane gauge symmetry becomes part of the 
four-dimensional gauge symmetry via the mechanism of \cite{DGS}. In fact, 
after a sequence of Higgsing, the gauge group in this model can be broken
down to the Pati-Salam gauge group. The charged matter contains 3 chiral
generations of quarks and leptons, the electroweak Higgs doublets, as well as
the Pati-Salam Higgs fields, which break the Pati-Salam gauge symmetry down
to that of the Standard Model. We also point out a pleasant bonus in this 
model - the superpotential is such that we have precisely one heavy (top-like)
generation. 

{}Given recent developments in brane world scenarios with infinite-volume
extra dimensions, it is reasonable to expect that the orientiworld approach,
which provides an explicit framework for consistently embedding such scenarios
in string theory, would have to be further developed. In particular, orbifold
examples are only a part of interesting starting Type IIB backgrounds. Thus,
one can consider more general backgrounds such as conifolds. It therefore
appears that orientiworld might provide a fruitful arena for embedding various
brane world scenarios in string theory. In fact, in the following we point
out that relevant (as opposed to marginal) blow-ups of the orbifold 
singularities, which turn the space transverse to the D3-branes into a 
conifold, play an important role in obtaining 4-dimensional gravity on 
D3-branes.

{}The rest of this paper is organized as follows. In section II we discuss
the orientiworld framework. In section III we discuss some general features
of the orientiworld scenario, in particular, we discuss how 4-dimensional
gravity is reproduced on D3-branes. In section IV we discuss how
the four-dimensional Einstein-Hilbert term arises in the orientiworld context. 
In section V we discuss concrete examples of orientiworld models.
In section VI we discuss some phenomenological aspects of 
the aforementioned 3-family supersymmetric model containing an 
$SU(6)$ subgroup.
We conclude with final remarks in section VII. Some details are relegated to
appendices.

\section{Framework}

{}In this section we discuss the orientiworld framework. First we describe the
underlying oriented Type IIB orbifold backgrounds. We then consider their
orientifolds. This illustrates why introduction of orientifold planes makes
a big difference, in particular, why in the orientiworld framework we can 
construct non-conformal gauge theories. Parts of 
our discussion here will closely 
follow \cite{BKV,orient}.

\subsection{Oriented Backgrounds}

{}Consider Type IIB string theory with $N$ parallel coincident D3-branes where
the space transverse to the
D-branes is ${\cal M}={\bf R}^6/\Gamma$. The orbifold group
$\Gamma= \left\{ g_a \mid a=1,\dots,|\Gamma| \right\}$ ($g_1=1$)
must be a finite discrete subgroup of $Spin(6)$ (it can be a subgroup of
$Spin(6)$ and not $SO(6)$ as we are dealing with a theory containing fermions).
If $\Gamma\subset SU(3)$ ($SU(2)$), we have
${\cal N}=1$ (${\cal N}=2$) unbroken supersymmetry,
and ${\cal N}=0$, otherwise.

{}Let us confine our attention to the cases where type IIB on ${\cal M}$ 
is a modular invariant 
theory\footnote{This is always the case if $\Gamma\subset SU(3)$. For 
non-supersymmetric cases this is also true provided that
$\not\exists{\bf Z}_2\subset\Gamma$. If $\exists{\bf Z}_2\subset\Gamma$,
then modular invariance requires that the set of points in ${\bf R}^6$
fixed under the ${\bf Z}_2$ twist has real dimension 2.}. The action of the
orbifold on
the coordinates $X_i$ ($i=1,\dots,6$) on ${\cal M}$ can be described
in terms of $SO(6)$ matrices:
$g_a:X_i\rightarrow \sum_j (g_a)_{ij} X_j$. (The action of $g_a$ on the
world-sheet superpartners of $X_i$ is the same.)
We also need to specify
the action of the orbifold group on the Chan-Paton charges carried by the
D3-branes. It is described by $N\times N$ matrices $\gamma_a$ that
form a representation of $\Gamma$. Note that $\gamma_1$ is an identity
matrix and ${\mbox {Tr}}(\gamma_1)=N$.

{}The D-brane sector of the theory is described by an {\it oriented} open
string theory. In particular, the world-sheet expansion corresponds
to summing over oriented Riemann surfaces with arbitrary genus $g$ and
arbitrary number of boundaries $b$, where the boundaries of the world-sheet 
correspond to the D3-branes. For example, 
consider the one-loop vacuum amplitude with no handles and two
boundaries ($g=0$, $b=2$). The
corresponding graph is an annulus whose boundaries lie on D3-branes.
The one-loop partition function in the
light-cone gauge is given by
\begin{equation}\label{partition}
 {\cal Z}={1\over 2|\Gamma|}\sum_a
 {\rm Tr}  \left( g_a (1-(-1)^F)
 e^{-2\pi tL_0}
 \right)~,
\end{equation}
where $F$ is the fermion number operator, $t$ is the real modular parameter
of the annulus, and the trace includes sum over the Chan-Paton factors.

{}The orbifold group $\Gamma$ acts on both ends of the open string.
The action of $g_a\in \Gamma$ on Chan-Paton charges is given by
$\gamma_a\otimes \gamma_a$. Therefore,
the individual terms in the sum in (\ref{partition})
have the following form:
\begin{equation}
 \left({\mbox {Tr}}(\gamma_a)\right)^2 {\cal Z}_a~,
\end{equation}
where ${\cal Z}_a$ are characters
corresponding to the world-sheet degrees of freedom. The ``untwisted''
character
${\cal Z}_1$ is the same as in the ${\cal N}=4$ theory for which
$\Gamma=\{1\}$. The information about the fact that the orbifold theory
has reduced supersymmetry is encoded in the ``twisted'' characters
${\cal Z}_a$, $a\not=1$.

{}In \cite{BKV} it was shown that the one-loop massless 
(and, in non-supersymmetric 
cases, tachyonic) tadpole cancellation conditions (that is, finiteness
of the corresponding string backgrounds) require that
\begin{equation}\label{tadpole}
 {\mbox {Tr}}(\gamma_a)=0~~~\forall a\not=1~.
\end{equation}
That is, all twisted characters ${\cal Z}_a$ contain divergences (which are
all massless tadpoles in the supersymmetric cases, while in non-supersymmetric
cases we also have tachyonic divergences).
In \cite{BKV} 
it was also shown that this condition implies that the Chan-Paton 
matrices $\gamma_a$
form an $n$-fold copy of the {\em regular} representation of $\Gamma$. 
The regular representation decomposes into a direct sum of all irreducible
representations ${\bf r}_i$ of $\Gamma$ with degeneracy factors
$n_i=|{\bf r}_i|$. The gauge group is ($N_i\equiv nn_i$)
\begin{equation}
 G=\otimes_i U(N_i)~. 
\end{equation}
The matter consists of Weyl fermions and scalars transforming in
bifundamentals $({\bf N}_i,{\overline {\bf N}}_j)$ 
(see \cite{LNV} for details). The overall center-of-mass $U(1)$ is free - 
matter fields are not charged under this $U(1)$. All other $U(1)$'s, however,
are running. If $\Gamma\subset SO(3)$, then these $U(1)$'s are anomaly free.
If, however, $\Gamma \subset SU(3)$ but $\Gamma\not\subset SO(3)$, then
some of such $U(1)$ factors are actually anomalous (in particular, we have
$U(1)_k SU(N_l)^2$ mixed anomalies), and are broken at the tree-level
via a generalized Green-Schwarz mechanism \cite{IRU,Poppitz}. Apart from the
$U(1)$ factors, all supersymmetric
theories of this type are conformal \cite{BKV}, in particular, their 
non-Abelian parts are conformal\footnote{We will discuss the $U(1)$ factors
in the next subsection.}.
On the other hand, in
non-supersymmetric cases to decouple bulk tachyons one considers the large
$N$ limit\footnote{In this limit we take the string coupling $g_s\rightarrow 
0$ together with $N\rightarrow \infty$ while keeping $Ng_s$ fixed.}. 
All (supersymmetric as well as non-supersymmetric) gauge theories
of this type are conformal in the large $N$ limit \cite{BKV}. The key reason
for this conformal property is the tadpole cancellation condition 
(\ref{tadpole}), which, as was explained in \cite{BKV}, implies that all
planar diagrams reduce to those of the parent ${\cal N}=4$ theory, which is
conformal. In the large $N$ limit non-planar diagrams are suppressed by
powers of $1/N$. For finite $N$ conformality of the non-Abelian parts of the
gauge theories follows for ${\cal N}=2$ cases (as ${\cal N}=2$ gauge
theories perturbatively are not renormalized beyond one loop), 
and can also be argued for
${\cal N}=1$ cases \cite{BKV}. 
In non-supersymmetric cases, however, bulk tachyons
prevent one from considering finite $N$ cases\footnote{In the large $N$
limit the bulk tachyons are harmless as the string coupling $g_s$
goes to zero.}.

\subsection{Unoriented Backgrounds}

{}Let us now consider a generalization of the above setup by including
orientifold planes. In the following we will mostly be interested in finite $N$
theories, so let us focus on theories with at least ${\cal N}=1$ unbroken
supersymmetry. Thus,
consider Type IIB string theory on ${\cal M}={\bf C}^3/\Gamma$ where
$\Gamma\subset SU(3)$. Consider the $\Omega J$ orientifold of this 
theory, where $\Omega$ is the world-sheet parity reversal, and $J$ 
is a ${\bf Z}_2$ element ($J^2=1$) acting on the complex coordinates $z_i$
($i=1,2,3$) on ${\bf C}^3$ such that the set of points in 
${\bf C}^3$ fixed under 
the action of $J$ has real dimension $\Delta=0$ or $4$. 

{}If $\Delta=0$ then we have an orientifold 3-plane. If $\Gamma$ has
a ${\bf Z}_2$ subgroup, then we also have an orientifold 7-plane.
If $\Delta=4$ then we have an orientifold 7-plane. We may also have
an orientifold 3-plane depending on whether $\Gamma$ has an appropriate
${\bf Z}_2$ subgroup. Regardless of whether we have an orientifold 3-plane,
we can {\em a priori} introduce an arbitrary number of 
D3-branes\footnote{In general, codimension-3 and higher objects (that is
D-branes and orientifold planes) do not introduce untwisted tadpoles.}. 
On the other hand, if we have an orientifold 7-plane we must 
introduce 8 of the corresponding D7-branes to cancel the 
Ramond-Ramond charge 
appropriately. (The number 8 of D7-branes is required by the corresponding 
untwisted tadpole cancellation conditions.) 

{}We need to specify the action of $\Gamma$ on the Chan-Paton factors
corresponding to the D3- and D7-branes (if the latter are present, which is
the case if we have an orientifold 7-plane). Just as in the previous 
subsection, 
these are given by Chan-Paton matrices which we collectively refer to
as $\gamma^\mu_a$, where the superscript $\mu$ refers to the corresponding
D3- or D7-branes. Note that ${\mbox{Tr}}(\gamma^\mu_1)=n^\mu$ where 
$n^\mu$ is the number of D-branes labelled by $\mu$. 

{} At the 
one-loop level there are three different sources for massless tadpoles:
the Klein bottle, annulus and M{\"o}bius strip amplitudes depicted in Fig.1. 
The Klein bottle amplitude corresponds to the contribution of unoriented 
closed strings into the one-loop vacuum diagram. It can be alternatively viewed
as a tree-level closed string amplitude where the closed strings propagate 
between two cross-caps. The latter are the (coherent Type IIB) states 
that describe the familiar orientifold planes. The annulus amplitude 
corresponds to the contribution of open strings stretched between two D-branes
into the 
one-loop vacuum amplitude. It can also be viewed as a tree-channel closed
string amplitude where the closed strings propagate between two D-branes.
Finally, the M{\"o}bius strip amplitude corresponds to the contribution of 
unoriented
open strings into the one-loop vacuum diagram. It can be viewed 
as a tree-channel closed 
string amplitude where the closed strings propagate between a 
D-brane and an orientifold plane.

{}Note that there are no Chan-Paton matrices associated with the Klein bottle 
amplitude since it 
corresponds to closed strings propagating between two cross-caps
which do not carry Chan-Paton charges. The M{\"o}bius strip has 
only one boundary.
This implies that the individual terms (corresponding to twists 
$g_a\in \Gamma$)
in the M{\"o}bius strip amplitude are proportional to 
${\mbox{Tr}}(\gamma^\mu_a)$. The annulus
amplitude is the same (up to an overall factor of $1/2$ due 
to the orientation reversal projection) 
as in the oriented case discussed in the previous subsection. 
Thus, the individual terms (corresponding to twists $g_a\in \Gamma$)
in the annulus amplitude are proportional to 
${\mbox{Tr}}(\gamma^\mu_a){\mbox{Tr}}(\gamma^\nu_a)$. 
The tadpoles can therefore be written as
\begin{equation}\label{KMA}
 \sum_a \left( K_a +\sum_\mu M^\mu_a {\mbox{Tr}}(\gamma^\mu_a)+
                         \sum_{\mu,\nu}A^{\mu\nu}_a
 {\mbox{Tr}}(\gamma^\mu_a){\mbox{Tr}}(\gamma^\nu_a)
 \right)~.
\end{equation}   
Here the terms with $K_a$, $M^\mu_a$ and $A^{\mu\nu}_a$ correspond to the
contributions of the Klein bottle, M{\"o}bius strip and annulus 
amplitudes, respectively.
In fact, the factorization property of string theory implies 
that the Klein bottle amplitude
should factorize into two cross-caps connected via a long thin tube. 
The M{\"o}bius strip
amplitude should factorize into a cross-cap and a disc connected via a 
long thin tube.
Similarly, the annulus amplitude should factorize into two discs connected 
via a long thin tube.
These factorizations are depicted in Fig.2. 
The implication of this for the tadpoles is that
they too factorize into a sum of perfect squares
\begin{equation}\label{tad}
 \sum_a \left(B_a+\sum_\mu C^\mu_a {\mbox{Tr}}(\gamma^\mu_a)\right)^2~,
\end{equation}  
where $B_a^2=K_a$, $2B_a C^\mu_a=M^\mu_a$ and $C^\mu_a C^\nu_a=A^{\mu\nu}_a$.
Thus, the {\em twisted} tadpole cancellation conditions read:
\begin{equation}
 B_a+\sum_\mu C^\mu_a {\mbox{Tr}}(\gamma^\mu_a)=0~,~~~a\not=1~.
\end{equation}
These should be contrasted with (\ref{tadpole}) in the oriented case.
In particular, since in certain cases some $B_a\not=0$, {\em a priori}
there is no
reason why the corresponding D3-brane world-volume gauge theories should
be conformal. That is, the presence of orientifold planes can indeed give rise
to non-conformal gauge theories.

\subsection{Orientiworld Models}

{}Let us see what kind of orientiworld models we can have. In particular, let
us discuss what kind of orientiworld models are non-conformal. For definiteness
let us focus on the cases where we do have an orientifold 3-plane. If there are
no orientifold 7-planes (that is, if $\Gamma$ does not contain a ${\bf Z}_2$
element), then the orientifold projection $\Omega$ can be either of the $SO$
or the $Sp$ type: the corresponding orientifold 3-plane is referred to as
O3$^-$ or O3$^+$, respectively. That is, before orbifolding, if we place
$2N$ D3-branes on top of the O3$^-$-plane (O3$^+$-plane), we have the ${\cal
N}=4$ super-Yang-Mills theory with the $SO(2N)$ ($Sp(2N)$) gauge 
group\footnote{Note that we can also place $2N+1$ D3-branes on top of
the O3$^-$-plane to obtain the $SO(2N+1)$ gauge group.}.
(We are using the convention where $Sp(2N)$ has rank $N$.) After the orbifold
projections the 33 (that is, the D3-brane) gauge group is a subgroup
of $SO(2N)$ ($Sp(2N)$), which can contain $U(N_k)$ factors as well as 
$SO$ ($Sp$) subgroups. The 33 matter can contain bifundamentals in any of these
subgroups as well as rank-2 antisymmetric (symmetric) representations in
the unitary subgroups. Next, if we have an O7-plane, the orientifold 
projection $\Omega$ 
must always be of the $SO$ type on the D7-branes - this is required
by the tadpole cancellation condition. This, in particular, implies that the
33 and 77 matter cannot contain rank-2 symmetric representations. Note that
we also have 37 matter in bifundamentals of the 33 and 77 gauge groups.

{}Let us start with ${\cal N}=2$ theories of this 
type\footnote{Orientifolds of non-compact Type IIB orbifolds were discussed
in detail in \cite{orient,FS,PU}.}. 
All such theories are
conformal (more precisely, their non-Abelian parts are - here we are
ignoring the $U(1)$ factors). 
The reason why can be understood as follows. First, note that
we can focus on the one-loop level as we have ${\cal N}=2$ supersymmetry.
It can be shown that cross-cap contributions to the twisted tadpoles 
corresponding to the twists of even order are all vanishing (see, for
instance, \cite{orient,PU} and references therein). This implies that the
corresponding $\gamma_a^\mu$ are traceless, and, for the same reason as
in the previous subsection, do not spoil conformality. Let us therefore
focus on orbifold groups $\Gamma\subset SU(2)$ that do not contain elements
of even order. These are given by ${\bf Z}_M$ subgroups of $SU(2)$, where
$M$ is odd. Let us see why all such theories are conformal (we can then
understand why theories with even order twists are also conformal). In these
theories the cross-cap contributions into twisted tadpoles, which are given by
the factors $B_a$, are non-vanishing, so neither are the corresponding 
${\rm Tr}(\gamma_a)$ (note that in these cases we have no D7-branes, so
we do not need to put an additional index $\mu$ on the Chan-Paton matrices).
However, the tadpole cancellation condition implies that
\begin{equation}\label{tadpole1}
 B_a+C_a{\rm Tr}(\gamma_a)=0~,~~~a\not=1~. 
\end{equation}
This, as we will see in a moment, implies that these ${\cal N}=2$ theories
are conformal.

{}To see this, let us consider renormalization of the gauge coupling for a
{\em non-Abelian} subgroup of the D3-brane gauge group. This can be deduced
from a 2-point function, where two gauge bosons corresponding to this subgroup
are attached to the {\em same} boundary in the corresponding one-loop graph.
It has to be the same boundary because these are non-Abelian gauge bosons,
so the Chan-Paton matrices $\lambda_r$, $r=1,2$, corresponding to the external
lines are traceless: ${\rm Tr}(\lambda_r)=0$. It then follows that
${\rm Tr}(\lambda_r\gamma_a)=0$ as well for by definition $\lambda_r$ are 
invariant under the orbifold group action as $\lambda_r$ correspond to the
gauge bosons of the gauge group left unbroken by the orbifold (in particular,
note that $\lambda_r$ commute with $\gamma_a$). This then implies that 
attaching only one such external line to a given boundary would produce a
vanishing amplitude. Note that this is correct for the non-Abelian gauge 
bosons, but does not hold for $U(1)$ factors. This, as we will see, 
is precisely the reason why $U(1)$'s can still run in these theories.

{}Thus, the {\em ultra-violet}
divergence structure for the one-loop amplitude containing two
external lines corresponding to non-Abelian gauge bosons is given by:
\begin{equation}\label{coupling}
 \sum_a \Big[2B_a C^\prime_a {\rm Tr}(\lambda_1\lambda_2\gamma_a)
 +2C_a C^\prime_a{\rm Tr}(\gamma_a){\rm Tr}(\lambda_1\lambda_2\gamma_a)
 \Big]~,
\end{equation}
where the first term comes from the M{\"o}bius strip, while the second
term comes from the annulus. In the second term the factor of 2 is due to
the two different ways we can attach external gauge bosons to the two
boundaries. The factors $B_a$ and $C_a$ are the same as before - they 
correspond to the {\em massless} 
twisted closed string states coupling to the cross-caps and 
D3-branes, respectively, in the factorization limit depicted in
Fig.2. The factors $C_a^\prime$ correspond to
these twisted closed string states coupling to the 
D3-branes with two external lines attached to them (in particular, 
$C_a^\prime$ are different from $C_a$). Note that (\ref{coupling}) vanishes
due to the tadpole cancellation conditions (\ref{tadpole1}). That is, we have
{\em no} ultra-violet divergences in the one-loop 2-point functions involving
non-Abelian gauge bosons.

{}The fact that there are no ultra-violet divergences in the non-Abelian
2-point functions does not by itself imply that the non-Abelian gauge couplings
do not run. Indeed, we must show that there are no {\em infra-red} divergences
in the corresponding 2-point functions either. Remarkably enough, however,
the ultra-violet and infra-red divergences in these ${\cal N}=2$ models are 
related to each other, in fact, they are {\em identical}. The reason for this
is the following. Consider the {\em loop-channel} one-loop 2-point function
involving non-Abelian or $U(1)$ gauge bosons (in the latter case we must 
also include the annulus amplitude with the external lines attached to two
different boundaries). Such an amplitude receives contributions only from
{\em massless} open string states running in the loop. This is due to the
${\cal N}=2$ supersymmetry - only BPS states (that is, those in short
supermultiplets) can contribute to the 
renormalization of the gauge couplings, and the massive open string excitations
in these backgrounds are all non-BPS \cite{BF,DL}. That is, there are no
massive string threshold contributions to the gauge couplings, and whatever
their field-theoretic logarithmic running, it continues above the string scale.
In other words, the massive open string states do {\em not} provide an 
ultra-violet cut-off for the gauge coupling running in these backgrounds.
This then implies that, if we had logarithmic infra-red divergences in the 
non-Abelian gauge couplings (that is, if the latter did run), we would have to
have the corresponding logarithmic ultra-violet divergences as well. However,
as we saw above, there are no ultra-violet divergences in the non-Abelian gauge
couplings. This then implies that there are no infra-red divergences either,
and the non-Abelian gauge couplings do {\em not} run - the non-Abelian
parts of these ${\cal N}=2$ gauge theories are, therefore, conformal.

{}What about the $U(1)$ gauge couplings? First, note that $U(1)$'s in these
backgrounds are non-anomalous - these theories are ${\cal N}=2$ supersymmetric,
and, therefore, non-chiral. So {\em a priori} they need not be broken. We can
then discuss the $U(1)$ gauge coupling running. In fact, all $U(1)$'s run as
there is always matter charged under them\footnote{Note that this is 
different from what happens in oriented backgrounds discussed in subsection
A, where we have one center-of-mass $U(1)$ which is free. In the oriented
case the gauge group before orbifolding is $U(N)$, and the free 
center-of-mass $U(1)$ is inherited from this $U(N)$ (while all the other 
$U(1)$'s come from breaking the $SU(N)$ part of $U(N)$, so there is always
matter charged under these $U(1)$'s, which, therefore, run). On the other hand,
in the unoriented case the gauge group before orbifolding is $SO(2N)$ or 
$Sp(2N)$ (as we have already mentioned, it can also be $SO(2N+1)$), and
in this case $U(1)$'s come from breaking these gauge groups, so there is
always matter charged under them.}. This implies that we have 
logarithmic {\em infra-red}
divergences in the {\em loop-channel} one-loop 2-point function involving
$U(1)$ gauge bosons. According to the above argument, this then implies  
that we must also have the corresponding logarithmic {\em ultra-violet}
divergences. And we do. These arise in the annulus amplitude with two
external $U(1)$ lines attached to {\em different} boundaries. In fact,
the corresponding infra-red divergence also arises in this amplitude, that is,
we have no infra-red divergences in the M{\"o}bius strip or annulus amplitudes
with two external $U(1)$ lines attached to the same boundary. On the other 
hand, the aforementioned ultra-violet divergence in the loop channel can be
understood as an infra-red divergence in the corresponding tree channel, which
arises in the factorization limit depicted in Fig.2 due to the exchange of
{\em massless} twisted states (note that 
massive states do not give rise to infra-red
divergences in the tree channel). Such a massless twisted state is a 
Ramond-Ramond twisted 2-form, call it $C$, which on top of the four dimensions
along the D3-brane world-volumes also propagates in the fixed point locus
of the corresponding twist in ${\bf C}^3$, which has real dimension 2. This
twisted 2-form has a coupling to a $U(1)$ gauge field strength of the following
form \cite{DM,DGM}:
\begin{equation}
 \int_{\rm D3} C\wedge F~.
\end{equation}
Exchanging this 2-form in the annulus amplitude with two $U(1)$ gauge bosons
attached to different boundaries then leads to a logarithmic divergence. 
The reason why this divergence is logarithmic is simply due to the fact that 
the massless 2-dimensional Euclidean propagator is logarithmic. 
We will come back to the running of the $U(1)$'s in a minute. 
However, let us first
understand the loop-channel infra-red divergences (in particular,
the absence thereof
for non-Abelian gauge couplings) in the tree-channel language.

{}The key point here is that, as we have already mentioned, in these 
backgrounds the ultra-violet and infra-red divergences in any given
one-loop 2-point function are {\em identical}. Thus, 
the ultra-violet and infra-red divergences in the annulus amplitude with
two non-Abelian gauge bosons attached to the same boundary are identical.
Similarly, they are identical in the M{\"o}bius strip amplitude with
two external lines corresponding to non-Abelian gauge bosons. Also, 
the ultra-violet and infra-red divergences are identical 
in the annulus amplitude with
two $U(1)$ gauge bosons attached to the same boundary. 
Similarly, they are identical in the M{\"o}bius strip amplitude with
two external lines corresponding to $U(1)$ gauge bosons. Finally, 
the ultra-violet and infra-red divergences are identical 
in the annulus amplitude with
two $U(1)$ gauge bosons attached to different boundaries. Note that for
twists with ${\rm Tr}(\gamma_a)\not=0$ these individual divergences
do {\em not} vanish. In fact, the divergences in the annulus amplitude with
two $U(1)$ gauge bosons attached to different boundaries do not
vanish even for the twists with ${\rm Tr}(\gamma_a)=0$.

{}To understand how this is possible, that is, how
individual infra-red and ultra-violet divergences in these backgrounds
can be identical,  
let us go to the tree channel. The
loop-channel ultra-violet divergences are then translated into the
tree-channel
infra-red divergences arising due to the exchange of massless twisted states
in the factorization limit depicted in Fig.2. These divergences are logarithmic
as the massless 2-dimensional Euclidean propagator is logarithmic. Note that
the exchange of massless twisted states in the tree channel also gives rise
to logarithmic ultra-violet divergences, which in the loop channel translate
into logarithmic infra-red divergences. In fact, precisely because 
the massless 2-dimensional Euclidean propagator is logarithmic, these 
tree-channel ultra-violet divergences (that is, the loop-channel infra-red
divergences) have the {\em exact same} structure as the 
tree-channel infra-red divergences (that is, the loop-channel ultra-violet
divergences). In particular, this structure is also given by (\ref{coupling}).
What about the exchange of massive twisted states in the tree channel? {\em 
A priori} these could also contribute into ultra-violet tree-channel
divergences. But they do not. This is also due to the 
${\cal N}=2$ supersymmetry.
In other words, in ${\cal N}=2$ backgrounds the couplings of massive twisted
states to the massless brane fields are such that their exchange in the 
tree channel does {\em not} give rise to tree-channel 
ultra-violet divergences. Indeed, such divergences, if non-vanishing, 
could only be logarithmic,
which would imply that the loop-channel infra-red divergences would be
different from the loop-channel ultra-violet divergences. This, however, is
not possible as the massive open string states do {\em not} affect the gauge
coupling running in the ultra-violet due to the ${\cal N}=2$ supersymmetry as
we discussed above\footnote{Here we note that 
these observations are the key ingredient of
the Brane-Bulk Duality recently discussed in \cite{radu,CIKL} in the context
of non-conformal D3-brane gauge theories from oriented Type IIB backgrounds
with uncanceled codimension-2 twisted tadpoles.}.   

{}Let us now go back to the running of the $U(1)$'s, which is 
due to the exchange of the massless twisted 2-form fields
$C$. In particular, we have the corresponding logarithmic ultra-violet 
divergences, which normally 
are not expected to be present in consistent string 
backgrounds\footnote{In particular, note that we have Landau poles for the
$U(1)$'s.}. Here we would like to discuss a resolution of this point.
The idea here is based on the above observation that the ultra-violet and 
infra-red divergences in these backgrounds go hand-by-hand. In particular,
we might expect that, if we remove the infra-red divergences, then this
should automatically take care of the ultra-violet divergences as well. 
This then gives us a hint that, if the $U(1)$'s become massive, then we 
should no longer have ultra-violet divergences. In fact, in a moment we 
will discuss a mechanism for this. The upshot of this mechanism is that these
$U(1)$'s are actually {\em not}
local gauge symmetries but {\em global} ones, and
we should treat them as such. This mechanism is essentially 
the same in the oriented as well as unoriented
backgrounds, so for simplicity we will discuss it in one of the oriented
${\cal N}=2$ backgrounds discussed in subsection A.

{}Thus, consider the oriented background with $\Gamma={\bf Z}_2$, 
where the generator $R$ of $\Gamma$ acts as follows on the complex
coordinates $z_1,z_2,z_3$ on ${\bf C}^3$:
\begin{equation}
 R:z_1\rightarrow z_1~,~~~R:z_{2,3}\rightarrow -z_{2,3}~.
\end{equation}
We must satisfy the condition ${\rm Tr}(\gamma_R)=0$. Thus, we can take
\begin{equation}
 \gamma_R={\rm diag}(I_N,-I_N)~,
\end{equation} 
where $I_N$ is the $N\times N$ unit matrix. The resulting D3-brane world-volume
theory is the ${\cal N}=2$ supersymmetric $U(N)\otimes U(N)$ gauge theory
with matter hypermultiplets transforming in
\begin{equation}
 ({\bf N},{\overline{\bf N}})(+1,-1)~,~~~
 ({\overline {\bf N}},{{\bf N}})(-1,+1)~,
\end{equation}
where the $U(1)$ charges are given in the parenthesis. Note that the diagonal
$U(1)$, which we will refer to as $U(1)_+$, is the free center-of-mass $U(1)$,
while the anti-diagonal $U(1)$, which we will refer to as $U(1)_-$, runs.

{}Consider the Chern-Simons 
part of the low energy action containing the twisted 2-form 
$C_R$ \cite{Doug,DM}:
\begin{equation}
 S_{\rm \small{CS}}={1\over 2\pi\alpha^\prime}\int_{\rm D3} C_R 
 \wedge {\rm Tr}\left(\gamma_R ~e^{2\pi\alpha^\prime{\cal F}}\right)~,
\end{equation} 
where ${\cal F}$ is the ($(2N)\times (2N)$ matrix valued) D3-brane gauge field
strength\footnote{Note that we are ignoring the NS-NS 2-form, which is not
important for our discussion here. At any rate, it does not
appear in the unoriented backgrounds.}. Let us decompose ${\cal F}$ into
($N\times N$ matrix valued) field strengths ${\cal F}_1$ and
${\cal F}_2$ for the two $U(N)$ factors. We then have
\begin{equation}\label{CS1}
 S_{\rm \small{CS}}={1\over 2\pi\alpha^\prime}\int_{\rm D3} C_R 
 \wedge \left[{\rm Tr}\left(e^{2\pi\alpha^\prime{\cal F}_1}\right)-
 {\rm Tr}\left(e^{2\pi\alpha^\prime{\cal F}_2}\right)\right]~.
\end{equation}
The linear term in the field strengths is given by:
\begin{equation}\label{CS}
 \int_{\rm D3} C_R \wedge F_-~,
\end{equation}
where
\begin{equation}
 F_-\equiv {\rm Tr}({\cal F}_1)-{\rm Tr}({\cal F}_2)
\end{equation}
is the (appropriately normalized) $U(1)_-$ field strength. In particular, note
that $C_R$ does not have such a coupling to the $U(1)_+$ field strength $F_+$.
This is consistent with the fact that the latter does not run.

{}We can consistently give a mass to the $U(1)_-$ gauge field as follows.
Let ${\overline C}_R$ be the pull-back of the 2-form $C_R$ onto the D3-brane
world-volume:
\begin{equation}
 {\overline C}^{\mu\nu}_R=\left.{\delta^\mu}_M {\delta^\nu}_N C^{MN}_R
 \right|_{\rm D3}~.
\end{equation}
Here $x^\mu$ are the coordinates along the D3-brane world-volume, while
$x^i$, $i=1,2$, are the real coordinates on the fixed point locus of the twist
$R$ ($z_1=x^1+ix^2$), and $x^M=(x^\mu,x^i)$. Note that the Chern-Simons 
coupling (\ref{CS}) is given by
\begin{equation}
 \int_{\rm D3} d^4x ~\epsilon_{\mu\nu\sigma\rho}{\overline C}^{\mu\nu}_R
 F_-^{\sigma\rho}~.
\end{equation}
Next, let ${\overline H}_R$ be the {\em four-dimensional} field 
strength of ${\overline C}_R$. That is,
\begin{equation}
 {\overline H}_R^{\mu\nu\sigma}=\partial^{\mu}{\overline C}^{\nu\sigma}_R+
 \partial^{\nu}{\overline C}^{\sigma\mu}_R+
 \partial^{\sigma}{\overline C}^{\mu\nu}_R~.
\end{equation}
Note that we can add a kinetic term for ${\overline C}_R$ in the D3-brane
world-volume, which is consistent with all symmetries of the system:
\begin{equation}\label{kinetic}
 -{L^2\over 12}\int_{\rm D3} {\overline H}_R^2~.
\end{equation}
Here $L$ has the dimension of length. The original action is restored in the
$L\rightarrow 0$ limit. In this limit, however, the $U(1)_-$ gauge boson
is infinitely heavy. Indeed, consider the action ($g_-$ is the 
$U(1)_-$ gauge coupling):
\begin{eqnarray}
 &&-{L^2\over 12}\int_{\rm D3} {\overline H}_R^2+\int_{\rm D3} {\overline
 C}_R \wedge F_- -{1\over 4g_-^2} \int_{\rm D3} F_-^2=\nonumber\\
 &&-{L^2\over 12}\int_{\rm D3} {\overline H}_R^2-{2\over 3}
 \int_{\rm D3} {\overline H}_R \wedge A_- 
 -{1\over 4g_-^2} \int_{\rm D3} F_-^2~,
\end{eqnarray} 
where $A_-$ is the $U(1)_-$ gauge field. The mass of the $U(1)_-$ gauge
boson therefore is
\begin{equation}
 m^2_-={16 g_-^2\over L^2}~.
\end{equation}
Thus, for non-zero $L$ the $U(1)_-$ gauge symmetry is broken. However, we still
have a {\em global} $U(1)_-$ symmetry. 

{}Since $U(1)_-$ is heavy for non-vanishing $L$, we have no infra-red 
divergences. Let us, however, check that our expectation about the absence of
the ultra-violet divergences is actually correct. What we need to do is to
compute the correction to the $F_-^2$ kinetic term due to the tree-channel
exchange of the 2-form $C_R$. The relevant action reads:
\begin{equation}
 -{L^2\over 12}\int_{\rm D3} {\overline H}_R^2+\int_{\rm D3} {\overline
 C}_R \wedge F_- -{1\over 4g_-^2} \int_{\rm D3} F_-^2-
 {1\over 12}\int_{{\rm D3}\times {\bf R}^2} H^2_R~,
\end{equation}
where $H_R$ is the field strength of the 2-form $C_R$ in the bulk,
which is the D3-brane world-volume times the fixed point locus of the
twist $R$. The computation of this tree-channel exchange is given
in Appendix A. As can be seen from this computation, we indeed do not have
logarithmic divergences in the $U(1)_-$ gauge coupling even if we take the
limit $L\rightarrow\infty$ where $m_-\rightarrow 0$. The physical 
interpretation of this is that $U(1)_-$ should {\em not} be interpreted as a
local gauge symmetry but as a {\em global} one. 
In particular, suppose we start from
the original background with a massless $U(1)_-$ gauge boson. Since this
$U(1)_-$ runs, we must introduce an ultra-violet cut-off $\Lambda$. Since the
2-form $C_R$ couples to the brane fields via (\ref{CS1}), already at the 
one-loop level a kinetic term (\ref{kinetic}) for $C_R$ will be generated
on the D3-branes.
At the one-loop order the relevant couplings in (\ref{CS1}) are those 
quadratic in the field strengths:
\begin{equation}
 \pi\alpha^\prime \int_{\rm D3} C_R\wedge\left[{\rm Tr}({\cal F}_1
 {\cal F}_1)-{\rm Tr}({\cal F}_2
 {\cal F}_2) \right]~.
\end{equation}
In fact, the coefficient $L^2$ in (\ref{kinetic}) is given by:
\begin{equation}
 L^2=b(\alpha^\prime)^2\Lambda^2~,
\end{equation} 
where $b$ is a dimensionless numerical coefficient. The computation of
Appendix A then shows that we do not have a logarithmic divergence in the
$U(1)_-$ gauge coupling even in the $\Lambda\rightarrow \infty$ limit.
This, as we have already mentioned, suggests that $U(1)_-$ should be treated
as a global symmetry and not as a local one. In fact, in Appendix A we 
point out that this is consistent with the case where the two extra dimensions
$x^i$ are compactified on a large 2-torus.

{}As we have already mentioned, the story in unoriented backgrounds is similar.
Once we interpret all $U(1)$'s as global symmetries, we are left with
${\cal N}=2$ superconformal non-Abelian gauge theories on D3-branes.
As should be clear from our discussions, 
the key reason why ${\cal N}=2$ theories are conformal lies in the
fact that all non-trivial twists have fixed point loci of real dimension 2.
In particular, the fact that the ultra-violet and infra-red divergences
in these backgrounds are identical is possible precisely because they
are identical in a massless 2-dimensional Euclidean 
propagator\footnote{As we discussed above, this {\em plus} 
${\cal N}=2$ supersymmetry are the key ingredients here - the latter ensures
that the massive states do not contribute.}.
This gives us a hint that all ${\cal N}=1$ theories with the orbifold groups
$\Gamma\subset SO(3)$ should also be conformal. Indeed, each individual twist
$g_a\in\Gamma$ in these cases has a fixed point locus of real dimension 2,
moreover, each individual twist $g_a\in SU(2)$, so it
preserves ${\cal N}=2$ supersymmetry (however, at least two different twists
belong to two {\em different} $SU(2)$'s, so altogether 
we have only ${\cal N}=1$ supersymmetry as $\Gamma\subset SO(3)\subset SU(3)$).
It is then not difficult to see that the one-loop non-Abelian
$\beta$-functions vanish in these theories. 
Clearly, since these theories are not ${\cal N}=2$ supersymmetric, the 
one-loop vanishing of the corresponding $\beta$-functions does not guarantee
that they are conformal at higher loops. Actually, it is not difficult to
show that in such theories even two-loop non-Abelian 
$\beta$-functions vanish. Indeed,
this would follow from the vanishing of the one-loop $\beta$-functions
plus the vanishing of the one-loop anomalous scaling dimensions for the
non-Abelian matter fields. The latter fact can be shown by repeating the above
arguments for matter fields. In fact, in the ${\cal N}=1$ cases one can argue 
that these theories are conformal (even at finite $N$) by adapting the
corresponding arguments given in \cite{BKV} in the case of oriented theories.

{}Before we move onto other ${\cal N}=1$ theories, let us note the following
about
${\cal N}=1$ theories with $\Gamma\in SO(3)$. First, the $U(1)$ factors are
all anomaly free in such theories (and should be treated as global
symmetries just as in the ${\cal N}=2$ cases). 
This, actually, follows from the fact that
all such theories are {\em non-chiral}. Since such theories are not
phenomenologically appealing, regardless of their conformality we would have
to move onto ${\cal N}=1$ theories with $\Gamma\not\subset SO(3)$. As we will
now discuss, such theories are {\rm not} conformal.

\subsection{Non-conformal Theories}

{}Thus, consider the cases where $\Gamma\subset SU(3)$ but $\Gamma\not
\subset SO(3)$. It then follows that $\exists g_a\in\Gamma$ such that the 
corresponding fixed point loci have real dimension 0. Let us focus on such
twists, and discuss their contributions to the one-loop running of the gauge
couplings. Let us first discuss the loop-channel ultra-violet divergences, 
which translate into the tree-channel infra-red divergences due to the exchange
of massless twisted states corresponding to such twists.
These massless divergences are
no longer logarithmic but {\em quadratic} (in energy or inverse length - in
terms of the real modular parameter on the annulus the corresponding divergence
is of the form $\int dt/t^2$ at $t\rightarrow 0$). This can be understood
from the fact that the twisted closed string states propagating in the tree
channel live at the fixed {\em point}. Then, if the external lines
correspond to {\em massless} on-shell states, the massless twisted state
in the tree-channel exchange is also on-shell, and its propagator $1/p^2$ is
quadratically divergent. Note, however, that these quadratic divergences
cancel not only in the vacuum amplitude but also in the two-point one-loop
amplitudes with two non-Abelian gauge bosons attached to the same boundary.
As in the previous subsection, this cancellation occurs between the
annulus and M{\"o}bius strip amplitudes. In fact, the quadratic divergence
structure is precisely of the form (\ref{coupling}). This then implies that
all such quadratic divergences cancel. This is just as 
well - in consistent backgrounds we should have no quadratic divergences in
a four-dimensional non-Abelian gauge boson propagator.  
In fact, such quadratic 
divergences also cancel in all other 2-point amplitudes involving
non-Abelian matter fields as they should in a consistent ${\cal N}=1$ 
supersymmetric theory.

{}As should be clear from our previous remarks, the same, however, 
does not hold for $U(1)$ factors. In fact, for the models with $\Gamma\subset
SU(3)$ but $\Gamma\not\subset SO(3)$ we always have at least one $U(1)$ 
factor in the D3-brane world-volume. Let us therefore consider 2-point
functions involving such a $U(1)$ factor.
Thus, at the one-loop level, for the diagrams where we 
attach both external lines corresponding to a $U(1)$ gauge boson to the same
boundary, we have a cancellation between the annulus and M{\"o}bius strip
amplitudes just as in the case of non-Abelian gauge bosons. There is, however,
another diagram we must consider, namely, the annulus diagram with the
two external lines attached to {\em different} boundaries. (Note that we do not
have the corresponding M{\"o}bius strip amplitude as the latter has only one
boundary.) This annulus amplitude {\em does} contain a quadratic
divergence due to a massless twisted closed string exchange.
Thus, we have at least one Ramond-Ramond twisted 2-form, call it $C$,
living at the fixed {\em point} in ${\bf C}^3$ (recall that in these models
we always have a twist with a fixed point locus of real dimension 0). This 
twisted 2-form has a coupling to the $U(1)$ field strength of the following
form:
\begin{equation}
 \int_{\rm D3} C\wedge F~.
\end{equation}
Exchanging this 2-form in the annulus amplitude with two $U(1)$ gauge bosons
attached to different boundaries
leads to a quadratic divergence. 

{}The presence of this quadratic divergence is due to the fact that at least
one $U(1)$ factor in such backgrounds is actually {\em anomalous} 
\cite{Sagnotti,KaSh,IRU}. In particular, we have 
a cubic $U(1)^3$ anomaly (note that all such theories are 
actually {\em chiral}).
The anomalous $U(1)$'s in such backgrounds are broken at the tree level
via a generalized Green-Schwarz mechanism involving twisted closed string
states \cite{IRU}. In particular, we now have a tree-level kinetic term
for the twisted 2-forms $C$, which propagate in the D3-brane world-volume.
The anomalous $U(1)$'s then become massive, and we have the corresponding 
global $U(1)$ symmetries at the orbifold point (that is, if we do not blow
up the orbifold\footnote{As was pointed out in \cite{KaSh,KST}, there appear
to be some subtleties in the world-sheet description in such backgrounds
at the orbifold point in the moduli space. However, as was discussed in
\cite{KaSh,KST}, the world-sheet description is adequate if we blow up
the orbifold fixed point. We will come back to this point in section IV.}). 
This then takes care of the aforementioned quadratic
divergences. In fact, to begin with these divergences appeared because of an
artificial separation of the physical degrees of freedom in massive $U(1)$'s.

{}Let us now turn to the loop-channel infra-red divergences, which translate 
into the tree-channel ultra-violet divergences. In particular, let us discuss
such divergences in the non-Abelian gauge couplings. The key point here is 
that, unlike in the ${\cal N}=2$ models discussed in the previous subsection, 
here the massive open string states do contribute into the gauge coupling 
renormalization (in particular, we have massive string threshold corrections).
In fact, they provide an effective ultra-violet cut-off
in the loop channel \cite{bachas,ABD}. There are, however, {\em infra-red}
divergences in the loop channel. Once again, unlike in the ${\cal N}=2$ models,
here the ultra-violet and infra-red divergence structures are {\em different}. 
In particular, in the tree channel the massive twisted states now do
contribute into ultra-violet divergences. In the loop channel this translates
into {\em logarithmic} infra-red divergences corresponding to the non-Abelian
gauge coupling running. That is, all models with $\Gamma\subset SU(3)$ but
$\Gamma\not\subset SO(3)$ are actually {\em non-conformal}. In particular,
the one-loop $\beta$-function coefficients in these theories
are non-vanishing \cite{orient,orient1}. Thus, we can obtain D3-brane
theories which are both chiral and non-conformal if we consider
orientiworld models with $\Gamma\subset SU(3)$ but $\Gamma\not\subset SO(3)$.

{}Before we end this subsection, let us make the following
observations. Let us assume that
we have an O3-plane. In the cases where $\Gamma$ contains a ${\bf Z}_2$ element
we also have the corresponding O7-plane as well as D7-branes. 
At the {\em tree level} the D7-brane
gauge symmetry serves as a {\em global} symmetry for the D3-brane gauge theory.
In particular, we have 37 bifundamental matter, which, from the 
four-dimensional viewpoint, appears as matter transforming in the fundamental
representations of various 33 gauge subgroups that also carries global
quantum numbers with the corresponding global symmetry identified with the 77
gauge symmetry. However, already at the one-loop level we have kinetic terms
for the 77 gauge bosons generated in the D3-brane world-volume. That is, the
full four-dimensional gauge group contains a subgroup corresponding to the
original 77 gauge group. More precisely, in chiral models 
with $\Gamma\subset SU(3)$ but $\Gamma\not\subset SO(3)$ we have anomalous
77 $U(1)$'s on the D3-branes. These $U(1)$'s are then broken. However, the
non-Abelian part of the 77 gauge group becomes part of the four-dimensional
gauge symmetry.

\section{Gravity in Orientiworld}

{}One interesting feature of the orientiworld framework we described in the
previous section is that we can construct models with non-conformal gauge 
theories in the D3-brane world-volumes. Following \cite{DGP,DG} we then 
expect that the Einstein-Hilbert term (among other terms) will be generated
on the D3-branes\footnote{The fact that the Einstein-Hilbert term can be
present on D-branes was originally suggested in \cite{alberto}.
String theory computations of such a term
were recently given in \cite{kiritsis,Lowe} in somewhat different contexts.} 
(we will discuss this point in more detail in the
next section). Then,
since D3-branes are codimension-6 objects, following \cite{DG},
we expect almost completely localized gravity on D3-branes with ultra-light
modes penetrating into the bulk. In this section we would like to discuss
this phenomenon in the orientiworld context.

{}To understand how such dramatic modification of gravity occurs in the
orientiworld context, we will first 
consider a simplified model. In this model the
3-brane on which the Standard Model fields are localized is assumed to be
tensionless. This assumption, however, is not expected to affect the main 
conclusions of our discussion. In fact, recently in \cite{codi2}
it was argued in the case of
non-zero tension codimension-2 branes that almost complete localization of
gravity still takes place on such branes. We then may 
expect the same to hold for
higher codimension branes as well\footnote{In the higher codimension case, if
we take a non-zero tension 3-brane to be $\delta$-function-like, the background
is badly singular at the location of the brane. However, as we will see in the
following, it appears that the brane thickness should be assumed to be
non-zero (albeit it can be small), which smoothes out such singularities - see
below.}.  

\subsection{A Toy Model}

{}In this subsection we discuss a toy 
brane world model, which captures the key features of gravity in the 
orientiworld framework, with a codimension-$d$ brane
($d\geq 2$) 
embedded in a $D$-dimensional bulk space-time. (For calculational convenience
we will keep the number of space-time dimensions $D$ unspecified, but we are
mostly interested in the case $D=10$ and $d=6$, 
where the brane is a 3-brane.) The
action for this model is given by:
\begin{equation}
 S={\widehat M}_P^{D-d-2}
 \int_{\Sigma}\ d^{D-d}x~\sqrt{-\widehat G}~{\widehat R}+
 M_P^{D-2}\int\ d^D x~\sqrt{-G}~
 R~.
 \label{model}
\end{equation}
Here $M_P$ is the (reduced) $D-$dimensional Planck mass, while ${\widehat M}_P$
is the (reduced) $(D-d)$-dimensional Planck mass; $\Sigma$ is a 
$\delta$-function-like codimension-$d$ source brane, which is 
a hypersurface $x^i=x^i_*$ ($x^i$, $i=1,\dots,d$, 
are the $d$ spatial coordinates 
transverse to the brane); the brane is assumed to be tensionless; also, 
\begin{equation}
 {\widehat G}_{\mu\nu}\equiv
 {\delta_\mu}^M{\delta_\nu}^N G_{MN}\Big|_\Sigma~,
\end{equation} 
where $x^\mu$ are the $(D-d)$ coordinates along the brane (the $D$-dimensional
coordinates are given by $x^M=(x^\mu,x^i)$, and the signature of the 
$D$-dimensional metric is $(-,+,\dots,+)$); finally, the $(D-d)$-dimensional
Ricci scalar ${\widehat R}$ is constructed from the $(D-d)$-dimensional
metric ${\widehat G}_{\mu\nu}$, and the Einstein-Hilbert term on the brane
is assumed to be generated via quantum loops of matter localized on the
brane \cite{DGP,DG}, 
which is assumed to be non-conformal. Note that this is nothing but the
Dvali-Gabadadze model \cite{DG}, 
except that we will not assume that the extra dimensions
transverse to the brane are flat.

{}The equations of motion following from the action (\ref{model}) are given by:
\begin{eqnarray}
 &&R_{MN}-\frac{1}{2} G_{MN}R+\displaystyle{
 {\sqrt{-{\widehat G}}\over\sqrt{-G}}}
 {\delta_M}^\mu {\delta_N}^\nu \left[{\widehat R}_{\mu\nu}-{1\over 2} 
 {\widehat G}_{\mu\nu}{\widehat R}\right] L^d \delta^{(d)}(x^i-x^i_*)=0~,
 \label{EoM}
\end{eqnarray}
where ${\widehat R}_{\mu\nu}$ is the $(D-d)$-dimensional Ricci tensor
constructed from the metric ${\widehat G}_{\mu\nu}$, and
$L^d\equiv {\widehat M}_P^{D-d-2}/M_P^{D-2}$.

{}Consider the following ans{\"a}tz for the 
metric
\begin{equation}
 ds^2=\eta_{\mu\nu}dx^\mu dx^\nu+{\widetilde G}_{ij}dx^i dx^j~,
\label{ansatz}
\end{equation}
where ${\widetilde G}_{ij}$ is a function of $x^i$ 
but is independent of $x^\mu$. Let ${\widetilde R}_{ij}$ and 
${\widetilde R}$ be the $d$-dimensional Ricci tensor respectively Ricci scalar
constructed from the $d$-dimensional metric ${\widetilde G}_{ij}$. 
With the above ans{\"a}tz we have that the $d$-dimensional manifold 
${\cal M}_d$ transverse to the brane must
be Ricci flat:
\begin{eqnarray}
 {\widetilde R}_{ij}=0~.
\end{eqnarray}
In particular, it need not be flat (${\cal M}_d={\bf R}^d$ 
is a special case of a Ricci-flat
space, in particular, for high enough $d$ there exist Ricci-flat manifolds
other than ${\bf R}^d$). 
We will, however, assume that ${\cal M}_d$ is non-compact (and has
infinite volume).

{}Ricci flat manifolds are precisely the spaces that arise in the orientiworld
construction (in the weak string coupling regime). 
Examples of such spaces are given by conifolds. Thus, let\footnote{In the 
$d=2$ case the manifold ${\cal M}_2$ is actually flat if the brane is
tensionless. In particular, if we choose ${\cal M}_2$ to be a two
dimensional ``wedge'' with a deficit angle, the singularity at the origin
of the wedge is $\delta$-function-like, and this solution corresponds to a
{\em non-zero} tension brane with the tension proportional
to the deficit angle. This solution was recently discussed in \cite{codi2}.} 
$d>2$, and let the metric on ${\cal M}_d$ be given by
\begin{equation}
 d{\widetilde s}^2=dr^2+r^2\gamma_{\alpha\beta}dy^\alpha dy^\beta~,
\end{equation}
where $\gamma_{\alpha\beta}$ is independent of the ``radial'' coordinate
$r$ ($r$ takes values between 0 and $\infty$), and is a metric on a (compact)
$(d-1)$-dimensional manifold ${\cal Y}_{d-1}$
of positive curvature. Then the manifold
${\cal M}_d$ is Ricci flat (it is also non-compact with infinite volume).
If ${\cal Y}_{d-1}$ is a unit $(d-1)$-sphere ${\bf S}^{d-1}$, then the
manifold ${\cal M}_d$ is flat. However, in all other cases ${\cal M}_d$ is
{\em not} flat, in fact it has a conifold singularity at $r=0$. For instance,
in the $d=6$ case if we take ${\cal Y}_5={\bf S}^5/\Gamma$ (where $\Gamma$ is
a subgroup of $Spin(6)$), then the space ${\cal M}_6$ is nothing but the
orbifold ${\bf R}^6/\Gamma$. On the other hand, we can take ${\cal Y}_5$ to
be other than ${\bf S}^5$ or an orbifold thereof. Then the resulting conifold
is no longer an orbifold of ${\bf R}^6$. We will discuss an orientiworld model
of this type in section IV.  

\subsection{Linearized Gravity}

{}In this subsection we study gravity in the brane world solution
discussed in the previous section. Thus, let us consider small fluctuations
around the solution
\begin{equation}
 G_{MN}=G^{(0)}_{MN}+h_{MN}
 \label{metric}
\end{equation}
where $G^{(0)}_{MN}$ is the background metric
\begin{eqnarray}
 G_{MN}^{(0)}=\left[
 \begin{array}{cc}
 \eta_{\mu\nu} & 0 \\
 0 & {\widetilde G}_{ij}\\
 \end{array}
 \right]~.
 \label{background}
\end{eqnarray}
The $(D-2)$-dimensional graviton $H_{\mu\nu}\equiv h_{\mu\nu}$
couples to the matter localized on the brane via
\begin{eqnarray}
 S_{\rm int}={1\over 2}\int_{\Sigma}d^{D-d}x\ T_{\mu\nu}H^{\mu\nu}~,
 \label{interaction}
\end{eqnarray}
where $T_{\mu\nu}$ is the conserved energy-momentum tensor for the
matter localized on the brane:
\begin{equation}
 \partial^\mu T_{\mu\nu}=0~.
\end{equation}
The linearized equations of motion read
\begin{eqnarray}
 &&-\nabla^L\nabla_L h_{MN}+2\nabla^L\nabla_{(M}h_{N)L}
 -\nabla_M\nabla_N h +G^{(0)}_{MN}\left(\nabla^L\nabla_L h
 -\nabla^L\nabla^K h_{LK}\right)=\nonumber\\
 &&{M_P^{2-D}\over
 \sqrt{\widetilde G}}{\widetilde T}_{MN}
 \delta^{(d)}(x^i-x^i_*)~,\label{EoML}
\end{eqnarray}
where the components of the ``effective'' energy-momentum tensor
${\widetilde T}_{MN}$ are given by
\begin{eqnarray}
 &&{\widetilde T}_{\mu\nu}=T_{\mu\nu}-
 {\widehat M}_P^{D-d-2}\biggl[-\partial^\lambda\partial_\lambda H_{\mu\nu}+
 2\partial^\lambda\partial_{(\mu} H_{\nu)\lambda}-\partial_\mu\partial_\nu H
 +\eta_{\mu\nu}\left(\partial^\lambda\partial_\lambda H
 -\partial^\lambda\partial^\sigma H_{\lambda\sigma}
 \right)\biggr]\label{T-mu-nu}~,\\
 &&{\widetilde T}_{\mu i}= {\widetilde T}_{ij} = 0~,
\end{eqnarray}
and we have defined $h\equiv h^M{}_M$ and $H\equiv H^\mu{}_\mu$.
Note that the fluctuations $h_{MN}$ are
assumed to vanish once we turn off the brane matter source $T_{\mu\nu}$.

{}Since the brane is tensionless, the $D$-dimensional diffeomorphisms are
intact:
\begin{equation}
 \delta h_{MN}=\nabla_M\xi_N+\nabla_N\xi_M~.
\end{equation}
We can, therefore, use the harmonic gauge:
\begin{equation}
 \nabla^M h_{MN}={1\over 2}\nabla_N h~.
\end{equation}
In the harmonic gauge the equations of motion simplify as follows:
\begin{equation}
 -\nabla^L\nabla_L h_{MN}={M_P^{2-D}\over\sqrt{\widetilde G}}
 \left[{\widetilde T}_{MN}-{1\over{D-2}}
 G^{(0)}_{MN} {\widetilde T}\right]
 \delta^{(d)}(x^i-x^i_*)~,
\end{equation}
where $T\equiv T_\mu{}^\mu$.
Note that the graviphoton components $h_{\mu i}$ vanish - indeed, they do not
couple to the conserved energy-momentum tensor on the brane. The graviscalar
components $\chi_{ij}\equiv h_{ij}$ can only couple to the trace  
of the energy-momentum tensor $T$, which implies that
\begin{equation}\label{gravisca}
 \chi_{ij}={1\over d} {\widetilde G}_{ij}\chi~,
\end{equation}
where $\chi\equiv{\widetilde G}^{ij}\chi_{ij}$. This gives:
\begin{eqnarray}
 &&-\nabla^L\nabla_L H_{\mu\nu}=
 {M_P^{2-D}\over\sqrt{\widetilde G}}\left[{\widetilde T}_{\mu\nu}-{1\over{D-2}}
 \eta_{\mu\nu} {\widetilde T}\right]
 \delta^{(d)}(x^i-x^i_*)~,\\
 &&\nabla^L\nabla_L\chi={d\over{D-2}}
 {M_P^{2-D}\over\sqrt{\widetilde G}}{\widetilde T}
 \delta^{(d)}(x^i-x^i_*)~.
\end{eqnarray}
Note that the harmonic gauge implies that
\begin{eqnarray}
 &&\partial^\mu H_{\mu\nu}={1\over 2}\partial_\nu H+{1\over 2}\partial_\nu
 \chi~,\\
 &&\partial_j H=-{{d-2}\over d}\partial_j\chi~.
\end{eqnarray}
In particular, taking into account the physically required behavior at
infinity, the last equation gives:
\begin{equation}\label{tracechi}
 H=-{{d-2}\over d}\chi~.
\end{equation}
We, therefore, have
\begin{equation}
 {\widetilde T}_{\mu\nu}=T_{\mu\nu}-
 {\widehat M}_P^{D-d-2}\biggl[-\partial^\lambda\partial_\lambda H_{\mu\nu}+
 \partial_\mu\partial_\nu \chi
 -{{d-1}\over d}\eta_{\mu\nu}
 \partial^\lambda\partial_\lambda\chi \biggr]~.
\end{equation}
As we will see, 
the presence of the term proportional to $\partial_\mu\partial_\nu \chi$ 
in ${\widetilde T}_{\mu\nu}$ has important implications\footnote{The 
importance of this term was originally pointed out in \cite{codi2}.}.

{}Let us Fourier transform the coordinates $x^\mu$ on the brane. Let the
corresponding momenta be $p^\mu$, and let $p^2\equiv p^\mu p_\mu$. Then we 
have:
\begin{eqnarray}
 &&-\left(
 \nabla^i\nabla_i-p^2\right) H_{\mu\nu}
 ={M_P^{2-D}\over
 \sqrt{\widetilde G}}\left[{\widetilde T}_{\mu\nu}(p)-
 {1\over{D-2}}\eta_{\mu\nu}
 {\widetilde T}(p)\right]
 \delta^{(d)}(x^i-x^i_*)~,\\
 &&\left[\nabla^i\nabla_i-p^2\right]\chi=
 {d\over{D-2}}{M_P^{2-D}\over
 \sqrt{\widetilde G}}{\widetilde T}(p)\delta^{(d)}(x^i-x^i_*)~,
\end{eqnarray}
where
\begin{equation}
 {\widetilde T}_{\mu\nu}(p)= T_{\mu\nu}(p)-
 {\widehat M}_P^{D-d-2}\biggl[p^2 H_{\mu\nu}-p_\mu p_\nu \chi
 +{{d-1}\over d}{p^2}\eta_{\mu\nu}\chi
 \biggr]~.
\end{equation}
Let us now solve these equations following \cite{DG}.

{}Let us first consider the $p^2\not=0$ modes. Then, due to the fact that
the $d$-dimensional propagator is divergent at the origin for $d\geq 2$, we
have:
\begin{equation}
 H_{\mu\nu}=\chi=0~,~~~r\not=0~,
\end{equation}
while on the brane we have 
\begin{eqnarray}
 &&H_{\mu\nu}(p_\lambda,r=0)={{\widehat M}_P^{2+d-D}\over p^2}\left[
 T_{\mu\nu}(p)-{T(p)\over{D-d-2}}\left(\eta_{\mu\nu}-{d\over {d-1}}
 {p_\mu p_\nu\over p^2} \right)\right]~,\\
 &&\chi(p_\lambda,r=0)={d\over (d-1)(D-d-2)}~
 {{\widehat M}_P^{2+d-D}\over p^2}~
 T(p)~.
\end{eqnarray}
Note that both the momentum as well as the tensor structures in the
graviton propagator are $(D-d)$-dimensional. In particular, the $p^2\not=0$
modes are completely localized on the brane.

{}Next, let us study the $p^2=0$ modes:
\begin{eqnarray}
 &&-
 \nabla^i\nabla_i H_{\mu\nu}
 ={M_P^{2-D}\over
 \sqrt{\widetilde G}}\left[T_{\mu\nu}(p)-
 {1\over{D-2}}\eta_{\mu\nu}
 T(p)+{\widehat M}_P^{D-d-2}p_\mu p_\nu\chi\right]
 \delta^{(d)}(x^i-x^i_*)~,\\
 &&\nabla^i\nabla_i \chi=
 {d\over{D-2}}{M_P^{2-D}\over
 \sqrt{\widetilde G}} T(p)\delta^{(d)}(x^i-x^i_*)~.
\end{eqnarray}
It is important to note that, if the brane matter is
not conformal, then the solution for $\chi$ blows up at the location of the
brane $x^i=x^i_*$, so that the equation for $H_{\mu\nu}$ is ill-defined
due to the term proportional to $p_\mu p_\nu\chi$. Note that this term 
{\em cannot} be removed by a gauge transformation\footnote{Note that
this term does not affect the coupling of the graviton $H_{\mu\nu}$
to the brane matter
as $p^\mu T_{\mu\nu}(p)=0$ for such matter. However, this term can be probed
by bulk matter as $p^\mu T^{\rm{\small bulk}}_{\mu\nu}(p)$ need {\em 
not} be zero.}. The aforementioned singularity is present even if the  
the manifold ${\cal M}_d$ is flat:
${\cal M}_d={\bf R}^d$. Indeed, in this case 
$\chi$ diverges as $\sim |x-x_*|^{2-d}$ for $d>2$, and as
$\sim \ln(|x-x_*|)$ for $d=2$. The same is true for a general 
smooth ${\cal M}_d$.
In the case of conifolds we also have a similar singularity. 

{}There are two points that arise due to the aforementioned singularity.
First, it is clear that the linearized approximation breaks down. However,
the second point might be relevant for addressing the former. Thus,
we expect that ultra-violet physics should smooth out singularities of this
type. One way this can happen is if the brane is not strictly 
$\delta$-function-like, but has finite width. Let us discuss this possibility
in a bit more detail. For simplicity 
we will focus on the case where the manifold ${\cal M}_d$
is flat: ${\cal M}_d={\bf R}^d$.

\subsection{A Non-zero Thickness Brane}

{}Smoothing out the brane in the presence of gravity is non-trivial.
Thus, as was originally pointed out in \cite{RSzura}, 
not only do we expect the Einstein-Hilbert term to be generated on the
brane, but also other terms, in particular, those involving the graviscalar
as well as graviphoton components. Moreover, we actually expect that an
infinite series of non-local terms will be present on the brane. Various
non-trivial issues arise in this context, which are outside of the scope of 
this paper, and will be discussed elsewhere \cite{smooth}\footnote{Smoothing
out the brane was also discussed in \cite{kiritsis}.}. 
However, 
a simple prescription for smoothing out the brane, which we will discuss in 
a moment, will suffice for our purposes here\footnote{One might expect
that the aforementioned
additional terms are suppressed by the brane thickness.
However, ignoring at least some of such terms might not be 
justified due to the singularities
arising in the $\delta$-function limit. In fact, in the following we will
argue that it appears to be necessary to include some of such terms for
consistency.}. 

{}Thus, we will simply replace the $\delta$-function-like brane $\Sigma$, 
which is a point-like object in ${\cal M}_d$, by an extended object 
${\widetilde \Sigma}$ in ${\cal M}_d$. The corresponding action now reads:
\begin{equation}
 {\widetilde S}={\widehat M}_P^{D-d-2}
 \int_{\widetilde \Sigma}\ d^{D}x~f^{(d)}(x^i-x^i_*)~
 \sqrt{-\widehat G}~{\widehat R}+
 M_P^{D-2}\int\ d^D x~\sqrt{-G}~
 R~.
 \label{model1}
\end{equation}
Here $f^{(d)}(x^i-x^i_*)$ is a $d$-dimensional distribution, which replaces
the $d$-dimensional $\delta$-function 
$\delta^{(d)}(x^i-x^i_*)$.
The $(D-d)$-dimensional Ricci scalar ${\widehat R}$ is constructed
from the $(D-d)$-dimensional metric ${\widehat G}_{\mu\nu}=G_{\mu\nu}(x^\sigma,
x^i)$, where the extra $d$-dimensional coordinates $x^i$ are treated as 
parameters (in particular, note that ${\widehat R}$ does {\em not} contain
derivatives w.r.t. $x^i$). Similarly, the coupling to brane matter is given by
the term
\begin{equation}
 S_{\rm{\small int}}={1\over 2}\int_{\widetilde \Sigma} d^Dx~
 f^{(d)}(x^i-x^i_*)~T_{\mu\nu}
 H^{\mu\nu}~, 
\end{equation}
where the energy-momentum tensor for the brane matter is conserved in the
$(D-d)$-dimensional sense
\begin{equation}
 \partial^\mu T_{\mu\nu}=0~,
\end{equation}
and is {\em independent} of the coordinates $x^i$.

{}A straightforward guess for smoothing out the brane would be to choose
${\widetilde \Sigma}$ to be a $d$-dimensional object in ${\cal M}_d$. This
is equivalent to replacing the $d$-dimensional $\delta$-function 
$\delta^{(d)}(x^i-x^i_*)$ by a smooth 
distribution $f^{(d)}(x^i-x^i_*)$. In particular, we could consider
a radially symmetric smooth distribution:
$f^{(d)}(x^i-x^i_*)=f_1(r)$, where $r\equiv |x-x_*|$. A simple example of such
a distribution would be (here $\theta(z)$ is the Heavyside step-function)
\begin{eqnarray}\label{dist1}
 &&f_1(r)={1\over v_d \epsilon^d}~\theta(\epsilon-r)~,
\end{eqnarray}
where $v_d$ is the volume of a $d$-dimensional ball ${\bf B}_d$ with unit
radius. Here $\epsilon$ plays the role of the brane thickness.
In this case ${\widetilde \Sigma}$ is a $d$-dimensional ball of radius 
$\epsilon$ in ${\cal M}_d$. However, it is not difficult to show that,
whatever our choice for the smooth $d$-dimensional distribution 
$f^{(d)}(x^i-x^i_*)$, the resulting system suffers from the presence of
an {\em infinite} tower of {\em tachyonic} modes. In fact, 
these modes correspond to the ``radion'' $\chi$. This can intuitively
be understood from the fact that $\chi$ (as well as the trace of $H_{\mu\nu}$)
couple to the interior (in ${\cal M}_d$) of such a smooth brane with
the sign opposite to that of the traceless part of $H_{\mu\nu}$. In Appendix
B we illustrate this in the case of the distribution $f_1(r)$ given by
(\ref{dist1}).

{}This implies that ${\widetilde\Sigma}$ cannot be a $d$-dimensional object
in ${\cal M}_d$. On the other hand, its dimension ${\widetilde d}$ in
${\cal M}_d$ should not be smaller than $(d-1)$ - indeed, if ${\widetilde
d}\leq (d-2)$, we still expect singularities of the type discussed in the 
previous subsection to be present. This suggests that
${\widetilde \Sigma}$ should be an object of dimension ${\widetilde d}=
(d-1)$ in ${\cal M}_d$. In fact, in this case we expect no singularities
to be present. This is suggested by the fact that there are no singularities
in the case of a codimension-1 brane in infinite-volume extra space \cite{DGP}.

{}We will therefore replace the $d$-dimensional $\delta$-function 
$\delta^{(d)}(x^i-x^i_*)$ by a {\em singular} distribution such that it is
smooth in $(d-1)$ dimensions, and has a $\delta$-function-like singularity
in the remaining one dimension. A simple example of such a distributions, which
we will adopt in the following, is given by:
\begin{equation}\label{dist}
 f(r)={1\over a_{d-1}\epsilon^{d-1}}~\delta(r-\epsilon)~,
\end{equation} 
where $a_{d-1}$ is the area of a $(d-1)$-sphere ${\bf S}^{d-1}$ with unit
radius. In this case ${\widetilde\Sigma}$ is a $(d-1)$-sphere of radius
$\epsilon$ in ${\cal M}_d$. As we will see in the following, such
partial smoothing out of the brane gives a consistent model in the present
context.

{}Thus, we can rewrite our action (\ref{model1}) as follows:
\begin{equation}\label{model2}
  {\widetilde S}={\widehat M}_P^{D-d-2}
 \int \ d^{D}x~f(r)~\sqrt{-\widehat G}~{\widehat R}+
 M_P^{D-2}\int\ d^D x~\sqrt{-G}~
 R~,
\end{equation}
and the coupling to brane matter is given by:
\begin{equation}
 S_{\rm{\small int}}={1\over 2}\int d^Dx~f(r)~T_{\mu\nu}
 H^{\mu\nu}~. 
\end{equation}
The background solution is unaffected, and we can still choose it so that
the entire $D$-dimensional space is flat. The linearized equations of
motion are given by (\ref{EoML}) with the $d$-dimensional $\delta$-function
$\delta^{(d)}(x^i-x^i_*)$ replaced by $f(r)$. It is straightforward to
solve these equations (as before, the graviphoton components are vanishing,
while the graviscalar components obey (\ref{gravisca})):
\begin{eqnarray}
 &&{\overline H}_{\mu\nu}\equiv H_{\mu\nu}-{1\over{D-d}}\eta_{\mu\nu}
 H=\nonumber\\
 &&-M_P^{2-D}\left[\left(T_{\mu\nu}(p)-{1\over {D-d}}
 \eta_{\mu\nu}T(p)\right)\Phi+{d\over{D-2}}L^d
 \left(p_\mu p_\nu-{1\over{D-d}}
 \eta_{\mu\nu}p^2\right)T(p)\Sigma\right]~,\\
 &&H=-{{d-2}\over d}\chi=-{{d-2}\over{D-2}}M_P^{2-D}T(p){\widetilde \Phi}~.
\end{eqnarray}
Here $\Phi$, $\Sigma$ and ${\widetilde\Phi}$ are the solutions to the
following equations
\begin{eqnarray}\label{Phi}
 &&\left(
 \partial^i\partial_i-p^2\right)\Phi=
 \left[1+L^dp^2\Phi\right]f(r)~,\\
 &&\left(
 \partial^i\partial_i-p^2\right)\Sigma=
 \left[{\widetilde \Phi}+L^dp^2\Sigma\right] f(r)~,\\
 &&\left(
 \partial^i\partial_i-p^2\right){\widetilde \Phi}=
 \left[1-\kappa L^d p^2{\widetilde \Phi}\right]f(r)
\end{eqnarray}
subject to the condition that $\Phi$, $\Sigma$ and ${\widetilde \Phi}$ 
decay to zero away from the brane (in the following we will focus on the
cases $d>2$, where this is true even for the $p^2=0$ modes). 
Here we have introduced the notation
\begin{equation}
 \kappa\equiv {(d-1)(D-d-2)\over{D-2}}
\end{equation}
to simplify expressions containing ${\widetilde\Phi}$. 

{}With the choice (\ref{dist}) for the distribution function $f(r)$, 
we can solve for $\Phi$, $\Sigma$ and ${\widetilde \Phi}$ 
in terms of elementary functions for $d=3$. 
Since in higher $d$ the
qualitative conclusions are the same as in $d=3$, let us focus on this case.
In Appendix C we treat the general $d\geq 2$ case.

\bigskip
\begin{center}
 {\em The $d=3$ Case}
\end{center}
\bigskip

{}To solve for $\Phi$, $\Sigma$ and ${\widetilde \Phi}$, it is convenient to
Wick rotate to the Euclidean space. Let $p\equiv\sqrt{p^2}$. We then have
the following radially symmetric solution for $\Phi$:
\begin{eqnarray}
 &&\Phi(r)=-{\epsilon\over{L^dp^2\sinh(p\epsilon)+a_d\epsilon^{d-1}p
 \exp(p\epsilon)}}~{\sinh(pr)\over r}~,~~~r\leq \epsilon~,\\
 &&\Phi(r)=-{\epsilon\sinh(p\epsilon)
 \over{L^dp^2\sinh(p\epsilon)+a_d\epsilon^{d-1}p
 \exp(p\epsilon)}}~{\exp\left(-p[r-\epsilon]\right)\over r}~,~~~r\geq 
 \epsilon~.
\end{eqnarray}
Next, the radially symmetric solution for $\Sigma$ is given by:
\begin{equation}
 \Sigma(r)={\widetilde \Phi}(\epsilon) ~\Phi(r)~.
\end{equation}
Finally, the radially symmetric solution for ${\widetilde \Phi}$ is given by:
\begin{eqnarray}
 &&{\widetilde \Phi}(r)=
 {\epsilon\over{\kappa L^dp^2\sinh(p\epsilon)-a_d\epsilon^{d-1}p
 \exp(p\epsilon)}}~{\sinh(pr)\over r}~,~~~r\leq \epsilon~,\\
 &&{\widetilde \Phi}(r)={\epsilon\sinh(p\epsilon)
 \over{\kappa L^dp^2\sinh(p\epsilon)-a_d\epsilon^{d-1}p
 \exp(p\epsilon)}}~{\exp\left(-p[r-\epsilon]\right)\over r}~,~~~r\geq 
 \epsilon~.
\end{eqnarray}
Unlike the solution for $\Phi(r)$, the solution for ${\widetilde\Phi}(r)$
has a pole. Here we would like to discuss this point in a bit more detail.

{}Thus, the pole in ${\widetilde \Phi}(r)$ is located at $p=p_*$, where
\begin{equation}
 p_*\left[1-\exp(-2p_*\epsilon)\right]=
 {2a_{d-1}\epsilon^{d-1}\over\kappa L^d}~.
\end{equation}
Suppose the brane thickness $\epsilon$ is small: $\epsilon\ll L$. Then we have:
\begin{equation}
 p_*^2\approx {a_{d-1}\epsilon^{d-2}\over\kappa L^d}~.
\end{equation}
As we will see in the following, $p_*$ is of order of the cross-over scale
$p_c$ above which gravity is $(D-d)$-dimensional, while below this scale it
is expected to become 
$D$-dimensional. Thus, around the cross-over scale we have a single
pole in the
radion propagator (and, consequently, in the graviton propagator). In fact, 
this pole is {\em tachyonic}. Thus, consider the equation of motion for 
$\chi$ in the absence of the brane matter:
\begin{equation}
 \left(
 \partial^i\partial_i-p^2\right)\chi=
 -\kappa L^d p^2\chi f(r)~.
\end{equation}
Let us work with Minkowski momenta $p^\mu$. We have a continuum of modes
with $p^2\leq 0$. These are massless and massive modes. However, we also have
an isolated quadratically normalizable tachyonic mode with mass squared
$m^2_*=-p_*^2$, which is given by:
\begin{eqnarray}
 &&\chi_*(r)=C~{\sinh(p_*r)\over r}~,~~~r\leq \epsilon~,\\
 &&\chi_*(r)=C~\sinh(p_*\epsilon)~{\exp\left(-p_*[r-\epsilon]\right)
 \over r}~,~~~r>\epsilon~,
\end{eqnarray}
where $C$ is a constant. 

{}At first it might seem that the presence of a tachyonic mode implies
an instability in the system. However, as we will now argue,
our approximations {\em cannot} be trusted at momenta 
$p^2{\ \lower-1.2pt\vbox{\hbox{\rlap{$<$}\lower5pt\vbox{\hbox{$\sim$}}}}\ }
p_c^2$, where we have defined the {\em cross-over} scale
\begin{equation}
 p_c^2\equiv {a_{d-1}\epsilon^{d-2}\over L^d}~.
\end{equation}
Indeed, as we have already mentioned, we have ignored an infinite series of
{\em non-local} terms on the brane, which are expected to be generated
once we smooth it out. In particular, we have ignored an infinite
number of non-local terms 
containing powers of the inverse momentum squared (in the momentum space):
\begin{equation}
 {1\over (p^2)^n}~.
\end{equation}
These terms are expected to be suppressed by the brane thickness $\epsilon$.
In fact, the expansion parameter is expected to be $p_c^2/p^2$:
\begin{equation}
 \left({p_c^2\over p^2}\right)^n~.
\end{equation}
{}For momenta $p^2\gg p_c^2$ these terms can be safely ignored. However,
in the infra-red, that is, for the momenta 
$p^2{\ \lower-1.2pt\vbox{\hbox{\rlap{$<$}\lower5pt\vbox{\hbox{$\sim$}}}}\ }
p_c^2$, an {\em infinite} number of such non-local terms becomes important.
We may then expect that the aforementioned tachyonic instability is an artifact
of dropping such non-local terms\footnote{Here we note that, 
had we chosen a smooth $d$-dimensional
distribution such as (\ref{dist1}), we would have had an infinite tower
of such tachyonic modes (see Appendix B for details). 
In this case it is less clear how the aforementioned
non-local terms could remove all the tachyonic states.}.

{}Let us discuss this point in a bit more detail. Thus,
consider the momenta $p^2\ll 1/\epsilon^2$. Then we have:
\begin{eqnarray}
 &&\Phi(\epsilon)\approx -{1\over L^d}~{1\over{p^2+p_c^2}}~,\\
 &&{\widetilde \Phi}(\epsilon)\approx {1\over \kappa 
 L^d}~{1\over{p^2-p_*^2}}~,\\
 &&\Sigma(\epsilon)\approx -{1\over \kappa L^{2d}}~{1\over{p^2+p_c^2}}~
 {1\over{p^2-p_*^2}}~,
\end{eqnarray}
where $p_*^2\approx p_c^2/\kappa$. On the brane we therefore have:
\begin{eqnarray}\label{graviton}
 &&H_{\mu\nu}(r=\epsilon)\approx {{\widehat M}_P^{2+d-D}\over {p^2+p_c^2}}
 \biggl[T_{\mu\nu}(p)-{1\over{D-d-2}}\eta_{\mu\nu} T(p)+\nonumber\\
 &&{d\over(d-1)(D-d-2)}\left(p_\mu p_\nu-\eta_{\mu\nu}
 {{(d-2)p_c^2+2(d-1)p_*^2}\over d(D-d)}\right){T(p)\over{p^2-p_*^2}}
 \biggr]~,\\
 &&\chi(r=\epsilon)\approx {d\over (d-1)(D-d-2)}~
 {{\widehat M}_P^{2+d-D}\over{p^2-p_*^2}}~T(p)~.\label{chi}
\end{eqnarray}
As we have already mentioned, 
we can trust our approximations for 
the momenta $p_c^2\ll p^2\ll 1/\epsilon^2$.
Then on the brane we have:
\begin{eqnarray}
 &&H_{\mu\nu}(r=\epsilon)\approx {{\widehat M}_P^{2+d-D}\over p^2}
 \left[T_{\mu\nu}(p)-{T(p)\over{D-d-2}}\left(\eta_{\mu\nu}-
 {d\over {d-1}}~{p_\mu p_\nu\over p^2}\right)\right]~,\\
 &&\chi(r=\epsilon)\approx {d\over (d-1)(D-d-2)}~
 {{\widehat M}_P^{2+d-D}\over p^2}~T(p)~.
\end{eqnarray}
This shows that both the momentum as well as the tensor structures of the
graviton propagator on the brane are indeed $(D-d)$-dimensional. 
On the other hand, for the momenta
$p^2{\ \lower-1.2pt\vbox{\hbox{\rlap{$<$}\lower5pt\vbox{\hbox{$\sim$}}}}\ }
p_c^2$ the above approximation is expected to break down. In fact,
we can get a hint
of what kind of non-local terms would be relevant for, in particular, removing
the tachyonic state by expanding (\ref{graviton}) and (\ref{chi})
in powers of 
$p_c^2/p^2$. Finally, 
note that the transverse-traceless components of $H_{\mu\nu}$ in 
(\ref{graviton}) behave as those of a {\em massive} $(D-d)$-dimensional
propagator with mass squared equal $p_c^2$. 
 
\subsection{Non-zero Tension Solutions}

{}For the following discussions it will be useful to understand the non-zero
tension solutions, which we would like to discuss in this subsection.
Thus, consider the following model: 
\begin{equation}
 {\widetilde S}={\widehat M}^{D-d-2}_P \int d^{D-d}x~f(r)~
 \sqrt{-{\widehat G}}\left[{\widehat R}-{\widehat\Lambda}\right]
 +{M}^{D-2}_P \int d^{D}x~
 \sqrt{-{G}}~{R}~.
\end{equation} 
Here
\begin{equation}
 \tau\equiv {\widehat M}^{D-d-2}_P{\widehat \Lambda}
\end{equation}
plays the role of the brane tension. That is, in the $\delta$-function limit
$\tau$ is the tension of the brane $\Sigma$, which is a point in the extra
space ${\cal M}_d$.

{}The equations of motion read
\begin{eqnarray}
 &&R_{MN}-\frac{1}{2} G_{MN}R+\displaystyle{
 {\sqrt{-{\widehat G}}\over\sqrt{-G}}}
 {\delta_M}^\mu {\delta_N}^\nu \left[{\widehat R}_{\mu\nu}-{1\over 2} 
 {\widehat G}_{\mu\nu}\left({\widehat R}-{\widehat \Lambda}\right)\right] 
 L^d f(r)=0~.
 \label{EoM4}
\end{eqnarray}
To solve these equations, let us use the following ans{\"a}tz for the 
background metric:
\begin{equation}
 ds^2=\exp(2A)~\eta_{\mu\nu}dx^\mu dx^\nu+\exp(2B)~\delta_{ij}dx^i dx^j~,
\end{equation}
where $A$ and $B$ are
functions of $x^i$ but are independent of $x^\mu$. We then have:
\begin{eqnarray}
 &&R_{\mu\nu}=-\eta_{\mu\nu}e^{2(A-B)}
 \left[\partial^i\partial_i A+
 (D-d)\partial^i A\partial_i A +(d-2)\partial^i A\partial_i B\right]~,\\
 &&R_{ij}=(d-2)\left[\partial_i B\partial_j B-\partial_i\partial_j B\right]
 +(D-d)\left[\partial_i A\partial_j B+\partial_j A\partial_i B-
 \partial_i A\partial_j A-\partial_i\partial_j A\right]-\nonumber\\
 &&\delta_{ij}\left[\partial^k\partial_k B +(d-2)\partial^kB\partial_k B +
 (D-d)\partial^k A\partial_k B\right]~,\\
 &&R=-e^{-2B}\Big[2(d-1)\partial^i\partial_i B+2(D-d)\partial^i\partial_i A
 +\nonumber\\
 &&(d-1)(d-2)\partial^i B\partial_i B +(D-d)(D-d+1)
 \partial^i A\partial_i A+2(D-d)(d-2)\partial^i A\partial_i B\Big]~. 
\end{eqnarray}
Here we are interested in radially symmetric solutions where $A$ and $B$
are functions of $r$. The equations of motion then read (here prime
denotes derivative w.r.t. $r$):
\begin{eqnarray}\label{AB1}
 &&(d-2)\left[(B^\prime)^2-B^{\prime\prime}+{1\over r} B^\prime\right]
 +(D-d)\left[2A^\prime B^\prime-(A^\prime)^2-A^{\prime\prime}+
 {1\over r}A^\prime\right]=0~,\\
 &&(d-2)\left[B^{\prime\prime}+{1\over 2}(d-3)(B^\prime)^2+(d-2){1\over r}
 B^\prime\right]+\nonumber\\
 &&(D-d)\left[A^{\prime\prime}+
 {1\over 2}(D-d+1)(A^\prime)^2+(d-2){1\over r}A^\prime+(d-3)
 A^\prime B^\prime\right]=0~,\label{AB2}\\
 &&(D-d-1)\left[A^{\prime\prime}+{1\over 2}(D-d)(A^\prime)^2+(d-1){1\over r}
 A^\prime+(d-2)A^\prime B^\prime\right]+\nonumber\\
 &&(d-1)\left[B^{\prime\prime}+{1\over 2}(d-2)(B^\prime)^2+(d-1){1\over r}
 B^\prime\right]+{1\over 2}e^{(2-d)B}{\widehat\tau}f(r)=0~,\label{AB3}
\end{eqnarray}
where ${\widehat\tau}\equiv \tau/M_P^{D-2}$.

{}Here we are interested in solutions with {\em infinite} volume extra 
dimensions\footnote{For any $d$ 
there exists a simple solution with $A\equiv 0$. However,
in this solution the extra space has finite volume for $d>2$.}. 
In $d=2$ there is a simple solution of this type with $A\equiv 0$. This 
solution was studied in \cite{codi2}. Here we will focus on the $d>2$ cases.
We can write down approximate solutions for small brane tension. Thus, let 
\begin{equation}
 {\widehat \tau}\equiv \gamma \epsilon^{d-2}~,
\end{equation}
where the dimensionless parameter $\gamma$ is small: $\gamma\ll 1$. Then it
is not difficult to see that we can drop terms non-linear in $A^\prime$ and
$B^\prime$ (assuming that we are looking for solutions non-singular at $r=0$
such that $A$ and $B$ asymptote to constant values at $r\rightarrow \infty$).
In particular, the corrections due to the non-linear terms in this case are
suppressed by powers of $\gamma$.
The above equations then simplify as follows:
\begin{eqnarray}
 &&B^\prime=-{{D-d}\over{d-2}}A^\prime~,\\
 &&A^{\prime\prime}+(d-1){1\over r}A^\prime=
 {{d-2}\over{D-2}}~{\gamma\over 2 a_{d-1}\epsilon} e^{(2-d)B}\delta(r-
 \epsilon)~. 
\end{eqnarray}
The solution is given by (here we have fixed the integration constants
so that both $A$ and $B$ vanish on the brane):
\begin{eqnarray}
 &&A(r)={\gamma\over 2(D-2)a_{d-1}}\left[1-{\epsilon^{d-2}\over r^{d-2}}\right]
 \theta(r-\epsilon)~,\\
 &&B(r)=-{{D-d}\over{d-2}}A(r)~.
\end{eqnarray}
Note that $A$ and $B$ are constant for $r<\epsilon$. In particular, these
solutions are non-singular, and the $D$-dimensional space-time becomes flat
as $r\rightarrow \infty$.

{}We would like to end this subsection with a few remarks. To find exact
solutions, one can proceed as follows. Consider the sum of (\ref{AB1}) and 
(\ref{AB2}):
\begin{equation}
 (d-2)\left[(B^\prime)^2+{2\over r}B^\prime\right]+
 (D-d)\left[{{D-d-1}\over{d-1}}(A^\prime)^2+{2\over r}A^\prime+
 2A^\prime B^\prime\right]=0~.
\end{equation} 
Note that this equation does not contain second derivatives of $A$ and $B$.
The solution for $B^\prime$ is given by:
\begin{equation}
 B^\prime=-{1\over r}-{{D-d}\over{d-2}} A^\prime\pm
 \sqrt{{1\over r^2}+{(D-d)(D-2)\over (d-1)(d-2)^2} (A^\prime)^2 }~.
\end{equation}
To obtain solutions with infinite-volume extra space we should choose the
plus root in the above solution. The non-singular solutions that asymptote
to a flat space then have the property that $A$ and $B$ are constant for
$r<\epsilon$. That is, the space inside of the $(d-1)$-sphere of radius
$\epsilon$ is actually flat. This will be important in the following.

{}Finally, let us note that in $d=2$ we have substantial simplifications, and
it is not difficult to solve the equations for $A$ and $B$ exactly. In 
particular, as was recently discussed in \cite{codi2}, in $d=2$ we have a 
solution where the $D$-dimensional space-time is a product of the 
$(D-2)$-dimensional Minkowski space-time and a 2-dimensional extra space with
a deficit angle (in the $\delta$-function limit this is a 2-dimensional 
``wedge'' with a conical singularity at the origin, while in the case of a
smoothed out brane the 2-dimensional extra space is non-singular). In this
example, which we discuss in Appendix D, one can explicitly see that, 
in the appropriate regime, we reproduce $(D-2)$-dimensional gravity on the 
brane even for non-vanishing brane tension\footnote{A similar analysis in
the $d>2$ cases will be presented elsewhere \cite{CIKLprep}.}. 

\subsection{Brane Thickness in Orientiworld}

{}Up to various subtleties mentioned in the beginning of
subsection C, following \cite{DG} we can expect that, above the cross-over
scale\footnote{This formula is correct for $d>2$. For $d=2$ we have a logarithm
of $\epsilon$ in the expression for $r_c$. See Appendix C for details.} 
\begin{equation}
 r_c\sim \sqrt{{{\widehat M_P}^{D-d-2}\over M_P^{D-2}}~
 {1\over \epsilon^{d-2}}}
\end{equation}
all the way down to the scales of order of the brane thickness
$\epsilon$, gravity on the brane is $(D-d)$-dimensional.

{}In the orientiworld context we have $D=10$, $d=6$, and ($g_s$ is the 
string coupling):
\begin{eqnarray}
 M^8_P\sim M_s^8/g_s^2~.
\end{eqnarray}
We, therefore, have:
\begin{equation}
 r_c\sim g_s ~{{\widehat M}_P \over M^2_s}~{1\over M_s^2 \epsilon^2}~.
\end{equation}
Phenomenologically we must require that ${\widehat M}_P$ is the 
four-dimensional Planck scale: ${\widehat M}_P\sim 10^{18}~{\rm GeV}$. 
Let us also require that the cross-over scale is at least as large as the
present size of the observable Universe, that is,
the Hubble size $r_H\sim 10^{28}~{\rm cm}$. Then $r_c{\widehat M}_P
{\ \lower-1.2pt\vbox{\hbox{\rlap{$>$}\lower5pt\vbox{\hbox{$\sim$}}}}\ }
10^{60}$. We will also assume that the string coupling $g_s\sim 1$. Then we
have the following upper bound on the brane thickness:
\begin{equation}
 \epsilon
 {\ \lower-1.2pt\vbox{\hbox{\rlap{$<$}\lower5pt\vbox{\hbox{$\sim$}}}}\ }
 10^{-30}~{{\widehat M}_P \over M^2_s}~.
\end{equation}
The phenomenological lower bound on $M_s$ is $M_s
{\ \lower-1.2pt\vbox{\hbox{\rlap{$>$}\lower5pt\vbox{\hbox{$\sim$}}}}\ }
{\rm TeV}$. This then implies that the lower bound on the brane thickness
is 
\begin{equation}
 \epsilon 
 {\ \lower-1.2pt\vbox{\hbox{\rlap{$<$}\lower5pt\vbox{\hbox{$\sim$}}}}\ }
 1/{\widehat M}_P~.
\end{equation}
This lower bound is saturated when
\begin{equation}\label{phenom}
 \epsilon\sim 1/{\widehat M}_P:~~~M_s\sim {\rm TeV}~,~~~r_c\sim r_H~. 
\end{equation}
If $M_s$ is much larger than TeV and/or if $r_c$ is much larger than $r_H$,
then $\epsilon$ must be much smaller than the inverse four-dimensional Planck 
scale.

{}If $\epsilon$ is non-zero, we expect that $1/\epsilon$ corresponds to some 
threshold scale in the theory. In particular, this ultra-violet threshold 
must be responsible for smoothing out the aforementioned singularities. It is
not unreasonable to expect that the ultra-violet physics that is responsible
for smoothing out these singularities should not be unrelated to the 
ultra-violet physics that is responsible for generating the four-dimensional
Einstein-Hilbert term\footnote{Otherwise we would have to have three different
threshold scales in the theory: $M_s \ll {\widehat M}_P \ll 1/\epsilon$.
Note that we could assume that $M_s\sim {\widehat M}_P$. Then we would have
to have an ultra-high threshold scale $1/\epsilon
{\ \lower-1.2pt\vbox{\hbox{\rlap{$>$}\lower5pt\vbox{\hbox{$\sim$}}}}\ }
10^{30} {\widehat M}_P$.}. If so, then the aforementioned bound should be
saturated, and we have (\ref{phenom}). That is, the string scale is
around TeV, and gravity becomes higher dimensional at the present Hubble
size. Clearly, 
such a scenario would have very important experimental implications. 
In particular, the string excitations could be observable at LHC. On the
other hand, the fact that gravity undergoes a transition from being 
4-dimensional to being higher dimensional would have important cosmological
implications. For instance, one might attempt to use this to explain an
accelerated Universe with vanishing 4-dimensional cosmological constant as
in \cite{DDG}.

{}Now we come to the key question. What is the ultra-violet physics responsible
for generating the four-dimensional Einstein Hilbert term and smoothing out
the brane? Moreover, how can the corresponding threshold be so much higher 
than the string scale? The next section is devoted to discussing these issues. 

\section{Four-Dimensional Einstein-Hilbert Term in Orientiworld}

{}Let us understand the origin of the four-dimensional Einstein-Hilbert term
in the orientiworld context. If we have a non-conformal field theory on 
D3-branes, and the field theory description is adequate up to some threshold
scale $\Lambda$, which we will refer to as the ultra-violet cut-off scale,
then from the four-dimensional effective field theory viewpoint we expect
that
\begin{equation}
 {\widehat M}_P^2\sim \Lambda^2~.
\end{equation} 
Thus, we must understand what $\Lambda$ is in a given orientiworld setup.
To do this, we will start from a simple ${\cal N}=2$ supersymmetric
model, and then
move onto non-conformal ${\cal N}=1$ models discussed in
subsection D of section II.

\subsection{A Simple ${\cal N}=2$ Example}

{}As we discussed in subsection C of section II, non-compact ${\cal N}=2$
orientiworld models are all conformal. Here we would like 
to discuss a non-conformal model. This
can be done in a setup where two of the six extra dimensions are compact.

{}Thus, consider Type IIB on $T^2\times ({\bf C}^2/\Gamma)$. For simplicity
we will take a square 2-torus $T^2$:
\begin{equation}
 x^i\sim x^i+2\pi r~,~~~i=1,2~.
\end{equation} 
We will assume that $r^2\gg\alpha^\prime$. The orbifold group $\Gamma$ can
be an appropriate subgroup of $SU(2)$. Here we would like to discuss
orientifolds of such backgrounds containing O3-planes. For simplicity we will
focus on the cases where we have no O7-planes. Then $\Gamma={\bf Z}_M$, 
where $M$ is
odd. The choice of $M$ is not going to be critical here, so for illustrative
purposes we will take the simplest example: $\Gamma={\bf Z}_3=\{1,\theta,
\theta^2\}$. The twist $\theta$ acts as follows on the complex coordinates
$z_2,z_3$ on ${\bf C}^2$:
\begin{equation}
 \theta:z_2\rightarrow\omega z_2~,~~~
 \theta:z_3\rightarrow\omega^{-1} z_3~,
\end{equation}
where $\omega\equiv\exp(2\pi i/3)$.

{}Now consider the $\Omega J$ orientifold of 
Type IIB on $T^2\times ({\bf C}^2/\Gamma)$, where $J$ acts on the complex
coordinates $z_1,z_2,z_3$ as follows ($z_1\equiv x^1+ix^2$ is the complex
coordinate on $T^2$):
\begin{equation}
 J:z_\alpha\rightarrow -z_\alpha~,~~~\alpha=1,2,3~.
\end{equation}
Note that $J$ has four fixed points in $T^2\times ({\bf C}^2/\Gamma)$. These
are located at $z_2=z_3=0$ and
\begin{equation}
 (x^1,x^2)=(0,0)~,~~~(\pi r,0)~,~~~(0,\pi r)~,~~~(\pi r,\pi r)~.
\end{equation}
We will place an O3$^-$-plane at each of these points.

{}The twisted closed string states have non-trivial couplings to the
orientifold planes, which gives rise to tadpoles. To cancel the tadpoles
we must introduce D3-branes. The number of these D3-branes is arbitrary.
We can cancel tadpoles {\em locally} at each O3$^-$-plane if we put
$(3n+2)$ D3-branes on top of each of the O3$^-$-planes and choose
the twisted Chan-Paton matrices as follows \cite{orient}:
\begin{equation}
 \gamma^I_\theta={\rm diag}(\omega I_n,\omega^{-1}I_n,I_{n+2})~,
\end{equation}  
where $I$ labels the four O3$^-$-planes. On each of the four sets of
D3-branes we then have a {\em conformal}\footnote{More precisely, these gauge
theories are conformal at low energies - see below.} 
${\cal N}=2$ gauge theory with the
gauge group $SU(n)\otimes SO(n+2)$ and matter hypermultiplets transforming in
$({\bf A},{\bf 1})$ and $({\bf n},{\bf {n+2}})$ (we have dropped the $U(1)$
factors for the reasons discussed in subsection C of section II), where
${\bf A}$ is the two-index anti-symmetric representation of $SU(n)$ (whose
dimension is $n(n-1)/2$). 

{}We can, however, obtain a non-conformal D3-brane gauge theory if we cancel
tadpoles only {\em globally} and not locally. Thus, let us place $(3N+8)$
D3-branes on top of the O3$^-$-plane located at the origin of $T^2$, and place
no D3-branes on top of the other three O3$^-$-planes.
We can cancel the twisted tadpoles globally if we choose the twisted Chan-Paton
matrix acting on the D3-branes as follows \cite{orient}:
\begin{equation}
 \gamma_\theta={\rm diag}(\omega I_N,\omega^{-1}I_N,I_{N+8})~.
\end{equation} 
The D3-brane gauge theory is then a non-conformal 
${\cal N}=2$ gauge theory with the
gauge group $SU(N)\otimes SO(N+8)$ and matter hypermultiplets transforming in
$({\bf A},{\bf 1})$ and $({\bf N},{\bf {N+8}})$, where 
${\bf A}$ is the two-index anti-symmetric representation of $SU(N)$.

{}As we discussed in subsection C of section II, the massive string thresholds
do not contribute into the four-dimensional gauge coupling renormalization. The
four-dimensional field theory description is therefore adequate up the the
open string {\em winding} threshold \cite{BF,bachas,ABD}
(see Appendix A for details), whose mass
is given by $r/\alpha^\prime$. The four-dimensional gauge theory
cut-off, therefore, is given by (up to a dimensionless numerical coefficient,
which is subtraction scheme dependent):
\begin{equation}\label{fieldth}
 \Lambda\sim {r\over \alpha^\prime}~.
\end{equation} 
Our field theory intuition then tells us that the induced
4-dimensional Planck scale should be
\begin{equation}
 {\widehat M}_P\sim \Lambda\sim {r\over\alpha^\prime}~.
\end{equation}
Let us now see this in the string theory language.

{}Note that 
the global tadpole cancellation guarantees that we have no {\em tree-channel}
infra-red divergences. However, unless the twisted tadpoles are canceled 
locally, we do have tree-channel {\em ultra-violet} divergences (which 
correspond to the loop-channel infra-red divergences). These divergences are
logarithmic in the renormalization of the gauge couplings for massless
D3-brane gauge bosons. That is, the loop-channel ultra-violet and infra-red 
divergences in the gauge coupling renormalization are {\em not} identical 
in this background. This is possible
because, even though the 
massive string excitations still do not contribute into
the corresponding amplitudes, the massive open string winding states do as they
{\em are} BPS states (this effectively introduces the ultra-violet cut-off
$\Lambda$ in the loop channel). 
In the tree channel this corresponds to non-vanishing
contributions from massive Kaluza-Klein modes of twisted closed string states,
which give rise to tree-channel ultra-violet divergences. 

{}What about the graviton kinetic term on the D3-branes?
Actually, the four-dimensional Einstein-Hilbert term receives 
contributions both from the open as well as {\em closed} string diagrams
with two external graviton lines. Let us discuss this in more detail at
the one-loop order. 
First consider the torus amplitude (this is the genus-1 amplitude with
no boundaries or cross-caps). Massless as well as massive
closed string states run in the loop. The untwisted closed string states
renormalize the 10-dimensional part of the effective action.
The twisted closed string states could {\em a priori} 
give rise to a {\em 
six-dimensional} Einstein-Hilbert term (among other terms) localized at the
orbifold fixed point:
\begin{equation}
 {\overline M}_P^4\int d^6x~\sqrt{-{\overline G}}~{\overline R}~.
\end{equation} 
However, the contribution to the 6-dimensional Planck scale 
${\overline M}_P\sim M_s$ due to the twisted closed string states propagating 
in the closed string loop actually vanishes. This follows from the general
result of \cite{Euler}, where it was shown that in Type II string theory on
${\bf R}^{1,3}\times {\cal M}_6$, where ${\cal M}_6$ is some 
(compact or non-compact) Ricci-flat manifold, the one-loop contribution to the
four-dimensional Planck scale is proportional to the Euler characteristic of 
${\cal M}_6$. In our case the latter vanishes, hence the above conclusion.  
This, in turn, can be related to the fact that before orientifolding
in the closed string sector we have 16 supercharges.

{}Next, let us consider the Klein bottle contribution to ${\overline M}_P^4$.
Let us first discuss 
contributions coming from the twisted closed string states running in the
{\em loop channel}. Such contributions actually {\em vanish}. This is because
only left-right symmetric states can contribute into the Klein bottle, while
the twisted closed string states are {\em not} left-right symmetric. In 
particular, note that the orientifold projection $\Omega$ does not
commute with the twist $\theta$ \cite{polchinski}:
\begin{equation}
 \Omega \theta \Omega^{-1}=\theta^{-1}~.
\end{equation} 
This, in particular, implies that the twisted closed string ground states
are not left-right symmetric, so the twisted closed string states do not
contribute to the Klein bottle amplitude. However, the untwisted closed string
states running in the loop channel do contribute into ${\overline M}_P^4$
\cite{Euler1}.
One way to see this is to go to the tree channel. In the tree channel these
contributions correspond to both untwisted as well as twisted exchanges.
The former only renormalize 10-dimensional terms. However, the latter give rise
to a six-dimensional Einstein-Hilbert term (among other terms). The 
corresponding contribution to the six-dimensional Planck scale ${\overline
M}_P^4$ is of order $M_s^4$. Upon compactification on $T^2$ this translates 
into the following contribution to the four-dimensional Planck scale 
${\widehat M}_P^2$:
\begin{equation}\label{six}
 {\overline M}_P^4 (2\pi r)^2\sim {r^2\over(\alpha^\prime)^2}~.
\end{equation}
In particular, this is consistent with our field theory intuition. 

{}Let us now discuss the open string one-loop amplitudes. An important point
here
is that we can attach the external graviton lines anywhere on the tube. In
particular, there is nothing singular about attaching these lines to the 
boundaries or cross-caps. 
Thus, if we do so, we do not change the Chan-Paton structure of
the amplitude. In fact, the correct way of thinking about these amplitudes is
that we consider all possible positions for the insertions, and then integrate
over these positions. It is then clear that the untwisted
annulus and M{\"o}bius strip amplitudes 
(that is, those with the untwisted Chan-Paton matrix $\gamma_1$ acting on a 
boundary) can only contribute to the renormalization of the 10-dimensional
terms. On the other hand, the twisted 
annulus and M{\"o}bius strip amplitudes 
(that is, those with the twisted Chan-Paton matrix $\gamma_\theta$ or
$\gamma_{\theta^2}$ acting on a boundary), can only contribute to the
6-dimensional Planck scale, and this contribution is of order $M_s^4$. 
In particular, there is no truly 4-dimensional
Einstein-Hilbert term coming from these contributions. Since the coordinates
$x^i$ are compact, we once again obtain a contribution of order (\ref{six}).  
That is, the 4-dimensional Planck scale is of order
\begin{equation}
 {\widehat M}_P^2\sim {r^2\over(\alpha^\prime)^2}~,
\end{equation}
which is consistent with the field theory expectation (\ref{fieldth}).

{}Note that if we decompactify the torus ($r\rightarrow\infty$), the 
4-dimensional Planck scale goes to {\em infinity}. 
The D3-brane field theory remains 
non-conformal in this case. This might appear to contradict our earlier claims
that all non-compact ${\cal N}=2$ models are conformal. Actually, there is no
contradiction here. Indeed, even though we took the other three O3$^-$-planes
to infinity, they do {\em not} decouple as the twisted closed string states 
that couple them to the D3-branes propagate in {\em two} extra dimensions,
and the massless 2-dimensional Euclidean propagator is logarithmically
divergent at large distances. Since the tadpoles are not canceled locally in
this background, the D3-brane theory becomes infinitely strongly
coupled. One can see this from the effective field theory viewpoint by noting
that, if we assume that the $SU(N)$ and $SO(N+8)$ gauge couplings are finite
at any finite energy scale, the former blows up at some higher but finite
scale (Landau pole) once we take the cut-off $\Lambda$ to infinity (which is
what happens when we take $r$ to infinity). 

{}Let us now go back to the case where we cancel all tadpoles locally. The low
energy theory on each set of D3-branes is then conformal. Note, however, that
at finite $r$ we have a finite 4-dimensional Planck scale. Thus,
we have a contribution from compactifying the 6-dimensional Einstein-Hilbert
term, which comes from the closed string Klein bottle amplitude as well as the 
open string annulus and M{\"o}bius strip amplitudes, 
on $T^2$. In particular, even though the D3-brane low energy effective field
theories are conformal, the open string annulus and M{\"o}bius strip amplitudes
do contribute into the six-dimensional Planck scale as we have a tower of
massive open string excitations\footnote{Note that having a non-vanishing
four-dimensional Planck scale upon compactification is {\em not}
inconsistent with the fact that the low energy effective field theories on
the branes are conformal. Indeed, we do have massive open string winding states
in this case, which determine the four-dimensional effective field theory
cut-off $\Lambda$ (as the
massive open string excitations do not contribute into the four-dimensional 
gauge coupling renormalization).}.
Now we come to one of the key points of our discussion here - 
in the decompactification limit, even in the conformal case, the 4-dimensional
Planck scale is not zero but {\em infinite}. This is due to the fact that,
in these backgrounds it actually comes from the compactification of the 
{\em six-dimensional} Einstein-Hilbert term localized at the orbifold fixed
point. Also, note that in the conformal case, once we decompactify the 2-torus,
the O3$^-$-planes and D3-branes at the other three fixed points decouple from
those at the origin. And gravity now decouples not because the 4-dimensional
Planck scale goes to zero but because it goes to infinity.  

{}Before we end this subsection, we would like to make the following 
remark. From the above discussion it might be tempting to consider a 
hybrid scenario with two extra dimensions (untouched by the orbifold) 
compactified on a large torus and four others non-compact. In particular,
in this case we could take $M_s\sim {\rm TeV}$, and obtain the desired 
four-dimensional Planck scale ${\widehat M}_P\sim 10^{18}~{\rm GeV}$ by
choosing the size $r$ of the two compact extra dimensions to be of order of
a millimeter as in \cite{TeV}. However, to obtain the desired cross-over scale
$r_c$, the brane thickness in this case would have to be extremely small.
Thus, let us focus on distance scales above the compactification size $r$.
Then we can integrate out the compact extra dimensions, which leaves us
with a model equivalent to having a 3-brane in 8-dimensional bulk. The
4-dimensional and 8-dimensional Planck scales are given by
\begin{eqnarray}
 &&{\widehat M}^2_P\sim M^4_s r^2~,\\
 &&M_P^6\sim M_s^8 r^2~.
\end{eqnarray} 
Following our discussion in subsection E of section III, we then have the 
following relation between the cross-over scale $r_c$ and the brane thickness
$\epsilon$:
\begin{equation}
 r_c\sim {1\over M_s^2\epsilon}~.
\end{equation}   
To have $r_c$ around or above the present Hubble size we must require that
$\epsilon$ is around or smaller than $10^{-30}{\widehat M}_P^{-1}$, which 
would require a presence of a new ultra-high threshold $1/\epsilon$ in the
theory.

\subsection{A Conifold Example}

{}From our discussion in the previous subsection it follows that, if we
consider an oriented background of Type IIB on ${\bf C}\times ({\bf C}^2/
{\bf Z}_2)$ in the presence of D3-branes, we have a {\em six-dimensional}
Einstein-Hilbert term localized at the orbifold fixed point. The
gauge theory on the D3-branes in this case is the ${\cal N}=2$ supersymmetric
theory with the $SU(N)\otimes SU(N)$ gauge group and matter hypermultiplets
transforming in $({\bf N},{\overline{\bf N}})$ and 
$({\overline {\bf N}},{{\bf N}})$. Note that this theory is conformal. 
Moreover, there is no ultra-violet cut-off in the D3-brane gauge theory. This
is because the string excitations do not contribute into the gauge coupling
renormalization as they are non-BPS states (and we have ${\cal N}=2$ 
supersymmetry). That is, the ultra-violet cut-off $\Lambda$ in the theory is 
actually infinite, so the four-dimensional Planck scale is also infinite.

{}Now we would like to deform this theory so that it is no longer conformal.
This can be done as follows. Note that the six-dimensional manifold 
${\cal M}_6$ transverse to the D3-branes is a Ricci-flat manifold with the
metric
\begin{equation}
 ds^2=dr^2+r^2\gamma_{\alpha\beta} dy^\alpha dy^\beta~,
\end{equation} 
where $\gamma_{\alpha\beta}$ is a metric on a compact 5-dimensional 
Einstein manifold ${\cal Y}_5={\bf S}^5/\Gamma$ (${\bf S}^5$ is a unit 
5-sphere, $\sum_{\alpha=1}^3 |z_\alpha|^2=1$), 
$\Gamma={\bf Z}_2$, and the action of the generator $R$ of
${\bf Z}_2$ on ${\bf S}^5$ is given by
\begin{equation}
 Rz_1=z_1~,~~~Rz_{2,3}=-z_{2,3}~.
\end{equation}
Note that the fixed point locus of $R$ in ${\bf S}^5$ is a unit circle 
${\bf S}^1$. We can deform the orbifold by blowing up this fixed circle.
The resulting space ${\widetilde {\cal M}}_6$ is Ricci-flat with the
metric
\begin{equation}
 ds^2=dr^2+r^2{\widetilde \gamma}_{\alpha\beta} dy^\alpha dy^\beta~,
\end{equation} 
where ${\widetilde \gamma}_{\alpha\beta}$ 
is a metric on a compact 5-dimensional 
Einstein manifold ${\cal Y}_5=T^{1,1}=(SU(2)\times SU(2))/U(1)$ \cite{KW}.
The blow-up breaks supersymmetry from ${\cal N}=2$ down to ${\cal N}=1$. 
In particular, in the low energy action it corresponds to adding a mass term
to the adjoint chiral ${\cal N}=1$ supermultiplet in the ${\cal N}=2$ vector
multiplet. This mass term is odd under the interchange of the two $SU(N)$ 
subgroups as the twisted closed string modes associated with the blow-up are
odd under the ${\bf Z}_2$ twist. We therefore have for this mass term 
\cite{KW}:
\begin{equation}
 {m\over 2}\left[{\rm Tr}\left(\Phi^2\right)-
 {\rm Tr}\left({\widetilde \Phi}^2\right)\right]~,
\end{equation}  
where $\Phi$ and ${\widetilde \Phi}$ are the aforementioned adjoint chiral
${\cal N}=1$ supermultiplets corresponding to the first and the second $SU(N)$
subgroups, respectively.

{}Let us decompose the original matter hypermultiplets in terms of the
${\cal N}=1$ chiral supermultiplets:
\begin{equation}
 A_a=2\times ({\bf N},{\overline {\bf N}})~,~~~B_a=
 2\times ({\overline {\bf N}},{{\bf N}})~,~~~a=1,2~.
\end{equation}
In the original ${\cal N}=2$ model we have the following superpotential
(in the ${\cal N}=1$ language):
\begin{equation}
 {\cal W}=g{\rm Tr}\left[\Phi\left(A_1B_1+A_2B_2\right)\right]+
 g{\rm Tr}\left[{\widetilde \Phi}\left(B_1A_1+B_2A_2\right)\right]~,
\end{equation}
where $g$ is the Yang-Mills gauge coupling. Integrating out the fields
$\Phi,{\widetilde \Phi}$, we obtain the following effective
superpotential
\begin{equation}
 {\cal W}_{\rm{\small eff}}={g^2\over 2m}\left[{\rm Tr}(A_1B_1A_2B_2)-
 {\rm Tr}(B_1A_1B_2A_2)\right]~,
\end{equation}
which is non-renormalizable. In the infra-red the ${\cal N}=1$ supersymmetric
$SU(N)\otimes SU(N)$ gauge theory with chiral matter supermultiplets $A_a$ and
$B_a$ flows into a superconformal field theory \cite{KW}, which is what we 
expect from the AdS/CFT correspondence 
\cite{malda,GKP,witten,KS,LNV,BKV,orient,KW,kehagias}. The above superpotential
at the infra-red fixed point becomes a marginal deformation giving rise to
a line of fixed points\cite{KW}\footnote{If we do not impose the ${\bf Z}_2$
symmetry under the interchange of the two gauge groups, we then have a 
fixed surface, where the second marginal deformation corresponds to the
difference between the kinetic energies of the two $SU(N)$'s \cite{KW}.}. 

{}Even though the above blown-up theory is conformal in the infra-red, it is
not a conformal theory. In particular, we have a massive threshold $m$ in the
theory. Note that at the energy scales much higher than $m$ (that is, in the
ultra-violet) the field theory once again becomes conformal - it is simply
described by the original ${\cal N}=2$ superconformal field theory. However,
since the blown-up theory is not conformal, as was originally pointed out in
\cite{alberto}, we expect the Einstein-Hilbert term to be generated on the
D3-branes. Here we would like to understand what the corresponding
four-dimensional Planck scale should be.

{}To begin with, note that the blown-up space ${\widetilde {\cal M}}_6$
is a {\em conifold}, and the D3-branes are located at the conifold
singularity. In particular, we no longer have an orbifold singularity. Let the
blow-up size be $\epsilon$. Then we have
\begin{equation}
 m\sim{\epsilon\over\alpha^\prime}~.
\end{equation} 
Let us assume that $\epsilon^2\ll\alpha^\prime$. Then we have $m^2\ll M_s^2$,
so that the massive adjoint chiral multiplets lie way below the string scale,
and we should be able to trust the world-sheet expansion. The contribution
of the low-energy field theory modes into the four-dimensional Planck scale
${\widehat M}_P^2$ is then expected to be of order $m^2$. This, however, is 
not the end of the story. The key point here is that we now have 
${\cal N}=1$ supersymmetry, so the massive string excitations do contribute
into the gauge coupling renormalization. Note, however, that each massive
level is expected to give a threshold contribution of order
\begin{equation}
 {m^2\over M^2_s}~.
\end{equation} 
Indeed, in the limit $m\rightarrow 0$ these threshold contributions should 
vanish as in this limit the ${\cal N}=2$ supersymmetry is restored. This, in
particular, implies that the gauge theory cut-off $\Lambda$ is {\em not} of
order $M_s$ (the lowest string excitation), but {\em higher}. In fact, we can
estimate this cut-off as follows. We expect the ultra-violet cut-off
in the gauge theory to be of order of the energy scale such that the
combined contribution of the massive string thresholds lying below this
scale becomes of order 1.
Let $n_*$ be the corresponding string level. Then we have:
\begin{equation}
 n_*~{m^2\over M^2_s}\sim 1~,
\end{equation} 
that is, $n_*\sim M_s^2/m^2$. The ultra-violet cut-off is then given by:
\begin{equation}
 \Lambda^2\sim n_* M_s^2\sim {M^4_s\over m^2}\sim {1\over\epsilon^2}~. 
\end{equation}
Thus, the four-dimensional Planck scale
\begin{equation}
 {\widehat M}_P\sim {1\over\epsilon}~.
\end{equation}
Note that this is consistent with our expectation that in the conformal limit
$\epsilon\rightarrow 0$ we must have ${\widehat M}_P\rightarrow\infty$.

{}We can understand this result intuitively in the following way. In the
conformal case (that is, before blowing up the orbifold) 
the annulus and M{\"o}bius strip amplitudes actually give finite
contributions to the 10-dimensional and 6-dimensional Planck scales depending
on whether a given amplitude is untwisted or twisted. But there is no truly
four-dimensional Einstein-Hilbert term. If, as in the previous subsection,  
we compactify the two extra dimensions untouched by the orbifold, then we have
a finite four-dimensional Planck scale which goes to infinity in the
decompactification limit.

{}Let us now discuss what happens 
once we blow up the orbifold (in the case where all six extra dimensions are
non-compact). Note that the transverse space
${\widetilde {\cal M}}_6$ is smooth everywhere except for the conifold
singularity (where the D3-branes are located). Attaching the external
graviton lines to boundaries in, say, the annulus amplitude now {\em is} 
special as this amounts to attaching these lines at the conifold singularity.
In particular, if the locations of the external insertions are away from the
conifold singularity, such diagrams only renormalize the bulk operators.
On the other hand,  
if the locations of the external insertions coincide with the
conifold singularity, the corresponding diagrams renormalize the 4-dimensional
Planck scale (among other terms).

{}Let us now understand the size of the cut-off $\Lambda$. Recall that 
in the orbifold background the infra-red and ultra-violet divergence structures
are identical. The blow-up modifies the background, but since the blow-up
size $\epsilon$ is small, at the leading order the relation between the
infra-red and ultra-violet effects remains intact. Thus, let us consider the
loop-channel infra-red physics. The blow-up introduces the infra-red scale $m$
in the loop channel. This translates into an ultra-violet scale $1/\epsilon$
in the tree channel. In particular, in the tree channel the ultra-violet 
effects become soft around this scale. 
In the tree channel this scale arises
due to the closed string Kaluza-Klein modes associated with the blow-up.
Since the tree-channel ultra-violet effects become soft 
around the scale $1/\epsilon$, the brane thickness is expected to be
of oder $\epsilon$ (recall that, if the brane
matter is non-conformal, the brane thickness is expected to be non-vanishing
so that the divergences discussed in section III are smoothed out). 

{}Similarly, in the tree channel we have an {\em infra-red} cut-off scale
$m$, which arises due to the fact that some of the twisted closed string 
states that are massless in the orbifold limit are now {\em massive} 
with masses of order $m$. In the loop channel this translates into the
{\em ultra-violet} scale $\Lambda\sim 1/\epsilon$. In particular, this implies
that we have a massive {\em open} string threshold associated with this scale
(which decouples in the orbifold limit). This open string threshold is due to
the fact that now the brane thickness is non-vanishing, so that we have a 
massive {\em breathing} mode with mass of order $1/\epsilon$.

{}Thus, as we see, the four-dimensional Planck scale in this example is
controlled by the size of the blow-up. The reason why this is possible is that
the blow up we are considering here is not a marginal deformation of the
orbifold singularity in ${\bf C}\times ({\bf C}^2/{\bf Z}_2)$ but a {\em 
relevant} one. Note that above we considered an oriented Type IIB background.
However, we can generalize our discussion to the orientifold of Type IIB
on ${\bf C}\times ({\bf C}^2/{\bf Z}_2)$. Note that in this case we also
have an O7-plane and 8 of the corresponding D7-branes. We will come back
to unoriented conifold backgrounds in section VI, where we discuss some
phenomenological applications of the orientiworld models.

{}Before we end this subsection, let us make the following remark.
To obtain the desired four-dimensional Planck scale ${\widehat M}_P\sim
10^{18}~{\rm GeV}$, we must choose $\epsilon\sim {\widehat M}_P^{-1}$. Then,
as we discussed in subsection E of section III, we must take 
$M_s\sim {\rm TeV}$ - in this case the cross-over scale $r_c\sim r_H$
(the present Hubble size), while for larger $M_s$ $r_c$ is smaller (and smaller
$M_s$ are phenomenologically disfavored). The infra-red scale $m$ is then
of order $10^{-3}~{\rm eV}$ (about an inverse millimeter). Here it is
difficult to judge whether this is a worrisome feature or a bonus - 
the model at hand is
not realistic, so we will have to postpone discussing this point until we have
a more realistic looking model.

\subsection{Non-conformal ${\cal N}=1$ Models}

{}It is clear that in the ${\cal N}=1$ orientiworld models with 
$\Gamma\subset SO(3)$ the situation is similar to that in the ${\cal N}=2$ 
theories discussed in the previous subsections. Let us therefore focus on the
cases with $\Gamma\subset SU(3)$ but $\Gamma\not\subset SO(3)$. 
As we discussed
in subsection D of section II, the D3-brane gauge theories in these models
are always chiral and non-conformal. Let us therefore understand the origin of
the four-dimensional Planck scale in these models.   

{}First, let us 
focus on the cases where all twists $g_a\in\Gamma$ have fixed point
loci of real dimension 0 (that is, $\Gamma$ does not have a non-trivial
subgroup which is also a subgroup of $SO(3)$, so $\Gamma={\bf Z}_M$, where
$M$ is odd).
Let us start with the closed string torus amplitude. The untwisted closed
string states running in the loop renormalize 10-dimensional 
terms. However, the {\em twisted}
closed string states give rise to a {\em four-dimensional} Einstein-Hilbert
term (among other terms) localized at the orbifold fixed point:
\begin{equation}
 {\widehat M}_P^2\int d^4x~\sqrt{-{\widehat G}}~{\widehat R}~.
\end{equation} 
The corresponding contribution to the 4-dimensional Planck scale 
${\widehat M}^2_P\sim M^2_s$. 
As in the previous subsection, the Klein bottle amplitude also gives a
contribution of the same order of magnitude
into ${\widehat M}^2_P$. This contribution is due to the untwisted
closed string states running in the loop channel, while the twisted closed
string states in the loop channel do not contribute
($\Omega\theta\Omega^{-1}=
\theta^{-1}$). 

{}Next, consider the one-loop open string amplitudes. The untwisted annulus
and M{\"o}bius strip amplitudes renormalize the bulk terms. However, the
{\em twisted} annulus and M{\"o}bius strip amplitudes now contribute to the
renormalization of the four-dimensional Einstein-Hilbert term (as well as other
terms). The corresponding contributions to the four-dimensional Planck scale
${\widehat M}^2_P$ are of order $M_s^2$. This is evident in the 
tree channel, while in the loop channel this can be understood from the fact
that the massive open string excitations in ${\cal N}=1$ backgrounds do
contribute into the renormalization of the D3-brane gauge couplings, so that
the ultra-violet cut-off $\Lambda$ in the D3-brane gauge theory is of order 
$M_s$. 

{}Here we come to an important point. The D3-brane gauge theory is 
non-conformal. At the orbifold point we have no ultra-violet divergences
in the open string loop channel (that is, we have no infra-red divergences
in the tree channel) as all twisted tadpoles are canceled. However, we do
have infra-red divergences in the loop channel, which translate into 
ultra-violet divergences in the tree channel. In fact,
as we discussed in section III, since the D3-brane gauge theory is 
non-conformal, and since the D3-branes are $\delta$-function-like, we expect
ultra-violet divergences to arise in the coupling between gravity and brane
matter at the tree level. On the other hand, we also expect these divergences
to be smoothed out. As we will now argue, blowing up the orbifold smoothes
out these divergences\footnote{The fact that we should blow up the orbifold
in these backgrounds was also argued in \cite{KaSh,KST} from a somewhat
different angle.}. 

{}Before we proceed further, let us 
discuss one important point concerning blowing up such orbifold singularities.
For definiteness, here we will focus on a simple example, but our
conclusions also apply to the general class of models we are discussing here.
Thus, consider the ${\cal M}_6={\bf C}^3/\Gamma$ orbifold, where $\Gamma=
{\bf Z}_3$, and the generator $\theta$ of ${\bf Z}_3$ acts as follows on the
complex coordinates $z_\alpha$, $\alpha=1,2,3$, on ${\bf C}^3$:
\begin{equation}\label{Z-orb}
 \theta: z_\alpha\rightarrow \omega z_\alpha~,
\end{equation} 
where $\omega\equiv \exp(2\pi i/3)$. We have a usual marginal blow-up in this
orbifold. However, we do not have the analog of the relevant blow-up
discussed in the previous subsection. Thus, we can view the above 
${\bf C}^3/{\bf Z}_3$ orbifold as follows. The metric on ${\cal M}_6$ is given
by  
\begin{equation}
 ds^2=dr^2+r^2\gamma_{\alpha\beta} dy^\alpha dy^\beta~,
\end{equation} 
where $\gamma_{\alpha\beta}$ is a metric on a compact 5-dimensional 
Einstein manifold ${\cal Y}_5={\bf S}^5/{\bf Z}_3$. The generator $\theta$ of
${\bf Z}_3$ has the same action on the coordinates $z_\alpha$ on ${\bf S}^5$,
$\sum_{\alpha=1}^3 |z_\alpha|^2=1$, as in (\ref{Z-orb}).
Note that $\theta$ does {\em not} have a fixed point set on ${\cal Y}_5$,
that is, the manifold ${\cal Y}_5$ is {\em smooth}. This is simply another
way of stating the fact that the orbifold singularity in ${\cal M}_6$ is the
same as the conifold singularity in ${\cal M}_6$, so blowing up the orbifold
singularity is equivalent to smoothing out the conifold. And the corresponding 
blow-up is marginal.

{}The fact that the orbifold blow-up is marginal in this case implies that
the blow-up does {\em not} introduce an infra-red cut-off in the tree channel,
that is, there is no new ultra-violet threshold in the loop channel associated 
with the blow-up. In particular, the four-dimensional Planck scale ${\widehat
M}_P$ remains of order $M_s$ even if we consider such a marginal blow-up.
Here we are assuming that the blow-up size, call it  $\Delta$, 
is small: $\Delta^2\ll\alpha^\prime$, so we can still trust the world-sheet
expansion. Note that (unlike 
in the ${\cal N}=2$ cases discussed above) 
here the infra-red and ultra-violet divergence structures are
{\em different}. This {\em a priori} allows to have an ultra-violet cut-off
in the tree channel without having an infra-red cut-off in the tree channel.
In particular, the blow-up provides such an ultra-violet cut-off in the tree
channel - in the closed string sector we have a heavy Kaluza-Klein threshold 
of order $1/\Delta$ associated with the blow-up (according to our 
expectations in section III, the brane thickness is then of order $\Delta$). 
In the loop channel this is
expected to translate into an infra-red cut-off of order
\begin{equation}
 \mu\sim {\Delta\over\alpha^\prime}~.
\end{equation}
That is, after the blow-up we must have light open string states with masses
of order $\mu$. And we do. Indeed, in the gauge theory blowing up the orbifold
corresponds to Higgsing along the F- and D-flat directions\footnote{In the
${\bf Z}_3$ orientifold model, which we review in the next section, the F- and
D-flat directions were discussed in detail in \cite{CVET}.} 
(this is because
the blow-up is marginal), that is, to an appropriate deformation of the gauge
bundle. The massive open string states resulting from this Higgsing then have
masses of order $\mu$. 

{}Note that in the above scenario we have ${\widehat M}_P\sim M_s$, so
both should be of order $10^{18}~{\rm GeV}$. Then, to have the cross-over
scale not smaller than the present Hubble size, the blow-up size should be
extremely small, not larger than $10^{-30}{\widehat M}_P^{-1}$. Then the scale 
$\mu$ is smaller than or of order $10^{-3}~{\rm eV}$ (about an 
inverse millimeter).

\subsection{More General Non-Conformal ${\cal N}=1$ Models}

{}In the previous subsection we focused on the simple case where all twists
in $\Gamma$ have fixed point loci of real dimension 0. In a more general
class of models we can also have twists with fixed point loci of real 
dimension 2. We would like to discuss this class of models next. In fact, as 
we will argue in the following, this class of models 
appears to be most suitable
for the phenomenological goals we have been pursuing.

{}Thus, consider the orbifold group $\Gamma\subset SU(3)$ but $\Gamma\not
\subset SO(3)$ such that some twists in $\Gamma$ have fixed point loci of real
dimension 2. This then implies that there exists a non-trivial subgroup of
$\Gamma$, call it ${\overline \Gamma}$, such that ${\overline \Gamma}\subset
SO(3)$ (in particular, ${\overline \Gamma}$ could be a subgroup of $SU(2)$).
The elements of $\Gamma$ that do not belong to ${\overline \Gamma}$ need
not commute with the elements of ${\overline \Gamma}$. However, for
our purposes here it will suffice to assume that they do (our conclusions, 
however, will be applicable in the general case). In this case $\Gamma=
Z_M\times {\overline \Gamma}$, where $M$ is odd, ${\bf Z}_M\subset SU(3)$ but 
${\bf Z}_M\not\subset SU(2)$ (it then also follows that 
${\bf Z}_M\not\subset SO(3)$). Let the generator of ${\bf Z}_M$ be $\theta$,
while ${\overline \Gamma}=\{{\overline g}_a|a=1,\dots,|{\overline \Gamma}|\}$.
Then $\Gamma=\{\theta^k,\theta^k{\overline g}_a|k=0,\dots, M;
a=1,\dots,|{\overline \Gamma}|\}$.
Note that in this case
we can view the orientifold of Type IIB on ${\bf C}^3/\Gamma$ as the
${\bf Z}_M$ orbifold of the orientifold of Type IIB on ${\bf C}^3/
{\overline \Gamma}$.

{}Let us first discuss gravity in such a model before any blow-ups. The
${\overline g}_a$ ($a\not=1$)
twisted sectors give rise to various 6-dimensional 
Einstein-Hilbert terms localized on the corresponding fixed points. More
precisely, if ${\overline \Gamma}\subset SU(2)$, then these twisted sectors
give rise to only one type of the 6-dimensional Einstein-Hilbert term, while
if ${\overline \Gamma}\not\subset SU(2)$, then we have more then one types
of 6-dimensional Einstein-Hilbert terms (with the extra two dimensions
identified with the corresponding fixed point loci). The $\theta^k {\overline
g}_a$, $k\not=0,a\not=1$, twisted sectors can also give rise to 
6-dimensional Einstein-Hilbert terms, but need not depending upon a particular
orbifold group. If they do not, then they give rise to a 
4-dimensional Einstein-Hilbert term. Finally, the $\theta^k$, $k\not=0$, 
twisted sectors contribute to the 4-dimensional Einstein-Hilbert term.
The important point here is that we always have at least one type of a
6-dimensional Einstein-Hilbert term present in such backgrounds. Note that
the 4-dimensional Planck scale ${\widehat M}_P\sim M_s$, and the 6-dimensional
Planck scale ${\overline M}_P\sim M_s$. 
Suppose we blow up the fixed point at the
origin of ${\bf C}^3/\Gamma$, which is associated
with the ${\bf Z}_M$ orbifold. The brane thickness is then of order of the
corresponding marginal blow-up size $\Delta$. 

{}Since we have 6-dimensional Einstein-Hilbert terms, before we discuss the
cross-over to the 10-dimensional gravity, we must discuss a possible
cross-over to the 6-dimensional gravity. However, since we do not have a
tree-level infra-red cut-off in these backgrounds, following our discussion in
subsection 2 of Appendix C, we conclude that we do not have a cross-over
to the 6-dimensional gravity. The cross-over to the 10-dimensional gravity
then goes as in the previous subsection. In particular, we must have an
extremely small $\Delta$ in this case.

{}The above scenario, however, is not the only possible one. In particular,
we can utilize the fact that, as we discussed in subsection B, 
in some cases we have relevant blow-ups of the orbifold singularities 
associated with the twists ${\overline g}_a$. For illustrative purposes let
us consider a simple example here. Thus, let ${\overline \Gamma}={\bf Z}_2$,
and let $\Gamma={\bf Z}_3\times {\bf Z}_2$, where the generators $\theta$ and
$R$ of ${\bf Z}_3$ respectively ${\bf Z}_2$ have the following action 
on the complex coordinates $z_\alpha$ on ${\bf C}^3$:
\begin{eqnarray}
 &&\theta z_\alpha=\omega z_\alpha~,~~~\alpha=1,2,3~,\\
 &&Rz_1=z_1~,~~~Rz_{2,3}=-z_{2,3}~.
\end{eqnarray}
The metric on ${\cal M}_6={\bf C}^3/\Gamma$ is given by:
\begin{equation}
 ds^2=dr^2+r^2\gamma_{\alpha\beta} dy^\alpha dy^\beta~,
\end{equation} 
where $\gamma_{\alpha\beta}$ is a metric on a compact 5-dimensional 
Einstein manifold ${\cal Y}_5={\bf S}^5/\Gamma$. The generators $\theta$ and 
$R$ have the same action on the coordinates $z_\alpha$ on ${\bf S}^5$,
$\sum_{\alpha=1}^3 |z_\alpha|^2=1$, as above. The fixed point locus of the
twist $R$ in ${\bf S}^5$ is a unit circle ${\bf S}^1$: $|z_1|^2=1$. The action
of the twist $\theta$ on this fixed circle is given by $\theta z_1=\omega z_1$,
so the ${\bf Z}_3$ orbifold simply reduces the size of this circle by a factor
of 3. We can therefore consider a relevant blow-up of this circle as in
subsection B. After this blow-up, whose size we will denote via $\epsilon$,
we no longer have a 6-dimensional Einstein-Hilbert term. Rather, both the ${\bf
Z}_3$ and the blown-up ${\bf Z}_2$ orbifolds contribute into the 4-dimensional
Planck scale. These contributions, however, are very different. Thus, from our
previous discussions it should be clear that at the
one-loop order the contributions of the ${\bf Z}_3$ (as well as ${\bf Z}_6$)
twisted sectors into the
4-dimensional Planck scale ${\widehat M}_P^2$ are of order $M_s^2$. However the
contributions from the blown-up ${\bf Z}_2$ orbifold are of order 
$1/\epsilon^2$, so assuming that $\epsilon^2\ll \alpha^\prime$, we have the
following estimate for the 4-dimensional Planck scale:
\begin{equation}
 {\widehat M}_P\sim {1\over \epsilon}~.
\end{equation}
Let us also blow up the ${\bf Z}_3$ singularities (this blow-up is marginal).
We will choose the corresponding blow-up size $\Delta\sim\epsilon$. It is then
clear that the brane thickness is of order $\epsilon$. Now we can take
$M_s\sim{\rm TeV}$ and obtain the cross-over scale (to the 10-dimensional 
gravity) of order of the present Hubble size. The infra-red scales
\begin{eqnarray}
 &&m\sim{\epsilon\over\alpha^\prime}~,\\
 &&\mu\sim{\Delta\over\alpha^\prime}~
\end{eqnarray}
in this case are both about an inverse millimeter.

{}Thus, as we see, in this case the situation is similar to that discussed in
the ${\cal N}=2$ models in subsection B except that here the D3-brane
gauge theories are non-conformal even in the orbifold limit. That is, one 
might hope to construct various realistic looking non-conformal chiral ${\cal
N}=1$ models in this context. We will discuss such models in section VI.

{}Before we end this subsection, the following remark is in order. Since
we are discussing ${\cal N}=1$ models, we expect that the D3-brane
gauge theory cut-off $\Lambda$ is or order $M_s$. It might then appear 
surprising that ${\widehat M}_P\gg M_s$. However, the key point here is that
there are actually {\em two} ultra-violet cut-offs in such backgrounds. One is
$\Lambda\sim M_s$, and another one is $\Lambda^\prime\sim 1/\epsilon$. In
particular, if we view the low energy modes as parts of ${\cal N}=2$ multiplets
corresponding to the ${\bf C}^3/{\bf Z}_2$ model, we expect that the massive
string thresholds will contribute to the gauge coupling renormalization up to
the scale $\Lambda^\prime$. However, the fact that we do not have complete
${\cal N}=2$ multiplets implies that it makes no sense to talk about the
gauge coupling renormalization beyond the scale $\Lambda$, which is determined
by the truly ${\cal N}=1$ supersymmetric massive string excitations that
contribute to the gauge coupling renormalization starting at the lowest such
excitation. In other words, the gauge theory cut-off is given by $\Lambda$ as 
we have a tower of massive open string states starting at around $M_s$, 
while the four-dimensional gravitational cut-off is given by $\Lambda^\prime$,
and gravity is weak all the way up to ${\widehat M}_P$. This is in some sense
analogous to what happens in little string theories \cite{LST}.   

\subsection{A ``Hybrid'' Scenario}

{}Before we end this section, as an aside
we would like to discuss a somewhat different
scenario, where 4 of the extra dimensions are compact, but we have 2 
non-compact extra dimensions. For illustrative purposes let us focus on the
orbifold group ${\bf Z}_3\times {\bf Z}_2$ 
discussed in the previous subsection. Thus, consider the case where the 
transverse space is given by ${\cal M}_6=({\bf C}
\times {\cal W}_4)/{\bf Z}_3$~, where ${\cal W}_4=T^4/{\bf Z}_2$ is 
(an orbifold limit of) a K3 surface, and the complex coordinates 
$z_1$ and $z_2,z_3$ now parametrize ${\bf C}$ respectively ${\cal W}_4$.
Since we have only two non-compact directions in ${\cal M}_6$, the number of
D3-branes is now fixed by the untwisted tadpole cancellation conditions to be
8. This is because we have 16 O3-planes located at the corresponding fixed 
points on $T^4$. We also have an O7-plane and 8 of the corresponding D7-branes.
Here we should point out that the corresponding ${\bf Z}_3\times {\bf Z}_2$
model is not realistic as the gauge group only contains $SU(2)$ factors
(see subsection E of section V), so we do
not have the $SU(3)_c$ subgroup of the Standard Model. However, here we would
like to use this model to discuss a point {\em a priori} unrelated to how
realistic this model is from the Particle Physics viewpoint. 

{}Thus, suppose the linear sizes of the K3 surface are all of order $1/M_s$.  
At low energies we can integrate out the K3 part and obtain a 
model which is equivalent to having a 3-brane in 6-dimensional bulk with
infinite volume extra-dimensions. Since we have no tree-level infra-red
divergences in this model (all tadpoles are canceled), following our 
discussion in the previous subsection, we could consider a model with
no cross-over from the 4-dimensional gravity to 6-dimensional gravity.
Here we should point out that the number of such ``hybrid'' 
models would be quite limited for the reasons mentioned in Introduction. 

\section{Explicit Examples}

{}In this section we would like to discuss explicit examples of orientiworld
models. As we have already mentioned in Introduction, the number of such models
is infinitely large, and we have a large number of potentially interesting
models to explore. Here we will not attempt to perform an exhaustive
search for phenomenologically most interesting models. Rather, we will discuss
some examples to get a flavor of what kind of features we can expect from the
orientiworld construction (we will discuss phenomenological applications in
the next section). 
Here we will focus on examples with the orbifold
group $\Gamma\subset SU(3)$ but $\Gamma\not \subset SO(3)$. Such examples
have been studied in \cite{orient,orient1,IRU}. For illustrative
purposes here we will discuss some of these models.

\subsection{The ${\bf Z}_3$ Models}

{}The simplest choice of the orbifold group $\Gamma$ in the present context
is $\Gamma={\bf Z}_3=\{1,\theta,\theta^2\}$, where the generator $\theta$ of
${\bf Z}_3$ acts on the complex coordinates $z_\alpha$, $\alpha=1,2,3$, 
on ${\bf C}^3$ as follows:
\begin{equation}
 \theta z_\alpha=\omega z_\alpha~,
\end{equation}
where $\omega\equiv\exp(2\pi i/3)$. At the origin of ${\bf C}^3/\Gamma$ we can
place an O3$^-$- or O3$^+$-plane. Let $\eta=-1$ in the former case, while
$\eta=+1$ in the latter case. The twisted tadpole cancellation requires that
\cite{orient}
\begin{equation}
 {\rm Tr}(\gamma_\theta)=4\eta~.
\end{equation} 
We can then place $(3N+4\eta)$ D3-branes on top of the O3-plane, and choose
\begin{equation}
 \gamma_\theta={\rm diag}(\omega I_N,\omega^{-1} I_N, I_{N+4\eta})~. 
\end{equation}
The massless spectrum of the resulting model is given in Table I. Note that 
the $U(1)$ factor is anomalous and is actually 
broken at the tree level as we discussed in subsection D of section II.
Let $\Phi_\alpha=3\times ({\bf R}_\eta,{\bf 1})$, and $P_\alpha=
3\times ({\overline {\bf N}},{\bf {N+4\eta}})$. Then the tree-level
superpotential in this model is given by:
\begin{equation}
 {\cal W}=\epsilon_{\alpha\beta\gamma} \Phi_\alpha P_\beta P_\gamma+\dots~,
\end{equation}
where the ellipses stand for non-renormalizable couplings.

{}An interesting case to consider is $\eta=-1$ and $N=5$ (in this case we have
11 D3-branes placed on top of the O3$^-$-plane). The gauge group of this
model is $SU(5)$, and we have 3 chiral generations transforming in 
${\bf 10}$ plus ${\overline {\bf 5}}$ of $SU(5)$. A compact version of this
model was originally constructed in \cite{LPT}. Even though we have three
generations and an appealing gauge group in this model, phenomenologically it
is still not very interesting as we have no way of breaking the 
``grand unified'' $SU(5)$ gauge group down to the Standard Model gauge group
$SU(3)_c\otimes SU(2)_w \otimes U(1)_Y$ - we have no Higgs field in an
appropriate higher dimensional representation (such as the adjoint Higgs).

\subsection{The ${\bf Z}_7$ Models}

{}Another simple choice of the orbifold group $\Gamma$ 
is $\Gamma={\bf Z}_7$, where the generator $g$ of
${\bf Z}_7$ acts on the complex coordinates $z_\alpha$, $\alpha=1,2,3$, 
on ${\bf C}^3$ as follows:
\begin{equation}
 gz_1=\alpha z_1~,~~~gz_2=\alpha^2 z_2~,~~~gz_3=\alpha^4 z_3~,
\end{equation}
where $\alpha\equiv\exp(2\pi i/7)$. At the origin of ${\bf C}^3/\Gamma$ we can
place an O3$^-$- or O3$^+$-plane. As before, 
let $\eta=-1$ in the former case, while
$\eta=+1$ in the latter case. The twisted tadpole cancellation requires that
\cite{orient}
\begin{equation}
 {\rm Tr}(\gamma_g)=-4\eta~.
\end{equation} 
We can then place $(7N-4\eta)$ D3-branes on top of the O3-plane, and choose
\begin{equation}
 \gamma_g={\rm diag}(\alpha I_N,\alpha^{-1} I_N,\alpha^2 I_N,\alpha^{-2} 
 I_N, \alpha^4 I_N,\alpha^{-4} I_N,I_{N-4\eta})~. 
\end{equation}
The massless spectrum of the resulting model is given in Table I. Note that 
the $U(1)$ factors are anomalous and are actually 
broken at the tree level as we discussed in subsection D of section II.
Let $\Phi_1=({\bf 1},{\bf 1},{\bf R}_\eta,{\bf 1})$, $P_1=({\bf N},{\bf 1},
{\bf 1},{\bf {N+4\eta}})$, $R_1=({\overline {\bf N}},{\bf N},{\bf 1},{\bf 1})$,
$Q_1=({\bf 1},{\overline{\bf N}},{\overline {\bf N}},{\bf 1})$,
plus cyclic permutations of the $SU(N)\otimes SU(N)\otimes SU(N)$ 
representations which define $\Phi_\alpha$, $P_\alpha$, $R_\alpha$, $Q_\alpha$
with $\alpha=2,3$. The tree-level superpotential in this model is given by:
\begin{equation}
 {\cal W}=\epsilon_{\alpha\beta\gamma} P_\alpha P_\beta Q_\gamma+
 \epsilon_{\alpha\beta\gamma} Q_\alpha R_\beta \Phi_\gamma+
 \epsilon_{\alpha\beta\gamma} R_\alpha R_\beta R_\gamma+\dots~,
\end{equation}
where the ellipses stand for non-renormalizable couplings.

{}Note that in this model we have a cyclic ${\bf Z}_3$ symmetry corresponding 
to the permutations of the $SU(N)\otimes SU(N)\otimes SU(N)$ quantum numbers.
In fact, this is a symmetry of the background itself. In the next subsection
we will consider further orbifolding the ${\bf Z}_7$ models by this symmetry.

\subsection{A Non-Abelian Orbifold}

{}Thus, consider the orbifold group ${\bf Z}_3=\{1,\theta,\theta^2\}$ 
whose generator $\theta$ has the following action on the complex coordinates
$z_\alpha$:
\begin{equation}\label{cyclic}
 \theta z_1=z_2~,~~~\theta z_2=z_3~,~~~\theta z_3=z_1~.
\end{equation}
That is, $\theta$ permutes the complex coordinates $z_\alpha$. Note that 
$\theta$ does not commute with the generator $g$ of the ${\bf Z}_7$ orbifold 
group discussed in the previous subsection:
\begin{eqnarray}
 &&\theta g\theta^{-1}=g^2~,~~~\theta g^2\theta^{-1}=g^4~,~~~
 \theta g^4\theta^{-1}=g~,\\
 &&\theta g^3\theta^{-1}=g^6~,~~~\theta g^6\theta^{-1}=g^5~,~~~
 \theta g^5\theta^{-1}=g^3~.
\end{eqnarray}
Also note that 
\begin{equation}
 (\theta g^k)^3=(\theta^2 g^k)^3=1~,~~~k=0,\dots,6~.
\end{equation}
Thus, $\theta$ and $g$ together generate a non-Abelian group $\Gamma_*$
with $|\Gamma_*|=21$. This orbifold group is a subgroup of $SU(3)$. 

{}Let us consider the orientiworld model corresponding to the ${\bf C}^3/
\Gamma_*$ orbifold. The eigenvalues of $\theta g^k$ and $\theta^2 g^k$, 
$k=0,\dots,6$, 
are $1,\omega,\omega^{-1}$. The twisted tadpole cancellation conditions
then imply that \cite{orient}
\begin{equation}
 {\rm Tr}(\gamma_\theta g^k)={\rm Tr}(\gamma_\theta^2 g^k)=-2\eta~,~~~
 k=0,\dots,6~.
\end{equation}
We can therefore choose:
\begin{eqnarray}
 &&\gamma_g={\rm diag}(\alpha I_N,\alpha^{-1} I_N,\alpha^2 I_N,\alpha^{-2} 
 I_N, \alpha^4 I_N,\alpha^{-4} I_N,I_{N-4\eta})~,\\
 &&\gamma_\theta={\rm diag}(I_{2N}\times {\cal P}_3,\omega I_n,\omega^{-1} I_n,
 I_{n-2\eta})~, 
\end{eqnarray}
where $N=3n+2\eta$, and ${\cal P}_3$ is a $3\times 3$ permutation matrix. 

{}The resulting ${\cal N}=1$ supersymmetric theory has the gauge group
$SU(N)\otimes SU(n)\otimes G_\eta (n-2\eta)$ (see Table I for notations),
where we have dropped the $U(1)$ factors. The charged matter is given by
\begin{eqnarray}
 && \Phi=({\bf R}_\eta,{\bf 1},{\bf 1})~,\\
 && \Sigma=({\bf N},{\bf n},{\bf 1})~,\\
 && {\widetilde \Sigma}=({\bf N},{\overline {\bf n}},{\bf 1})~,\\
 && P=({\bf N},{\bf 1},{\bf n-2\eta})~,\\ 
 && R=({\bf Adj},{\bf 1},{\bf 1})~,\\
 && Q_\eta=({\bf R}_\eta,{\bf 1},{\bf 1})~,\\
 && Q_{-\eta}=({\bf R}_{-\eta},{\bf 1},{\bf 1})~,
\end{eqnarray}
where ${\bf Adj}$ stands for the adjoint representation of $SU(N)$. The
tree-level superpotential is given by:
\begin{equation}
 {\cal W}=\Phi Q_\eta R+PPQ_{-\eta}+\Sigma{\widetilde \Sigma}Q_\eta+
 \Sigma{\widetilde \Sigma}Q_{-\eta}+\dots~,
\end{equation}
where the ellipses stand for non-renormalizable couplings.

{}An interesting feature of this model is that we have matter in the 
{\em adjoint} of the $SU(N)$ subgroup. Such higher dimensional representations
are necessary in grand unified models to break the grand unified gauge group 
down to that of the Standard Model. Note, however, that in this model the
appearance of the adjoint is achieved at the cost of having only one 
``generation'' instead of three\footnote{Compact models with adjoint 
representations were recently discussed in \cite{CSU} in the context of 
Type IIA orientifolds. In particular, an $SU(5)$ model with 2 chiral 
generations was constructed in\cite{CSU}, but, to the best of our knowledge, 
a grand unified orientifold model with 3 
chiral generations has not been constructed yet.}.

\subsection{The ${\bf Z}_3\times {\bf Z}_3$ Models}

{}Next, let us consider the case with 
$\Gamma={\bf Z}_3\times {\bf Z}_3$. 
Let $g_1$ and $g_2$ be the generators of the two ${\bf Z}_3$ subgroups. 
Their action
on the complex coordinates $z_\alpha$ is given by: 
\begin{eqnarray}
 &&g_1z_1=\omega z_1~,~~~g_1z_2=\omega^{-1} z_2~,~~~g_1z_3= z_3~,\\ 
 &&g_2z_1= z_1~,~~~ 
 g_2z_2=\omega z_2~,~~~g_2z_3=\omega^{-1} z_3~. 
\end{eqnarray}  
The twisted tadpole cancellation conditions read \cite{orient1}:
\begin{eqnarray}
 &&{\mbox{Tr}}(\gamma_{g_1})= {\mbox{Tr}}(\gamma_{g_2})=
 {\mbox{Tr}}(\gamma_{g_1g_2})=-2\eta~,\\
 &&{\mbox{Tr}}(\gamma_{g_1g^2_2})={\mbox{Tr}}(\gamma_{g^2_1g_2})=4\eta~.
\end{eqnarray} 
We can place $(9N+4\eta)$ D3-branes on top of the O3-plane located at the
origin of ${\bf C}^3/\Gamma$, and choose
\begin{eqnarray}
 &&\gamma_{g_1}={\rm diag}(
 \omega I_{3N+2\eta},\omega^{-1}I_{3N+2\eta},I_{3N})~,\\
 &&\gamma_{g_2}={\rm diag}(
 \omega I_{N+2\eta},\omega^{-1}I_N,I_N,
 \omega^{-1}I_{N+2\eta},
 \omega I_N,I_N,\omega I_N, \omega^{-1}
 I_N,I_N)~.
\end{eqnarray}
The massless spectrum of the resulting model is given in
Table I (the $U(1)$ factors should be dropped). 
Let 
$\chi_1= ({\bf 1}, {\overline {\bf N}},{\bf 1},{\bf 1},{\bf N})$,
$P_1=({\bf 1},{\overline {\bf N}},{\bf 1}, {\bf N+2\eta},{\bf 1})$, 
$Q_1=({\bf 1},{\bf 1}, {\overline {\bf N}},{\overline {\bf N+2\eta}},
{\bf 1})$, 
$R_1=({\bf 1},{\bf N},{\bf N},{\bf 1},{\bf 1})$,
plus cyclic permutations
of the $SU(N)\otimes SU(N)\otimes SU(N)$ representations 
which define $\chi_\alpha, P_\alpha,R_\alpha, Q_\alpha$ with $\alpha=2,3$.
The tree-level superpotential in this model is given by:
\begin{equation}
 {\cal W}=\epsilon_{\alpha\beta\gamma} \chi_\alpha \chi_\beta R_\gamma+
 \epsilon_{\alpha\beta\gamma} P_\alpha Q_\beta R_\gamma+\dots~,
\end{equation}
where the ellipses stand for non-renormalizable couplings.

{}Note that in this model we also have a cyclic ${\bf Z}_3$ symmetry whose 
generator $\theta$ acts on the complex coordinates $z_\alpha$
as in (\ref{cyclic}).
We can therefore consider further orbifolding by this symmetry. Note that
$\theta$ does not commute with the original ${\bf Z}_3\times {\bf Z}_3$
generators $g_1,g_2$. In fact, $g_1$ and $\theta$ generate a non-Abelian
discrete group\footnote{The group $\Delta(3\cdot 3^2)$ is a semi-direct product
of ${\bf Z}_3$ and ${\bf Z}_3\times {\bf Z}_3$. It is the $n=3$ member of
the infinite series referred to as $\Delta(3\cdot n^2)$, where
$\Delta(3\cdot n^2)$ is a semi-direct product of ${\bf Z}_3$ and 
${\bf Z}_n\times {\bf Z}_n$.} 
$\Gamma=\Delta(3\cdot 3^2)$ with $|\Gamma|=27$, 
which is a subgroup of $SU(3)$. The resulting model has the gauge group
$SU(3N-2\eta)\times SU(N)^4\times G_\eta(N-2\eta)$, and the matter content
can be found in \cite{class}. In particular, we have matter in 
${\bf R}_\eta$ and ${\bf R}_{-\eta}$ of $SU(3N-2\eta)$ 
as well as bifundamentals (but no adjoints). 

\subsection{The ${\bf Z}_3\times {\bf Z}_2$ Models}

{}So far we have discussed cases without D7-branes. Let us now consider cases
with D7-branes. Thus, consider the 
case with $\Gamma={\bf Z}_3\times {\bf Z}_2$. Here
the action of the generators $\theta$ 
of ${\bf Z}_3\subset \Gamma$ and $R$ of ${\bf Z}_2\subset \Gamma$ 
on the complex coordinates $z_\alpha$ 
is given by: 
\begin{eqnarray}
 &&\theta z_\alpha=\omega z_\alpha~,\\
 &&Rz_1=-z_1~,~~~Rz_2=-z_2~,~~~Rz_3=z_3~. 
\end{eqnarray}
Note that now we have an O3-plane as well as a single O7-plane (the O7-plane
is located at the origin of the $z_3$ complex plane). 
The untwisted tadpole cancellation conditions require the presence 
of 8 of the corresponding D7-branes. The twisted tadpole cancellation 
conditions read \cite{orient1} ($k=1,2$):
\begin{eqnarray}
 &&{\mbox{Tr}}(\gamma_{R,3})={\mbox{Tr}}(\gamma_{R,7})=0~,\\
 &&{\mbox{Tr}}(\gamma_{\theta^kR,3})={\mbox{Tr}}(\gamma_{\theta^kR,7})=0~\\
 &&{\mbox{Tr}}(\gamma_{\theta^k,3})={\mbox{Tr}}(\gamma_{\theta^k,7})=-4~.
\end{eqnarray} 
We can place $(6N-4)$ D3-branes on top of the O3-plane, and choose
\begin{eqnarray}
 &&\gamma_{\theta,3}={\mbox{diag}}(\omega {\bf I}_{2N},
 \omega^2 {\bf I}_{2N},{\bf I}_{2N-4})~,\\
 &&\gamma_{R,3}={\mbox{diag}}(i,-i)\times {\bf I}_{3N-2}~,\\
 &&\gamma_{\theta,7}={\mbox{diag}}(\omega {\bf I}_{4},\omega^2 {\bf I}_{4})~,\\
 &&\gamma_{R,7}={\mbox{diag}}(i,-i)\times {\bf I}_{4}~.
\end{eqnarray}
The spectrum of this model is given in Table II (the 33 $U(1)$ factors should
be dropped). Let 
$\Phi_a=2\times ({\bf A}, {\bf 1},
{\bf 1})_{33}$,
${\widetilde \Phi}_a=2\times ({\bf 1},{\overline {\bf A}}, 
{\bf 1})_{33}$, 
$Q_a=2\times ({\overline {\bf N}}, {\bf 1},
{\overline {\bf N-2}})_{33}$, ${\widetilde Q}_a=2\times ({\bf 1},{\bf N}, 
{\bf N-2})_{33}$. Next, let $P=({\bf N},{\overline {\bf N}},
{\bf 1})_{33}$, $R= ({\overline {\bf N}}, {\bf 1},
{\bf N-2})_{33}$, ${\widetilde R}= ( {\bf 1},{\bf N},
{\overline {\bf N-2}})_{33}$, 
$S= ({\bf N}, {\bf 1},{\bf 1};{\bf 2},{\bf 1})_{37}$,
${\widetilde S}=({\bf 1},{\overline {\bf N}},{\bf 1};{\bf 1},{\bf 2})_{37}$. 
We will also use
$T=({\bf 1},{\bf 1},{\bf N-2};{\bf 1},{\bf 2})_{37}$,
${\widetilde T}=({\bf 1},{\bf 1},{\overline {\bf N-2}};{\bf 2},{\bf 1})_{37}$,
$U=({\bf 2},{\bf 2})_{77}$ (see Table II).
Here $a=1,2$ labels the multiplicity of states in the
33 open string sector. The tree-level superpotential in this model is given by:
\begin{eqnarray}
 {\cal W}=&&\Phi_1 Q_2R+\Phi_2 Q_1 R+{\widetilde\Phi_1} {\widetilde Q_2}{\widetilde R}+
 {\widetilde \Phi}_2 {\widetilde Q}_1 {\widetilde R}+\nonumber\\
 &&Q_1{\widetilde Q_2}P+Q_2{\widetilde Q_1}P+T{\widetilde T}U+
 T{\widetilde S}{\widetilde R}+{\widetilde T} S R+\dots~,
\end{eqnarray}
where the ellipses stand for non-renormalizable couplings.

\subsection{The ${\bf Z}_3\times {\bf Z}_2\times {\bf Z}_2$ Models}

{}Let us consider one more case with D7-branes. In particular, here we would 
like to discuss a model with three sets of D7-branes. 
Thus, consider the case with $\Gamma={\bf Z}_3\times 
{\bf Z}_2\times {\bf Z}_2$, where the generators $\theta$ of 
${\bf Z}_3$, $R_1$ of the first ${\bf Z}_2$ subgroup and $R_2$ of the second
${\bf Z}_2$ subgroup act as follows:
\begin{eqnarray}
 &&\theta z_\alpha=\omega z_\alpha~,\\
 &&R_\alpha z_\beta=-(-1)^{\delta_{\alpha\beta}} z_\beta~,
\end{eqnarray}
where $R_3\equiv R_1 R_2$. Note that now we have an O3-plane as well as three
different O7$_\alpha$-planes (the O7$_\alpha$-plane is located at the
origin of the $z_\alpha$ complex plane). 
The untwisted tadpole cancellation conditions require 
the presence of three sets of the corresponding  
D7$_\alpha$-branes with 8 D7-branes in each set. The twisted
tadpole cancellation conditions read \cite{orient1} ($k=1,2$):
\begin{eqnarray}
 &&{\mbox{Tr}}(\gamma_{R_\alpha,3})={\mbox{Tr}}
 (\gamma_{R_\alpha,7_\beta})=0~,\\
 &&{\mbox{Tr}}(\gamma_{\theta^k R_\alpha,3})=
 {\mbox{Tr}}(\gamma_{\theta^kR_\alpha,7_\beta})=0~\\
 &&{\mbox{Tr}}(\gamma_{\theta^k,3})={\mbox{Tr}}(\gamma_{\theta^k,
 7_\alpha})=-4~.
\end{eqnarray}
We can place $(6N-4)$ D3-branes on top of the O3-plane, and choose
\begin{eqnarray}
 &&\gamma_{\theta,3}={\mbox{diag}}(\omega {\bf I}_{2N},
 \omega^2 {\bf I}_{2N},{\bf I}_{2N-4})~,\\
 &&\gamma_{R_\alpha,3}=i\sigma_\alpha\times {\bf I}_{3N-2}~,
\end{eqnarray}
where $\sigma_\alpha$ are the $2\times 2$ Pauli matrices.
The action on the D$7_\alpha$ 
Chan-Paton charges is similar except that their number is fixed.
The spectrum of this model is given in Table II. The 33 $U(1)$ factor is 
anomalous, and is broken at the tree level. Let
$\Phi_\alpha=3\times ({\bf A},{\bf 1})_{33}$,
$\chi_\alpha=3\times ({\overline {\bf N}}, {\bf N-2})_{33}$, 
$P^\alpha=({\bf N},{\bf 1};{\bf 2}_\alpha)_{37_\alpha}$,
$R^\alpha=({\bf 1},{\bf N-2};{\bf 2}_\alpha)_{37_\alpha}$ 
(see Table II).
The tree-level superpotential in this model
is given by:
\begin{equation}
 {\cal W}=\epsilon_{\alpha\beta\gamma}\Phi_\alpha\chi_\beta\chi_\gamma + 
 \chi_\alpha P^\alpha R^\alpha+\dots~,
\end{equation}
where the ellipses stand for non-renormalizable couplings.

\section{A Three-family $SU(6)$ Model}

{}In the ${\bf Z}_3\times {\bf Z}_2\times {\bf Z}_2$ orbifold models
we discussed in the previous section it is particularly interesting to
discuss the case $N=6$ (see subsection F of section V). 
The corresponding ${\cal N}=1$ supersymmetric
model has the gauge group $[SU(6)\otimes Sp(4)]_{33}\otimes
\bigotimes_{\alpha=1}^3 SU(2)_{7_\alpha 7_\alpha}$. As we pointed out at the 
end of section II, at the quantum level 
the 77 gauge symmetry becomes part of the four-dimensional
gauge symmetry. Note that we have deleted the 77 $U(1)$ factors, which are
anomalous and are broken via the generalized Green-Schwarz mechanism.

{}The charged chiral matter multiplets in this model are given by:
\begin{eqnarray}
 &&\Phi_\alpha=3\times ({\bf 15},{\bf 1})_{33}~,\\
 &&\chi_\alpha=3\times ({\overline {\bf 6}}, {\bf 4})_{33}~,\\ 
 &&P^\alpha=({\bf 6},{\bf 1};{\bf 2}_\alpha)_{37_\alpha}~,\\
 &&R^\alpha=({\bf 1},{\bf 4};{\bf 2}_\alpha)_{37_\alpha}~,\\
 &&T^{\alpha\beta}=({\bf 2}_\alpha;{\bf 2}_\beta)_{7_\alpha 7_\beta}~.
\end{eqnarray}
The tree-level superpotential is given by
\begin{equation}\label{supSU(6)}
 {\cal W}=\epsilon_{\alpha\beta\gamma}\Phi_\alpha\chi_\beta\chi_\gamma + 
 \chi_\alpha P^\alpha R^\alpha+\dots~,
\end{equation}
where the ellipses stand for non-renormalizable couplings.

{}We can Higgs the 77 gauge group down to the $SU(2)$ diagonal subgroup using
the $T^{\alpha\beta}$ matter multiplets. The resulting model now has the gauge
group $SU(6)\otimes Sp(4)\otimes SU(2)$ and the charged matter multiplets
\begin{eqnarray}
 &&\Phi_\alpha=3\times ({\bf 15},{\bf 1},{\bf 1})~,\\
 &&\chi_\alpha=3\times ({\overline {\bf 6}}, {\bf 4},{\bf 1})~,\\ 
 &&P^\alpha=3\times ({\bf 6},{\bf 1},{\bf 2})~,\\
 &&R^\alpha=3\times ({\bf 1},{\bf 4},{\bf 2})~,\\
 &&T=({\bf 1},{\bf 1},{\bf 3})~,
\end{eqnarray}
and the tree-level superpotential is still given by (\ref{supSU(6)}).

{}Note that we cannot break the $SU(6)$ gauge group down to that of the 
Standard Model as we have no appropriate higher dimensional Higgs fields such
as an adjoint Higgs. However, following \cite{KT}, we can still break the 
$SU(6)\otimes Sp(4)\otimes SU(2)$ gauge group down to $SU(3)_c\otimes SU(2)_w
\otimes U(1)_Y$. This can be done as follows. Note that under the breaking
$SU(6)\otimes Sp(4)\supset [SU(4)\otimes SU(2)\otimes U(1)]\times[SU(2)\otimes
SU(2)]$ we have the following branchings (the $U(1)$ charges are given in 
parenthesis):
\begin{eqnarray}
 &&({\bf 15},{\bf 1})=({\bf 6},{\bf 1},{\bf 1},{\bf 1})(+2)\oplus
 ({\bf 1},{\bf 1},{\bf 1},{\bf 1})(-4)\oplus({{\bf 4}},{\bf 2},
 {\bf 1},{\bf 1})(-1)~,\\
 &&({\overline {\bf 6}},{\bf 4})=({\overline {\bf 4}},{\bf 1},{\bf 2},{\bf 1})
 (-1)\oplus({\overline {\bf 4}},{\bf 1},{\bf 1},{\bf 2})(-1)\oplus
 ({\bf 1},{\bf 2},{\bf 2},{\bf 1})(+2)\oplus({\bf 1},{\bf 2},{\bf 1},{\bf 2})
 (+2)~.
\end{eqnarray}
Now consider giving non-zero vacuum expectation values to $({\bf 1},{\bf 1},
{\bf 1},{\bf 1})(-4)$ in $\Phi_1$ and $({\bf 1},{\bf 2},{\bf 2},{\bf 1})(+2)$
in $\chi_1$ (it is not difficult to see that we have the required F- and 
D-flat directions). This Higgsing breaks the $SU(6)\otimes Sp(4)\otimes SU(2)$
gauge symmetry down to $SU(4)\otimes SU(2)\otimes SU(2)\otimes SU(2)$. The
charged matter is given by\footnote{Determining the matter content after 
Higgsing requires some care as we have non-trivial couplings in the 
superpotential.}:
\begin{eqnarray}
 &&\Phi_\alpha=3\times ({\bf 6},{\bf 1},{\bf 1},{\bf 1})~,\\
 &&{\widetilde \Phi}_1=({\bf 4},{\bf 1},{\bf 2},{\bf 1})~,\\
 &&\chi_\alpha=3\times ({\overline {\bf 4}},{\bf 2},{\bf 1},{\bf 1})~,\\
 &&{\widetilde \chi}_1=({\overline{\bf 4}},{\bf 1},{\bf 2},{\bf 1})~,\\
 &&{\widehat \chi}_1=({\bf 1},{\bf 2},{\bf 2},{\bf 1})~,\\
 &&P^\alpha=3\times ({\bf 4},{\bf 1},{\bf 1},{\bf 2})~,\\
 &&{\widehat P}^1=({\bf 1},{\bf 2},{\bf 1},{\bf 2})~,\\
 &&R^\alpha=3\times ({\bf 1},{\bf 1},{\bf 2},{\bf 2})~,\\
 &&{\widehat R}^1=({\bf 1},{\bf 2},{\bf 1},{\bf 2})~,\\   
 &&T=({\bf 1},{\bf 1},{\bf 1},{\bf 3})~,
\end{eqnarray}
and the superpotential is given by:
\begin{equation}\label{supHiggs}
 {\cal W}=\epsilon_{\alpha\beta\gamma}\Phi_\alpha\chi_\beta\chi_\gamma+
 \chi_1 P^1 {\widehat R}^1+{\widetilde\chi}_1 P^1 R^1+{\widehat \chi}_1
 {\widehat P}^1 R^1+\dots~.
\end{equation}
Note that this superpotential does not contain $R^2$ or $R^3$. Let us give
a non-vanishing vacuum expectation value to $R^2$. This breaks the gauge group
down to the {\em Pati-Salam} 
gauge group $SU(4)_c\otimes SU(2)_w\otimes SU(2)_R$. The
charged matter is given by (we have interchanged ${\bf 4}$ and ${\overline
{\bf 4}}$ of $SU(4)$):
\begin{eqnarray}
 &&\Phi_\alpha=3\times ({\bf 6},{\bf 1},{\bf 1})~,\\
 &&{\widetilde \Phi}_1=({\overline{\bf 4}},{\bf 1},{\bf 2})~,\\
 &&\chi_\alpha=3\times ({{\bf 4}},{\bf 2},{\bf 1})~,\\
 &&{\widetilde \chi}_1=({{\bf 4}},{\bf 1},{\bf 2})~,\\
 &&{\widehat \chi}_1=({\bf 1},{\bf 2},{\bf 2})~,\\
 &&P^\alpha=3\times ({\overline {\bf 4}},{\bf 1},{\bf 2})~,\\
 &&{\widehat P}^1=({\bf 1},{\bf 2},{\bf 2})~,\\
 &&R^1,R^3=2\times ({\bf 1},{\bf 1},{\bf 3})~,\\
 &&{\widehat R}^1=({\bf 1},{\bf 2},{\bf 2})~,\\   
 &&T=({\bf 1},{\bf 1},{\bf 3})~,
\end{eqnarray}
and the superpotential is given by (\ref{supHiggs}).

{}We can identify $\chi_\alpha$ and $P^\alpha$ as three generations of quarks
and leptons in the Pati-Salam model. The fields ${\widetilde\chi}_1$ and
${\widetilde\Phi}_1$ are the Pati-Salam Higgs fields, 
which serve the purpose of
breaking the Pati-Salam gauge group down to $SU(3)_c\otimes SU(2)_w\otimes
U(1)_Y$ (that is, non-vanishing vacuum expectation values of 
${\widetilde\chi}_1$ and ${\widetilde\Phi}_1$ break $SU(4)_c\otimes SU(2)_R$
down to $SU(3)_c\otimes U(1)_Y$). The field ${\widehat R}^1$ gives rise to the
electroweak Higgs doublets. In fact, as can be seen from the superpotential
(\ref{supHiggs}), the third generation (that is, the top quark generation)
is identified with $\chi_1$ and $P^1$. The extra triplets $R^1,R^3,T$ should
acquire soft masses upon supersymmetry breaking (these masses are presumably
of order $M_s\sim {\rm TeV}$). Here we note that, upon integrating out these
states, it might be possible to generate higher dimensional operators
responsible for giving masses to the other two chiral generations.
Potentially dangerous states are the
anti-symmetric states $\Phi_\alpha$. They too could acquire masses upon 
supersymmetry breaking. However, it needs to be checked whether there are
no dangerous baryon number violating couplings due to these states. This is,
however, outside of the scope of this paper.

{}Thus, as we see, in the orientiworld context we have constructed a truly
3-family Pati-Salam model with the correct Higgs content. In fact, we even have
the correct couplings in the superpotential for one generation to be much 
heavier then the other two. To the best of our knowledge, this model is the 
closest to the Minimal Supersymmetric Standard Model constructed via the 
orientifold construction. 

{}Before we end this section, let us make the following remark. Here we 
discussed this model at the orbifold point. To obtain four-dimensional 
gravity, as we discussed in section IV, we need to consider relevant blow-ups
of the ${\bf Z}_2$ twists, that is, we need to consider the corresponding 
conifold background. In the process of blowing up these twists, some of the
massless states acquire non-zero masses. However, these masses are about
$10^{-3}~{\rm eV}$, so the above discussion is not expected to be modified
substantially. It would be interesting to understand whether this scale 
$10^{-3}~{\rm eV}$ has anything to do with the neutrino masses (note that 
in the Pati-Salam model we have right-handed neutrinos).

\section{Discussion}

{}In the orientiworld framework we discussed in this paper the Standard Model
gauge and matter fields are localized on (a collection of) D3-branes embedded
in infinite-volume extra space. This transverse space can be an orbifold or
a more general conifold. The four-dimensional gravity on D3-branes is 
reproduced due to a four-dimensional Einstein-Hilbert term induced at the 
quantum level. In particular, gravity can be four-dimensional up to the
cross-over distance scale $r_c$, which can be as large as the present Hubble
size. Not surprisingly, this requires a hierarchy of scales. Here we would like
to comment on this hierarchy.

{}For definiteness let us focus on the scenario we discussed in subsection D
of section IV. In this scenario, where the transverse space is a conifold
obtained by relevant blow-ups of certain orbifold singularities, the string 
scale $M_s$ is of order TeV, while the four-dimensional Planck scale 
${\widehat M}_P\sim 1/\epsilon$, where $\epsilon$ is the size of these 
blow-ups. The aforementioned hierarchy then is in having two very different
distance scales, namely, the string length $1/M_s$ and the blow-up size 
$\epsilon$. One possibility is that the size of the blow-up is 
determined dynamically upon supersymmetry breaking. So we must understand 
supersymmetry breaking in the orientiworld framework.

{}One possibility for obtaining non-supersymmetric orientiworld models is
to start from ${\cal N}=1$ supersymmetric models, and break supersymmetry
dynamically. This would require constructing orientiworld models with certain
additional sectors that would dynamically break supersymmetry and then mediate
it to the Standard Model sector. Alternatively, especially in scenarios with
$M_s$ of order TeV, we could consider constructing orientiworld backgrounds 
that are non-supersymmetric already at the tree level. Such models were
originally discussed in \cite{non-susy} (for subsequent developments see
\cite{non-susy1}). Thus, one can construct
anomaly-free chiral models where all tadpoles are canceled. In particular,
these models contain anti-D3-branes (or anti-O3-planes). However, in all of
such models we have {\em twisted} 
closed string tachyons. Let us point out that in the models
of \cite{non-susy} all divergences due to the closed string tachyons coupling 
to the D-branes and orientifold planes are canceled just as 
all the massless tadpoles. So the presence of the closed string tachyons does
not destabilize the gauge theories on D3-branes. As to the bulk,
there is a possibility that, once all the tachyons condense, we obtain a
supersymmetric closed string background \cite{SilPol}. If so, this would mean
that we could construct orientiworld models where the bulk is supersymmetric,
while the brane supersymmetry is broken at the tree level. In other words,
the branes would be non-BPS, and would break all supersymmetries.

{}Such orientiworld models could also be interesting for the cosmological
constant problem. In particular, 
if the bulk supersymmetry is intact, one might hope
that it might protect the brane cosmological constant 
\cite{DGP0,witten1,DVALI,zura}. Thus, if we consider a codimension-2 or higher
brane in infinite-volume extra space, we have flat brane solutions for
a continuous range of the brane tension (that is, unlike in the codimension-1
case, positive brane tension does not necessarily imply inflation on the 
brane). This by itself, however, is not sufficient to successfully address the
cosmological constant problem. Indeed, even in higher codimension cases we do 
have solutions with inflation on the brane. The idea then is the 
following \cite{DGP0,witten1}.
Suppose the bulk is supersymmetric, and we break brane supersymmetry.
Since the volume of the extra space is infinite, the bulk supersymmetry is,
nonetheless, intact. Now, the cosmological constant is an infra-red quantity - 
it is measured at large distances. In the scenarios we discussed above,
gravity becomes higher dimensional around the Hubble size, so, at least
naively, we would then be measuring the higher-dimensional cosmological
constant (and not the four-dimensional one), which vanishes\footnote{Note that
we could still obtain the accelerated Universe as in \cite{DDG}.} 
due to the unbroken
bulk supersymmetry. It would be interesting to test this idea in higher 
codimension models\footnote{This idea was tested in \cite{zura} in
codimension-1 cases. It appears that in these cases either the brane 
world-volume theory is conformal, or gravity actually becomes four-dimensional
again at the Hubble size in the presence of non-vanishing cosmological constant
on the brane.}.

{}Another point we would like to comment on here is gauge coupling unification.
Thus, in the scenario where $M_s$ is of order TeV we might expect that the
extrapolated unification of gauge couplings in the MSSM would have to be an
accident. Here we would like to point out that this might not necessarily be 
the case. First, one might consider a scenario where the Standard Model fields
live on D4- or D5-branes with the corresponding compactification sizes along
the branes of order
of $1/M_s$. Then we could have accelerated unification as in 
\cite{dienes,TSSM,TSSM1}. 
However, {\em a priori} there is another possibility here,
which need {\em not} involve compactification.
The general idea was originally discussed in \cite{BW} (also see
\cite{A-H}), and here we would like
to tailor it for the orientiworld context. As we have already mentioned, in
the scenario discussed in subsection D of section IV we have two different
cut-offs on D3-branes, namely, $\Lambda\sim M_s$ and $\Lambda^\prime\sim 
{\widehat M}_P$. The key reason why this is possible is that we have {\em two}
extra dimensions transverse to the D3-branes that are untouched by ${\cal N}=2$
supersymmetric orbifold twists. If these dimensions were compact, then we could
have large logarithmic threshold corrections due to the corresponding open 
string windings \cite{bachas}. The compactification in this case plays
the role of an ultra-violet cut-off. In the orientiworld context, once we 
deform to a conifold, we effectively
have such a cut-off, namely, $\Lambda^\prime$, even without compactification.
Then one might hope that we could achieve unification along the lines of
\cite{BW} by having the dilaton and/or the relevant
twisted field vacuum expectation values varying from point to point 
in the bulk
(in some cases such backgrounds could be studied in the 
F-theory \cite{vafa} language \cite{BW}). 
It would be interesting to see if this idea can be made more concrete.

{}Finally, let us mention that various issues such as proton stability, flavor
violation as well as neutrino masses might need to be addressed in the 
orientiworld context with $M_s$ of order TeV. It is important to note that
gravity in such scenarios is weak. Nonetheless, one might have to worry
about the open string excitations in this context. Some of the ideas 
discussed in the context of \cite{TeV} might be applicable in the orientiworld
context as well. In particular, proton stabilization \cite{proton,proton1},
neutrino masses \cite{neutrino,proton1} and flavor violation 
\cite{flavor,flavor1,flavor2}
can be addressed in theories with $M_s$ of order TeV. 

\acknowledgments

{}I would like to thank Gia Dvali, Gregory Gabadadze and Ignatios Antoniadis
for valuable discussions.
This work was supported in part by the National Science Foundation and
an Alfred P. Sloan Fellowship. Parts of this work were completed during
my visit at New York University.
I would also like to thank Albert and Ribena Yu for financial support.

\appendix

\section{The $U(1)$ Factors in Orientiworld Models}

{}In this appendix we give the calculation of the tree-level exchange
of a twisted 2-form in the cylinder amplitude that leads to renormalization
of the $U(1)$ gauge couplings in orientiworld models. More precisely, as we
already mentioned in subsection C of section II, for our purposes here it
suffices to discuss this calculation in the oriented Type IIB setup. In fact,
for simplicity we will focus on the ${\bf Z}_2$ model discussed in 
subsection C of section II.  

{}The relevant action reads (see subsection C of section II for details):
\begin{equation}
 -{L^2\over 12}\int_{\rm D3} {\overline H}_R^2+\int_{\rm D3} {\overline
 C}_R \wedge F_- -{1\over 4g_-^2} \int_{\rm D3} F_-^2-
 {1\over 12}\int_{{\rm D3}\times {\bf R}^2} H^2_R~.
\end{equation}
What we need to compute is the exchange of the 2-form $C_R$ between two
boundaries with a $U(1)_-$ gauge boson attached to each boundary. To do this,
we can simply compute the $C_R$ field created by a $U(1)_-$ background field,
and then couple it back to this background field. That is, we are simply
computing the classical self-interaction of the $U(1)_-$ field strength via
the $C_R$ field.

{}We need to fix the gauge for the $C_R$ field. A convenient gauge choice is
\begin{equation}
 \partial_M C^{MN}_R=0~.
\end{equation}
The gauge-fixed equation of motion for $C_R$ is then given by (here we are
taking into account the fact that $C^{\mu i}_R=C^{ij}_R=0$):
\begin{equation}
 -\partial^K\partial_K C_R^{MN}={\delta^M}_\mu{\delta^N}_\nu
 \left[2\epsilon^{\mu\nu}{}_{\sigma\rho}F_-^{\sigma\rho}+
 L^2 \partial^\sigma\partial_\sigma C^{\mu\nu}_R\right]\delta^{(2)}(x^i)~.
\end{equation}
It is convenient to Fourier transform the coordinates $x^\mu$ and $x^i$.
Let the corresponding momenta be $p^\mu$ and $k^i$, respectively. We then 
have the following equation for $C_R^{MN}(p,x^i)$:
\begin{equation}
 C_R^{MN}(p,x^i)={\delta^M}_\mu{\delta^N}_\nu
 \left[2\epsilon^{\mu\nu}{}_{\sigma\rho}F_-^{\sigma\rho}(p)-
 L^2 p^2 C^{\mu\nu}_R(p,0)\right]
 \int {d^2k\over(2\pi)^2}~{e^{-ik\cdot x}\over{k^2+p^2}}~.
\end{equation}
In particular, we have:
\begin{equation}
 {\overline C}^{\mu\nu}_R(p)=
 \left[2\epsilon^{\mu\nu}{}_{\sigma\rho}F_-^{\sigma\rho}(p)-
 L^2 p^2 {\overline C}^{\mu\nu}_R(p)\right]
 \int {d^2k\over(2\pi)^2}~{1\over{k^2+p^2}}~.
\end{equation}
We must regularize the integral over $k^i$ in the ultra-violet. We will use
the following regularization:
\begin{equation}
 \int {d^2k\over(2\pi)^2}~{1\over{k^2+p^2}}=
 \int^{k^2={\widetilde \Lambda}^2} 
 {d^2k\over(2\pi)^2}~{1\over{k^2+p^2}}={1\over 4\pi}
 \ln\left(1+{{\widetilde \Lambda}^2\over p^2}\right)~.
\end{equation}
Note that ${\widetilde \Lambda}$ here is the ultra-violet cut-off in the
{\em tree channel}, which corresponds to an infra-red cut-off in the {\em loop
channel}. We then have:
\begin{equation}
 {\overline C}^{\mu\nu}_R(p)={1\over 2\pi}
 \epsilon^{\mu\nu}{}_{\sigma\rho}F_-^{\sigma\rho}(p)~
 {\ln\left(1+{{\widetilde \Lambda}^2\over p^2}\right)
 \over{1+{L^2 p^2\over 4\pi}
 \ln\left(1+{{\widetilde \Lambda}^2\over p^2}\right)}}~.
\end{equation}
The self-interaction term is given by:
\begin{equation}
 \int_{\rm D3} {\overline C}_R \wedge F_-=-{2\over\pi}
 \int {d^4p\over(2\pi)^4}~F_-^{\mu\nu}(p)F_{-\mu\nu}(-p)~
 {\ln\left(1+{{\widetilde \Lambda}^2\over p^2}\right)
 \over{1+{L^2 p^2\over 4\pi}
 \ln\left(1+{{\widetilde \Lambda}^2\over p^2}\right)}}~.
\end{equation}
This results in the renormalization of the $U(1)_-$ gauge coupling. The
renormalized kinetic term for the $U(1)_-$ gauge field is now given by:
\begin{equation}\label{renorm}
 -\int {d^4p\over(2\pi)^4}~{1\over 4{\widetilde g}_-^2(p)}~
 F_-^{\mu\nu}(p)F_{-\mu\nu}(-p)~,
\end{equation}
where 
\begin{equation}
 {1\over {\widetilde g}_-^2(p)}={1\over {g}_-^2} +{8\over\pi}~
 {\ln\left(1+{{\widetilde \Lambda}^2\over p^2}\right)
 \over{1+{L^2 p^2\over 4\pi}
 \ln\left(1+{{\widetilde \Lambda}^2\over p^2}\right)}}~.
\end{equation}
For $L=0$ this reproduces the usual logarithmic $U(1)_-$ gauge coupling
running (in fact, it is not difficult to check that with the normalization
for the bare $U(1)_-$ gauge coupling $g_-$ we have adopted, we have the correct
one-loop $\beta$-function coefficient). In this case we have a logarithmic
divergence, which must be regularized by introducing the tree-channel
ultra-violet cut-off ${\widetilde \Lambda}$. We also have a logarithmic
tree-channel infra-red divergence at $p^2\rightarrow 0$, which translates into
the corresponding loop-channel ultra-violet divergence. 

{}However, for non-vanishing $L$
the gauge coupling does {\em not} run. In fact, we only have a {\em finite
non-local} correction even if we take ${\widetilde \Lambda}
\rightarrow \infty$ limit (that 
is, if we remove the tree-channel ultra-violet cut-off):
\begin{equation}
 {1\over {\widetilde g}_-^2(p)}={1\over {g}_-^2} \left[1+{2m_-^2\over p^2}
 \right]~.
\end{equation}
Here it is important to note that we do {\em not} have a tree-channel infra-red
divergence either. In particular, the integral over $p^\mu$ in (\ref{renorm})
is cut off in the infra-red at $p^2=m_-^2$, where
\begin{equation}
 m_-^2={16 g_-^2\over L^2}
\end{equation}
is the mass of the $U(1)_-$ gauge boson. That is, we do not have a loop-channel
ultra-violet divergence either.

{}Thus, as we see, once we introduce the kinetic term for the $C_R$ field
on the D3-branes, we have no ultra-violet or infra-red divergences 
in the $U(1)_-$ coupling.
As we argued in subsection C of section II, such a kinetic term would be
generated at the one-loop level with
\begin{equation}\label{L}
 L^2=b(\alpha^\prime)^2\Lambda^2~,
\end{equation}
where $b$ is a dimensionless numerical coefficient, and
$\Lambda$ is the loop-channel ultra-violet cut-off. We then interpret the above
result as the $U(1)_-$ symmetry being {\em global} and not local. 

{}In fact, here we would like to argue that this is consistent with the case 
where the two extra dimensions $x^i$ are compactified on a large 2-torus.
Thus, consider a model where we compactify $x^i$ on a square 2-torus $T^2$:
\begin{equation}
 x^i\sim x^i+2\pi r~.
\end{equation} 
Then in the absence of Wilson lines on $T^2$ we actually have the same
four-dimensional spectrum. Let us assume that $r^2\gg\alpha^\prime$. Then we 
actually have a {\em tree-level} 4-dimensional kinetic term (\ref{kinetic})
for the $C_R$ field with 
\begin{equation}
 L^2=(2\pi r)^2~.
\end{equation}
Thus, the $U(1)_-$ 
gauge field is massive. In fact, we can now see why (\ref{L})
should be 
the case. Note that in the D3-brane language we have {\em winding} open
string modes with masses:
\begin{equation}
 M_w^2=w^2 {r^2\over(\alpha^\prime)^2}~,
\end{equation}
where $w^2=(w_1)^2 +(w_2)^2$, 
and $w_i\in {\bf Z}$. This can also be seen by T-dualizing
along the $T^2$. The D3-branes then go into D5-branes, and the 2-torus $T^2$
goes into a dual 2-torus ${\widetilde T}^2$ with the coordinates
\begin{equation}
 {\widetilde x}^i\sim{\widetilde x}^i+2\pi {\widetilde r}~,
\end{equation}
where 
\begin{equation}
 {\widetilde r}={\alpha^\prime\over r}~.
\end{equation}
Note that ${\widetilde r}^2\ll\alpha^\prime$. Now we have Kaluza-Klein modes
with masses
\begin{equation}
 M_m^2={m^2\over {\widetilde r}^2}~,
\end{equation}
where $m^2=(m_1)^2+(m_2)^2$, and $m_i\in{\bf Z}$. These Kaluza-Klein modes in
the D5-brane language correspond to the winding modes in the D3-brane language.

{}Note that the theory on the D5-branes is a 4-dimensional gauge theory
at the momenta $p^2\ll 1/{\widetilde r}^2$ (that is, below the Kaluza-Klein
threshold). Even though the string excitations lie way below the Kaluza-Klein
threshold, they do not contribute to the gauge coupling renormalization as
we discussed in subsection C of section II. This implies that the 4-dimensional
gauge theory cut-off is given by the Kaluza-Klein threshold 
\cite{BF,bachas,ABD} (that is, compactification on $T^2$ effectively
introduces an ultra-violet cut-off into the D3-brane gauge theory):
\begin{equation}
 \Lambda={\xi\over {\widetilde r}}={\xi r\over\alpha^\prime}~,
\end{equation}
where a dimensionless numerical coefficient $\xi$ parametrizes the
subtraction scheme dependence. This confirms our intuition that we should have
(\ref{L}). Note that at the momenta $p^2\gg 1/{\widetilde r}^2$ the D5-brane
theory becomes 6-dimensional. In this 6-dimensional theory (which can be 
thought about in the language of a model where we place non-compact D5-branes
at the ${\bf Z}_2$ orbifold singularity) the $U(1)_-$ gauge symmetry is 
anomalous (note that in six dimensions $U(1)$ anomalies come from box 
diagrams). It is broken by a Chern-Simons 
mixing with the twisted 4-form, call it $C^\prime_R$:
\begin{equation}
 \int_{\rm D5} C^\prime_R\wedge F_- ~.
\end{equation}
Note that this twisted 4-form propagates in six dimensions along the D5-brane
world-volume. Its exchange in the cylinder amplitude with a $U(1)_-$ gauge
boson attached to each boundary gives rise to a quadratic divergence. 
However, this quadratic divergence is removed once $U(1)_-$ acquires 
mass, which breaks the $U(1)_-$ gauge symmetry to a global one.

\section{A Completely Smooth Brane}

{}In this Appendix we would like to discuss the case where we smooth out the
brane completely, that is, when we replace the $d$-dimensional 
$\delta$-function $\delta^{(d)}(x^i-x^i_*)$ by a smooth distribution
$f^{(d)}(x^i-x^i_*)$. In fact, for 
our purposes here it will suffice to study the simple distribution
$f_1(r)$ given by (\ref{dist1}).

{}Thus, consider the equation of motion for $\chi$ in the absence of the brane
matter (see subsection C of section III for details):
\begin{equation}
 \left(\partial^i\partial_i-p^2\right)\chi=-\kappa L^d p^2\chi f_1(r)~.
\end{equation}
We have a continuum of modes with $p^2\leq 0$ corresponding to massless and
massive modes (here we are working with Minkowski momenta $p^\mu$). However,
we also have an infinite tower of quadratically normalizable tachyonic modes.
Thus, let us assume that $p^2>0$. Let
\begin{equation}
 q^2\equiv p^2\left[{\kappa L^d\over v_d\epsilon^d}-1\right]\equiv 
 \gamma^2 p^2>0~.
\end{equation}
Then the equation of motion for $\chi$ reads:
\begin{eqnarray}
 &&\left(\partial^i\partial_i+q^2\right)\chi=0~,~~~r\leq \epsilon~,\\
 &&\left(\partial^i\partial_i-p^2\right)\chi=0~,~~~r>\epsilon~.
\end{eqnarray}
As before, for simplicity let us focus on the $d=3$ case. Then the 
radially symmetric solution is given by:
\begin{eqnarray}
 &&\chi(r)=C~{\sin(qr)\over r}~,~~~r\leq \epsilon~,\\
 &&\chi(r)=C~\sin(p\epsilon)~{\exp\left(-p[r-\epsilon]\right)\over r}~,~~~
 r>\epsilon~,
\end{eqnarray}
where $C$ is a constant, and the matching of $\partial_r\chi(r)$ at
$r=\epsilon$ gives the following condition on $p$:
\begin{equation}
 \tan(q\epsilon)=-{q\over p}~.
\end{equation}
This equation has infinitely many roots (here we must keep those corresponding
to $p>0$):
\begin{equation}
 p_n={\pi\over 2\gamma\epsilon}~(2n-1)+{1\over\gamma\epsilon}\arctan
 \left({1\over\gamma}\right)~,~~~n\in {\bf N}~.
\end{equation}
If the brane thickness is small ($\epsilon\ll L$), then $\gamma\gg 1$, and
we have
\begin{equation}
 p_n\approx{\pi\over 2\gamma\epsilon}~(2n-1)~,~~~n\in {\bf N}~.
\end{equation} 
Thus, we have an infinite tower of quadratically normalizable tachyonic modes
in this case.

\section{Graviton Propagator on The Brane}

{}In the main text we studied the $d=3$ case in detail. However, we can
equally easily determine the graviton propagator on the brane in general $d$.
Because of the logarithmic property of the 2-dimensional propagator, the
$d=2$ case requires a separate treatment, while the $d>2$ cases can be 
discussed together. In this appendix we provide the corresponding details.

\subsection{The $d>2$ Cases}

{}For the reader's convenience, here we repeat the relevant equations: 
\begin{eqnarray}
 &&{\overline H}_{\mu\nu}\equiv H_{\mu\nu}-{1\over{D-d}}\eta_{\mu\nu}
 H=\nonumber\\
 &&-M_P^{2-D}\left[\left(T_{\mu\nu}(p)-{1\over {D-d}}
 \eta_{\mu\nu}T(p)\right)\Phi+{d\over{D-2}}L^d
 \left(p_\mu p_\nu-{1\over{D-d}}
 \eta_{\mu\nu}p^2\right)T(p)\Sigma\right]~,\\
 &&H=-{{d-2}\over d}\chi=-{{d-2}\over{D-2}}M_P^{2-D}T(p){\widetilde \Phi}~.
\end{eqnarray}
Here $\Phi$, $\Sigma$ and ${\widetilde\Phi}$ are the solutions to the
following equations
\begin{eqnarray}\label{Phi1}
 &&\left(
 \partial^i\partial_i-p^2\right)\Phi=
 \left[1+L^dp^2\Phi\right]f(r)~,\\
 &&\left(
 \partial^i\partial_i-p^2\right)\Sigma=
 \left[{\widetilde \Phi}+L^dp^2\Sigma\right] f(r)~,\\
 &&\left(
 \partial^i\partial_i-p^2\right){\widetilde \Phi}=
 \left[1-\kappa L^d p^2{\widetilde \Phi}\right]f(r)
\end{eqnarray}
subject to the condition that $\Phi$, $\Sigma$ and ${\widetilde \Phi}$ 
decay to zero away from the brane (here we are focusing on the
cases $d>2$, where this is true even for the $p^2=0$ modes). 
As before, we are using the notation
\begin{equation}
 \kappa\equiv {(d-1)(D-d-2)\over{D-2}}
\end{equation}
to simplify expressions containing ${\widetilde\Phi}$.

{}Let us solve for $H_{\mu\nu}$ and $\chi$ on the brane, that is, at 
$r=\epsilon$. Here we will assume that $\epsilon\ll L$. Consider the
momenta with $p^2\ll 1/\epsilon^2$. Then it is not difficult to see that,
up to ${\cal O}(p\epsilon)$ corrections, we have:
\begin{eqnarray}
 &&\Phi(r)\approx A~,~~~r\leq\epsilon~,\\
 &&\Phi(r)\approx {B\over r^{d-2}}~,~~~r
 {\ \lower-1.2pt\vbox{\hbox{\rlap{$>$}\lower5pt\vbox{\hbox{$\sim$}}}}\ }
 \epsilon~,
\end{eqnarray}
where $A$ and $B$ are constants.
Matching the values of $\Phi(r)$ at $r=\epsilon$ gives
\begin{equation}
 B\approx \epsilon^{d-2} A~. 
\end{equation}
On the other hand, from (\ref{Phi1}) we obtain:
\begin{equation}
 -(d-2){B\over\epsilon^{d-1}}\approx \left[1+L^dp^2 A\right]{1\over 
 a_{d-1}\epsilon^{d-1}}~.
\end{equation}
We, therefore, have
\begin{equation}
 \Phi(\epsilon)\approx -{1\over L^d}~{1\over{p^2+p_c^2}}~,
\end{equation}
where
\begin{equation}
 p_c^2\equiv {(d-2)a_{d-1}\epsilon^{d-2}\over L^d}~.
\end{equation}
Similarly, for ${\widetilde\Phi}$ and $\Sigma$ we have:
\begin{eqnarray}
 &&{\widetilde \Phi}(\epsilon)\approx {1\over \kappa L^d}~
 {1\over {p^2-p_*^2}}~,\\
 &&\Sigma(\epsilon)\approx -{1\over\kappa L^{2d}}~{1\over{p^2+p_c^2}}~
 {1\over {p^2-p_*^2}}~,
\end{eqnarray}
where 
\begin{equation}
 p_*^2\equiv {(d-2)a_{d-1}\epsilon^{d-2}\over\kappa L^d}~.
\end{equation}
On the brane we then have:
\begin{eqnarray}\label{graviton4}
 &&H_{\mu\nu}(r=\epsilon)\approx {{\widehat M}_P^{2+d-D}\over {p^2+p_c^2}}
 \biggl[T_{\mu\nu}(p)-{1\over{D-d-2}}\eta_{\mu\nu} T(p)+\nonumber\\
 &&{d\over(d-1)(D-d-2)}\left(p_\mu p_\nu-\eta_{\mu\nu}
 {{(d-2)p_c^2+2(d-1)p_*^2}\over d(D-d)}\right){T(p)\over{p^2-p_*^2}}
 \biggr]~,\\
 &&\chi(r=\epsilon)\approx {d\over (d-1)(D-d-2)}~
 {{\widehat M}_P^{2+d-D}\over{p^2-p_*^2}}~T(p)~.\label{chi4}
\end{eqnarray}
Note that at the momenta $p^2\gg p_c^2$ the graviton propagator on the brane is
$(D-d)$-dimensional. 

\subsection{The $d=2$ Case}

{}Once again, let us assume that $\epsilon\ll L$, and consider the
momenta with $p^2\ll 1/\epsilon^2$. Then it is not difficult to see that,
up to ${\cal O}(p\epsilon)$ corrections, we have:
\begin{eqnarray}
 &&\Phi(r)\approx A~,~~~r\leq\epsilon~,\\
 &&\Phi(r)\approx B\ln\left(r\over R\right)~,~~~r
 {\ \lower-1.2pt\vbox{\hbox{\rlap{$>$}\lower5pt\vbox{\hbox{$\sim$}}}}\ }
 \epsilon~,
\end{eqnarray}
where $A$, $B$ and $R$ are constants.
Matching the values of $\Phi(r)$ at $r=\epsilon$ gives
\begin{equation}
 B\approx A \ln^{-1}\left(\epsilon\over R\right)~. 
\end{equation}
On the other hand, from (\ref{Phi1}) we obtain:
\begin{equation}
 {B\over\epsilon}\approx \left[1+L^2 p^2 A\right]{1\over 
 2\pi \epsilon }~.
\end{equation}
We, therefore, have
\begin{equation}
 \Phi(\epsilon)\approx -{1\over L^2}~{1\over{p^2+p_c^2}}~,
\end{equation}
where
\begin{equation}
 p_c^2\equiv {2\pi \over L^2}~\ln^{-1}\left(R\over\epsilon\right)~.
\end{equation}
Similarly, for ${\widetilde\Phi}$ we have:
\begin{eqnarray}
 &&{\widetilde \Phi}(\epsilon)\approx {1\over \kappa L^2}~
 {1\over {p^2-p_*^2}}~,\\
 &&\Sigma(\epsilon)\approx -{1\over\kappa L^{4}}~{1\over{p^2+p_c^2}}~
 {1\over {p^2-p_*^2}}~,
\end{eqnarray}
where 
\begin{equation}
 p_*^2\equiv {2\pi\over\kappa L^2}~\ln^{-1}\left(\xi R\over\epsilon\right)~,
\end{equation}
where $\xi$ is a dimensionless parameter.
On the brane we then have:
\begin{eqnarray}\label{graviton5}
 &&H_{\mu\nu}(r=\epsilon)\approx {{\widehat M}_P^{4-D}\over {p^2+p_c^2}}
 \biggl[T_{\mu\nu}(p)-{1\over{D-4}}\eta_{\mu\nu} T(p)+\nonumber\\
 &&{2\over{D-4}}\left(p_\mu p_\nu-\eta_{\mu\nu}
 {{p_*^2}\over {D-2}}\right){T(p)\over{p^2-p_*^2}}
 \biggr]~,\\
 &&\chi(r=\epsilon)\approx {2\over {D-4}}~
 {{\widehat M}_P^{4-D}\over{p^2-p_*^2}}~T(p)~.\label{chi5}
\end{eqnarray}
Note that $H_{\mu\nu}$ is traceless in this case. Also,
note that at the momenta $p^2\gg p_c^2$ the graviton propagator on the brane is
$(D-2)$-dimensional. If we take $R\sim L$, then we can have 
$p_c\ll 1/L$. Note, however, that in this case 
$p_c$ is only logarithmically small compared with $1/L$. On the other hand, if
we take $R\rightarrow\infty$, then $p_c,p_*\rightarrow 0$, and there is no
cross-over between the $(D-2)$-dimensional gravity and the 
$D$-dimensional gravity
(that is, gravity on the brane is always $(D-2)$-dimensional). Note that
$R$ is a tree-level infra-red cut-off. We can then take the 
$R\rightarrow\infty$ limit in the cases where such a cut-off is absent.

\section{A Non-zero Tension Codimension-2 Brane}

{}In this appendix we discuss the case of a codimension-2 brane with
positive tension. In $d=2$ the equations of motion for the warp factors $A$ and
$B$ discussed in subsection D of section III simplify as follows:
\begin{eqnarray}
 &&2A^\prime B^\prime -(A^\prime)^2-A^{\prime\prime}+{1\over r}A^\prime=0~,\\
 &&A^{\prime\prime}+{1\over 2}(D-1)(A^\prime)^2-A^\prime B^\prime=0~,\\
 &&(D-3)\left[A^{\prime\prime}+{1\over 2}(D-2)(A^\prime)^2+{1\over r}A^\prime
 \right]+B^{\prime\prime}+{1\over r}B^\prime+{1\over 2}{\widehat\tau}
 f(r)=0~.
\end{eqnarray}
It is then not difficult to see that for non-zero ${\widehat \tau}$ the only
solution is given by $A\equiv 0$ (more precisely, $A$ should be constant, 
which we can set to zero), and
\begin{equation}
 B^{\prime\prime}+{1\over r}B^\prime+{1\over 2}{\widehat\tau}
 f(r)=0~.
\end{equation}
The smooth
solution to this equation is given by (we have set the integration 
constant so that $B$ vanishes on the brane):
\begin{equation}
 B(r)=-\lambda\theta(r-\epsilon)\ln\left(r\over\epsilon\right)~,
\end{equation}
where
\begin{equation}
 \lambda\equiv{{\widehat\tau}\over 4\pi}~.
\end{equation}
Note that the space is locally flat, but for $r>\epsilon$ we have a deficit
angle equal
\begin{equation}
 2\pi\lambda={1\over 2}{\widehat \tau}~.
\end{equation}
We will therefore assume that $0<{\widehat \tau}<4\pi$.

{}In this background we can also solve the linearized equations of motion
for the graviton and the graviscalars \cite{codi2}. Thus, we have (note that
$H_{\mu\nu}$ is traceless):
\begin{eqnarray}
 &&H_{\mu\nu}=\nonumber\\
 &&-M_P^{2-D}\left[\left(T_{\mu\nu}(p)-{1\over {D-2}}
 \eta_{\mu\nu}T(p)\right)\Phi+{2\over{D-2}}L^2
 \left(p_\mu p_\nu-{1\over{D-2}}
 \eta_{\mu\nu}p^2\right)T(p)\Sigma\right]~,\\
 &&\chi={2\over{D-2}}M_P^{2-D}T(p){\widetilde \Phi}~.
\end{eqnarray}
Here $\Phi$, $\Sigma$ and ${\widetilde\Phi}$ are the solutions to the
following equations ($\kappa=(D-4)/(D-2)$):
\begin{eqnarray}
 &&\left(
 \partial^i\partial_i-e^{2B}p^2\right)\Phi=
 \left[1+L^2p^2\Phi\right]f(r)~,\\
 &&\left(
 \partial^i\partial_i-e^{2B}p^2\right)\Sigma=
 \left[{\widetilde \Phi}+L^2p^2\Sigma\right] f(r)~,\\
 &&\left(
 \partial^i\partial_i-e^{2B}p^2\right){\widetilde \Phi}=
 \left[1-\kappa L^2 p^2{\widetilde \Phi}\right]f(r)~.
\end{eqnarray}
It is not difficult to see that the $e^{2B}$ factor does not affect the
discussion of subsection 2 of Appendix C for momenta $p^2\ll 1/\epsilon^2$
(assuming that $\epsilon\ll L$). In particular, all the equations and 
conclusions given there are unmodified up to ${\cal O}(p\epsilon)$
corrections.

\newpage
\begin{figure}[t]
\hspace*{3 cm} 
\epsfxsize=10 cm
\epsfbox{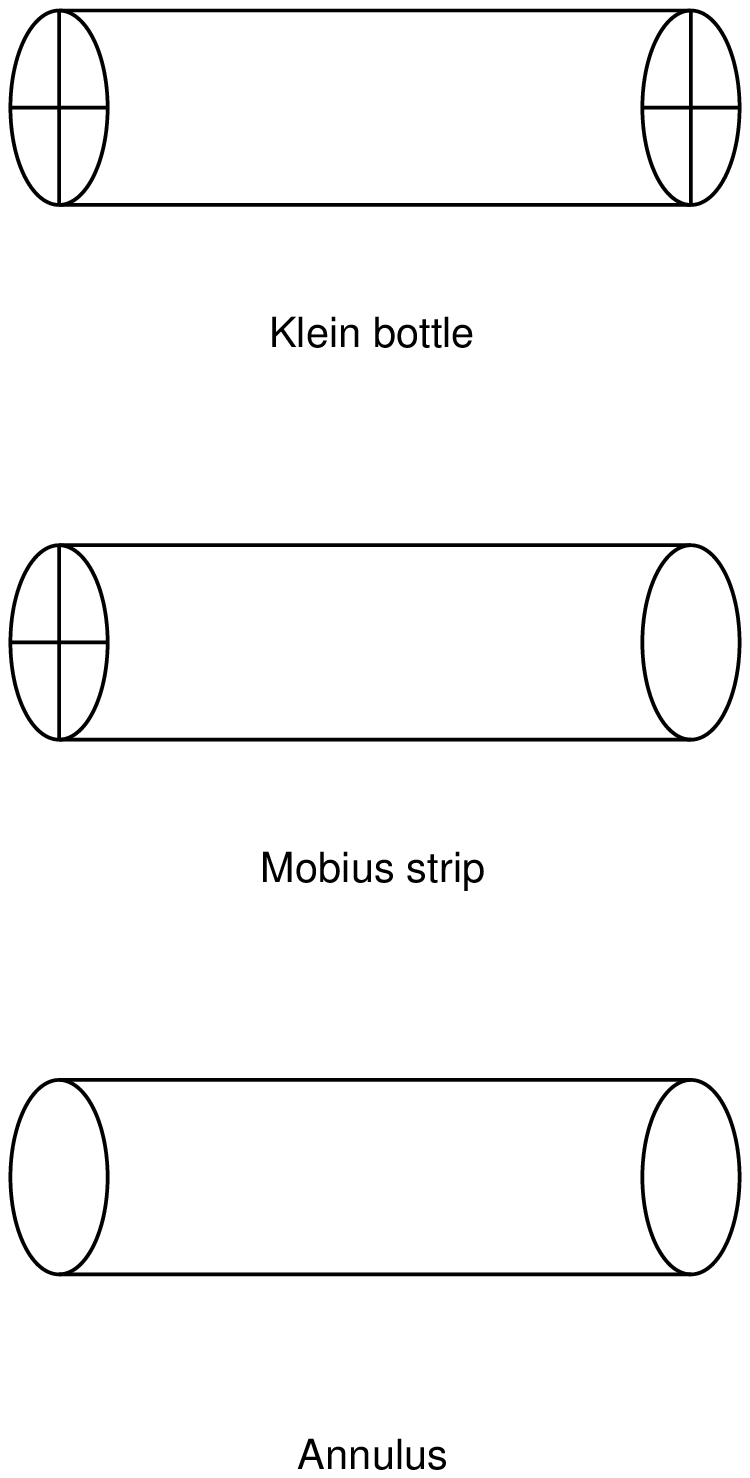}
\caption{Tree-channel Klein bottle, M{\"o}bius strip and annulus amplitudes.}
\end{figure}

\newpage
\begin{figure}[t]
\hspace*{3 cm}
\epsfxsize=10 cm
\epsfbox{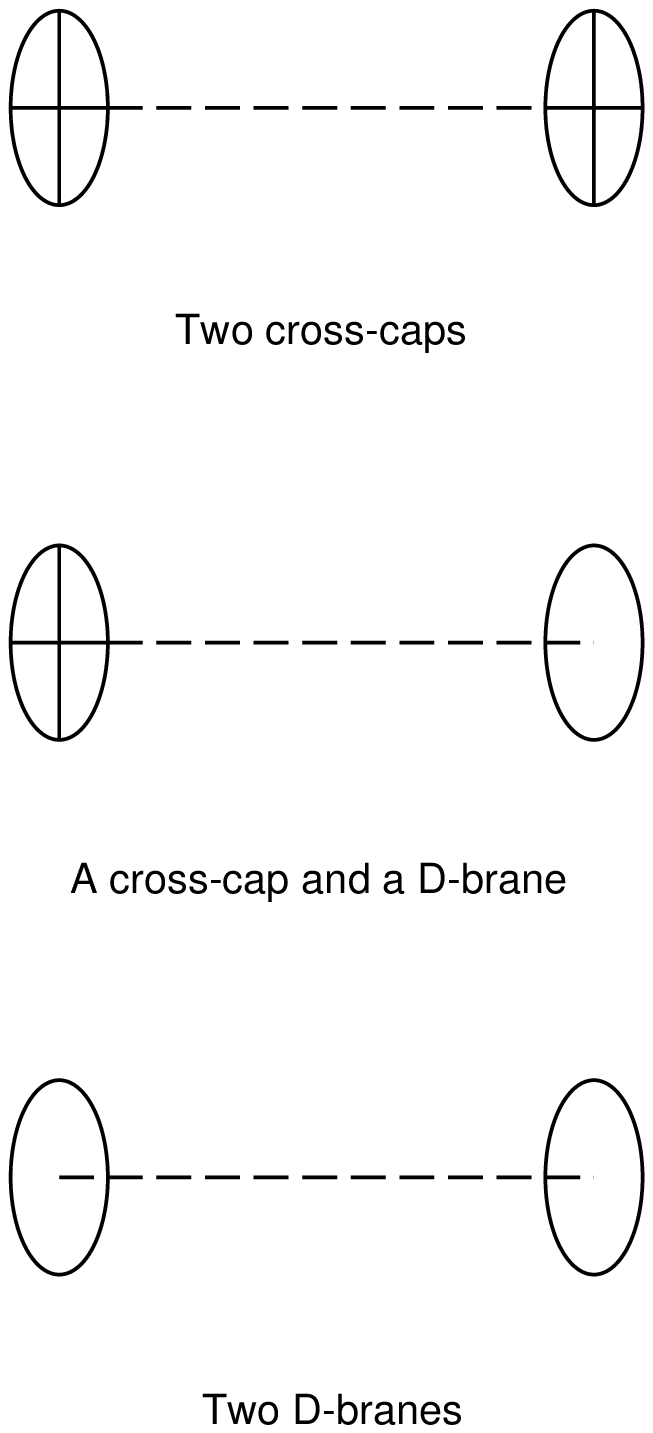}
\caption{Factorization of the Klein bottle, M{\"o}bius strip and annulus 
amplitudes.}
\end{figure}

\newpage

\begin{table}[t]
\begin{tabular}{|c|c|l|c|}
 Model & Gauge Group & \phantom{Hy} Charged  & Twisted Sector 
  \\
       &                &Chiral Multiplets & Chiral Multiplets
 \\
\hline
${\bf Z}_3$ & $U(N)\otimes G_\eta(N+4\eta)$  & 
 $3\times ({\bf R}_\eta,{\bf 1})(+2)$ & $1$
 \\
                &  & $3\times ({\overline {\bf N}},{\bf N+4\eta})(-1)$ &  \\
\hline
${\bf Z}_7$ & $U(N)\otimes U(N)\otimes U(N)\otimes$  & 
 $({\bf R}_\eta,{\bf 1},{\bf 1},{\bf 1})(+2,0,0)$ & $3$
  \\
            &  $G_\eta(N-4\eta)$ & $({\bf N},{\bf 1},{\bf 1},{\bf N-4\eta})
 (+1,0,0)$ &  \\
               &  & $({\overline {\bf N}},{\bf N},{\bf 1},{\bf 1})
 (-1,+1,0)$ &  \\
              &  & $({\overline {\bf N}},{\overline {\bf N}},{\bf 1},{\bf 1})
 (-1,-1,0)$ &  \\
 & & plus cyclic permutations of the& \\
 & & $U(N)\otimes U(N)\otimes U(N)$ representations &\\  
\hline
${\bf Z}_3\times {\bf Z}_3$ & $U(N)\otimes U(N)\otimes U(N)\otimes$  & 
 $({\bf R}_\eta,{\bf 1},{\bf 1},{\bf 1},{\bf 1})(+2,0,0,0)$ & $7$
 \\
                & $U(N+2\eta)\otimes 
 G_\eta(N)$ & $({\overline {\bf N}},{\bf 1},{\bf 1},{\bf 1},{\bf N})
 (-1,0,0,0)$ &  \\
 & & $({\bf 1},{\overline {\bf N}},{\bf 1},{\bf N+2\eta},{\bf 1})
 (0,-1,0,+1)$ &\\
 & & $({\bf 1},{\bf 1},{\overline {\bf N}},{\overline {\bf N+2\eta}},{\bf 1})
 (0,0,-1,-1)$ &\\
 & & $({\bf 1},{\bf N},{\bf N},{\bf 1},{\bf 1})(0,+1,+1,0)$ &\\
 & & plus cyclic permutations of the& \\
 & & $U(N)\otimes U(N)\otimes U(N)$ representations &\\
\hline
\end{tabular}
\caption{The massless spectra of ${\cal N}=1$ orientifolds of Type IIB on 
${\bf C}^3/{\bf Z}_3$, ${\bf C}^3/{\bf Z}_7$ and ${\bf C}^3/({\bf Z}_3\times
{\bf Z}_3)$. Here ${\bf R}_\eta={\bf A}$ (two-index $N(N-1)/2$ 
dimensional anti-symmetric representation of $U(N)$) for $\eta=-1$,
and ${\bf R}_\eta={\bf S}$ (two-index $N(N+1)/2$ dimensional symmetric 
representation of $U(N)$) for $\eta=+1$.
The $U(1)$ charges of the states in the 33 open string sector are
given in parentheses. The untwisted closed string sector states are not shown.}
\label{spectrum3} 
\end{table}

\begin{table}[t]
\begin{tabular}{|c|c|l|c|}
 Model & Gauge Group & \phantom{Hy} Charged  & Twisted Sector 
  \\
       &                &Chiral Multiplets & Chiral Multiplets
 \\
\hline
${\bf Z}_3\times{\bf Z}_2$ & $[U(N)\otimes U(N)\otimes U(N-2)]_{33}$  & 
 $2 \times ({\bf A},{\bf 1},{\bf 1})(+2,0,0)_{33}$ & $3$
 \\
              &  $ \otimes[U(2)\otimes U(2)]_{77}$
                      & $2 \times ({\bf 1},{\overline {\bf A}},{\bf 1})
 (0,-2,0)_{33}$ &  \\
              &    & $2 \times ({\overline {\bf N}},{\bf 1},{\overline 
 {\bf N-2}})(-1,0,-1)_{33}$ &  \\
              &    & $2 \times ({\bf 1},{\bf N},{\bf N-2})
 (0,+1,+1)_{33}$ &  \\
              &    & $ ({\bf N},{\overline {\bf N}},{\bf 1})
 (+1,-1,0)_{33}$ &  \\
              &    & $ ({\overline {\bf N}},{\bf 1},{\bf N-2})
 (-1,0,+1)_{33}$ &  \\
              &    & $ ({\bf 1},{\bf N},{\overline {\bf N-2}})
 (0+1,-1)_{33}$ &  \\
               &   & $2 \times ({\bf 1},{\bf 1})(+2,0)_{77}$ &  \\
              &    & $2 \times ({\bf 1},{\bf 1})(0,-2)_{77}$ &  \\
              &    & $ ({\bf 2},{\bf 2})(+1,-1)_{77}$ &  \\
              &    & $ ({\bf N}, {\bf 1},{\bf 1};{\bf 2},{\bf 1})
 (+1,0,0;+1,0)_{37}$ &  \\
              &    & $ ({\bf 1},{\bf 1},{\bf N-2};{\bf 1},{\bf 2})
 (0,0,+1;0,+1)_{37}$ &  \\
              &    & $ ({\bf 1},{\overline {\bf N}},{\bf 1};{\bf 1},{\bf 2})
 (0,-1,0;0,-1)_{37}$ &  \\
              &    & $ ({\bf 1},{\bf 1},{\overline {\bf N-2}};{\bf 2},{\bf 1})
 (0,0,-1;-1,0)_{37}$ &  \\
\hline
${\bf Z}_3\times{\bf Z}_2\times {\bf Z}_2$ & $[U(N)\otimes Sp(N-2)]_{33} 
 \otimes$  & 
 $3 \times ({\bf A},{\bf 1})(+2)_{33}$ & $7$
 \\
               &  $\bigotimes_{\alpha=1}^3 U(2)_{7_\alpha 7_\alpha}$   
                  & $3 \times ({\overline {\bf N}},{\bf N-2})(-1)_{33}$ &  \\
           &&       $3 \times ({\bf 1}_\alpha)(+2_\alpha)_{7_\alpha7_\alpha}$ 
 &\\
     & & $({\bf N},{\bf 1};{\bf 2}_\alpha)(+1;+1_\alpha)_{3 7_\alpha}$ &  \\
                 & & $({\bf 1},{\bf N-2};{\bf 2}_\alpha)(0;-1_\alpha)_{3 
 7_\alpha}$ &  \\
                &  & $({\bf 2}_\alpha;{\bf 2}_\beta)
 (+1_\alpha;+1_\beta)_{7_\alpha 7_\beta}$ &  \\
\hline
\end{tabular}
\caption{The massless spectra of the ${\cal N}=1$ orientifolds of Type IIB 
on ${\bf C}^3/
({\bf Z}_3\times {\bf Z}_2)$ and
${\bf C}^3/
({\bf Z}_3\times {\bf Z}_2\times {\bf Z}_2)$. 
 The semi-colon in the column ``Charged Chiral Multiplets'' 
separates the $33$ and $7_\alpha 7_\alpha$ 
representations. We are also using some of
the notations from Table I.  
The untwisted closed string sector states are not shown.}
\label{Z2} 
\end{table}

\end{document}